\documentclass[fleqn,usenatbib]{mnras}

\defcitealias{McVenn2020}{MV20a}
\defcitealias{McVenn2020_updated}{MV20b}

\usepackage[T1]{fontenc}
\usepackage{hyperref}

\DeclareRobustCommand{\VAN}[3]{#2}
\let\VANthebibliography\thebibliography
\def\thebibliography{\DeclareRobustCommand{\VAN}[3]{##3}\VANthebibliography}

\usepackage{graphicx}	
\usepackage{amsmath}	
\usepackage{amssymb}	
\usepackage{newtxtext,newtxmath}

\usepackage{booktabs,siunitx}
\usepackage{array}
\usepackage{graphicx} 
\usepackage{multirow}
\usepackage{hhline,booktabs}
\usepackage{makecell}
\usepackage{longtable}
\usepackage{ltablex}
\usepackage{tabularray}
\usepackage{lscape}
\usepackage{rotating}
\usepackage{pdflscape}
\usepackage{afterpage}
\usepackage{subfig}

\title[Signatures of Extended Structures in MW Dwarfs]{Small-scale stellar haloes: detecting low surface brightness features in the outskirts of Milky Way dwarf satellites}

\author[J. Jensen et al.]{Jaclyn Jensen$^{1}$\thanks{E-mail: jaclynjensen@uvic.ca},
Christian R. Hayes$^{2}$,
Federico Sestito$^{1}$,
Alan W. McConnachie$^{2,1}$,
Fletcher Waller$^{1}$,
\newauthor Simon E. T. Smith$^{1}$,
Julio Navarro$^{1}$,
Kim A. Venn$^{1}$
\\
$^{1}$Department of Physics \& Astronomy, University of Victoria, Victoria, BC, V8P 1A1, Canada \\
$^{2}$NRC Herzberg Astronomy \& Astrophysics, 5071 West Saanich Road, Victoria, BC, V9E 2E7, Canada \\
}

\date{Submitted XXX. Received YYY; in original form ZZZ}

\pubyear{2022}

\begin{document}
\label{firstpage}
\pagerange{\pageref{firstpage}--\pageref{lastpage}}
\maketitle

\begin{abstract}

Dwarf galaxies are valuable laboratories for dynamical studies related to dark matter and galaxy evolution, yet it is currently unknown just how physically extended their stellar components are. Satellites orbiting the Galaxy’s potential may undergo tidal stripping by the host, or alternatively, may themselves have accreted smaller systems whose debris populates the dwarf’s own stellar halo. Evidence of these past interactions, if present, is best searched for in the outskirts of the satellite. However, foreground contamination dominates the signal at these large radial distances, making observation of stars in these regions difficult. In this work, we introduce an updated algorithm for application to Gaia data that identifies candidate member stars of dwarf galaxies, based on spatial, color-magnitude and proper motion information, and which allows for an outer component to the stellar distribution. Our method shows excellent consistency with spectroscopically confirmed members from the literature despite having no requirement for radial velocity information. We apply the algorithm to all $\sim$60 Milky Way dwarf galaxy satellites, and we find 9 dwarfs (Bo\"otes~1, Bo\"otes~3, Draco~2, Grus~2, Segue~1, Sculptor, Tucana~2, Tucana~3, and Ursa Minor) that exhibit evidence for a secondary, low-density outer profile. We identify many member stars which are located beyond 5 half-light radii (and in some cases, beyond 10). We argue these distant stars are likely tracers of dwarf stellar haloes or tidal streams, though ongoing spectroscopic follow-up will be required to determine the origin of these extended stellar populations. 

\end{abstract}




\section{Introduction}
\label{sect:intro}

The early stages of galaxy formation involve the accumulation of baryonic and non-baryonic matter at conjunctions of the cosmic web, where the first proto-galaxies are suggested to have formed (\citealt{mo2010}). Baryons are accreted into dark matter haloes, and these halos grow hierarchically (\citealt{white_rees1978}; \citealt{frenk1998}). This general framework applies to galaxies of all sizes, including typical galaxies like the Milky Way (MW; $M_{*} \approx 5 \times 10^{10} M_{\odot}$) and Andromeda (M31), down to the least massive systems (ultra-faint dwarfs, UFDs; $M_{*} \lesssim~10^{2~-~5} M_{\odot}$; \citealt{simon2019}). 

The debris of past accretions can be traced in the stellar haloes of galaxies, where gravitational accelerations are low and dynamical timescales are long. Substantial observational evidence of accreted substructures is present in our own MW, as our stellar halo is populated with a multitude of surviving dwarf galaxies ($\sim$60 to date; \citealt{mcconnachie2012}) and disrupted satellites in the form of stellar streams (see \citealt{carlin2016} and \citealt{helmi2020} for recent reviews, and \citealt{mateu2023} for the most up-to-date compilation of known stellar streams in the MW halo). 

While these hierarchical processes are also expected to apply to smaller scale systems, little observational evidence for stellar haloes around dwarf galaxies currently exists (\citealt{deason2022}), and their observational study is complicated by their intrinsically low masses. They are particularly feeble systems: their shallow potential wells are sensitive to a multitude of internal (star formation, stellar feedback) and external processes (mergers, ram pressure stripping, tidal stripping) that can strongly influence their individual morphologies (e.g., see discussion in \citealt{higgs2021}). Over a dwarf's lifetime, these environmental and internal factors make it difficult to disentangle the origin of the stars which are present in the dwarf's outskirts. In addition, the intrinsic faintness of dwarfs means observations of the extended substructure in the dwarf's outskirts are challenging, especially when viewed against the dominant MW foreground populations. 

With the advent of the Gaia satellite (\citealt{gaia_collab_2016}) and its wealth of kinematic data for $\sim$1.5 billion stars in the most recent data release (DR3; \citealt{gaia_collab_2023}), major progress has been made to study the individual resolved stars of MW dwarf galaxy satellites. A significant challenge when studying MW satellites is the ability to effectively and appreciably remove MW foreground contamination. However, Gaia proper motions provide one way of overcoming this challenge, as stars moving with the bulk proper motion of the dwarf galaxy can be separated from MW stars moving in the foreground. Recent works using Bayesian techniques (applied to Gaia photometry and astrometry) have found high success in obtaining dwarf member samples, thereby enabling the measurement of systemic proper motions for most of these dwarf galaxies (e.g., \citealt{pace_li2019}, \citealt{McVenn2020, McVenn2020_updated}, \citealt{pace2022}, \citealt{qi2022}, and \citealt{battaglia2022}).

In some of these Bayesian framework examples, a primary assumption for the dwarf model is the spatial distribution of stars. In most cases, the stellar density of the dwarf is assumed to be an exponentially decaying function (e.g., \citealt{McVenn2020}; \citealt{battaglia2022}; \citealt{pace2022}). However, as the stellar distribution in the outskirts of dwarf galaxies remains unknown, this presumption may unduly restrict the detectability of stars in the dwarf’s stellar halo, should one exist. For example, satellite systems (as a consequence of their accretion into the MW's halo) are under the influence of the MW potential and susceptible to tidal forces, which can act to strip stars from the dwarf satellite. Additionally, recent simulations conducted by \citet{goater2024} and \citet{deason2022} showed that dwarf-dwarf mergers may exhibit a secondary, extended radial profile beyond the primary component of the dwarf galaxy. The extended secondary component in this case (depending on the mass ratio of the merger itself) may contain a large fraction of stars which originated from a past accretion event and may even exhibit lengthy stellar distributions similar to tidal tails.

Recently, an exciting observation was made by \citet{chiti2021} who uncovered evidence of an extended stellar halo in the MW satellite, Tucana 2 (Tuc2). Spectroscopically confirmed members of this UFD were identified out to significantly large radial distances ($\sim$9 circularized half-light radii, equating to a physical distance of 1 kpc). Given (i) the outskirt stars’ lower metallicity compared to stars in the main body of the dwarf, (ii) a lack of a velocity gradient across the body of Tuc2, and (iii) the extended orientation of stars which are perpendicular to Tuc2's orbit, the authors argue that these radially distant stars are likely evidence of a past accretion event (i.e., dwarf-dwarf merger). 

Motivated by this recent discovery, we initiated a search using Gaia to hunt for evidence of extended stellar structures around all of the known MW dwarf galaxies (and candidate systems). In this work, we expand the Bayesian methods of \citet[hereafter \citetalias{McVenn2020}]{McVenn2020} to include extended multi-component substructure in the membership criteria, thus allowing us to locate members at much larger radii than in previous studies. This paper is organized as follows. Section \ref{sect:data} describes the relevant properties of the MW dwarfs we examine and our quality cuts in the Gaia dataset. In Section \ref{sect:methods}, we discuss the primary differences in our analysis method between this work and the implementation of \citetalias{McVenn2020}. In Section \ref{sect:results}, we apply the algorithm and estimate the contamination rates. Section \ref{sect:discussion} provides a discussion of the systems for which we find evidence of extended populations (i.e., “haloes”) of stars. We conclude with a summary in Section \ref{sect:summary}.


\section{Data}
\label{sect:data}

\begin{figure*}
 \centering
 \includegraphics[width=0.75\textwidth]{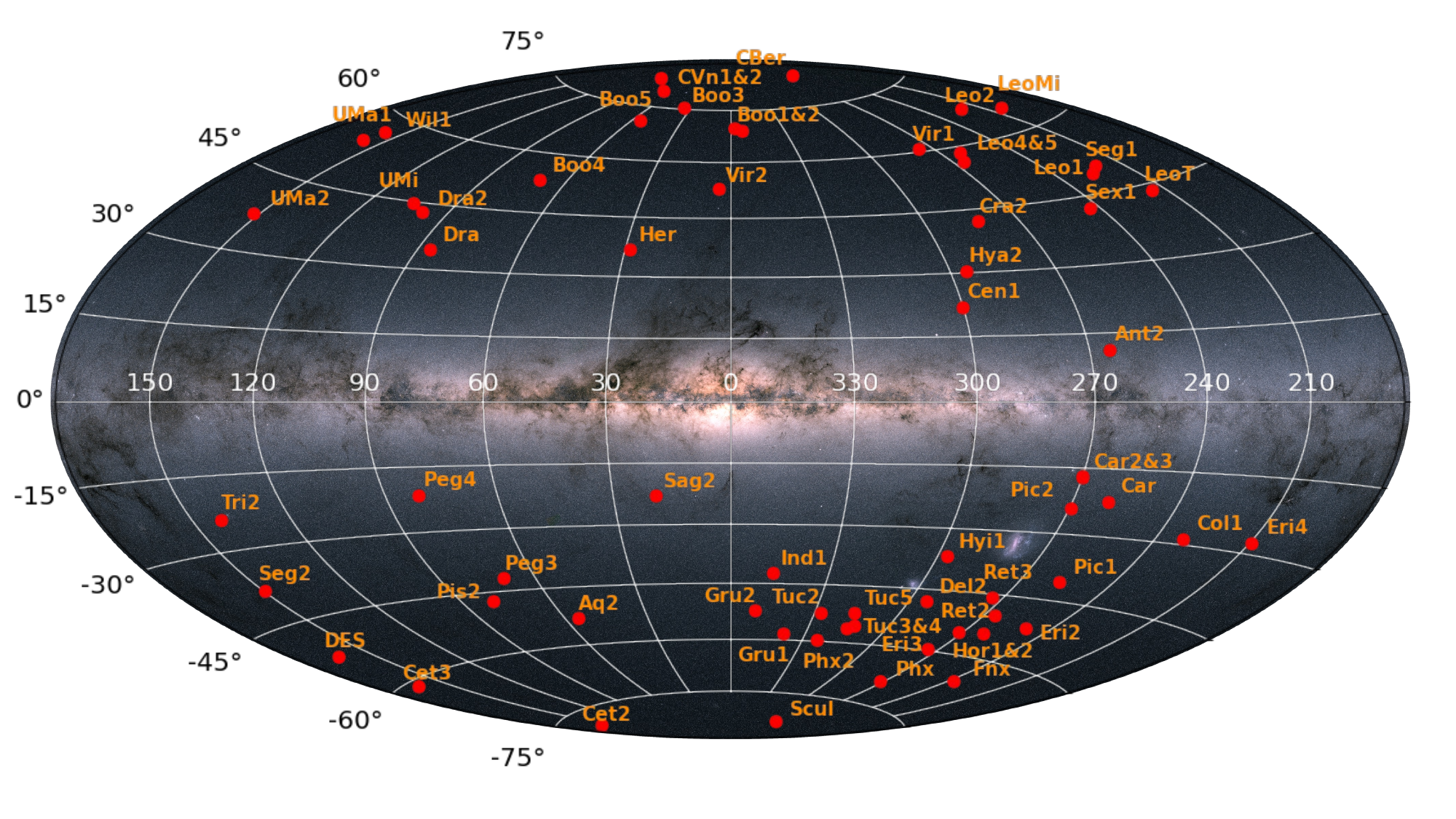}
 \caption{Galactic coordinates of the MW dwarf galaxy satellites used in this work, overlaid on the \href{https://www.esa.int/ESA_Multimedia/Images/2018/04/Gaia_s_sky_in_colour2}{Gaia stellar density map} (image credit: ESA/Gaia/DPAC).}
 \label{fig:galactic_plot}
\end{figure*}

We target all Milky Way (MW) dwarf galaxy satellites and likely candidates, as maintained in the tables originally published in \cite{mcconnachie2012}\footnote{\url{ https://www.cadc-ccda.hia-iha.nrc-cnrc.gc.ca/en/community/nearby/}}. However, we intentionally exclude the relatively massive Magellanic clouds, as well as the Sagittarius (Sgr) dwarf galaxy whose outskirts are so extended that they form the basis of an extensive literature by themselves (for the most recent Gaia-based work, see \citealt{gaia_collab_2018_kinematics} and \citealt{kallivayalil2013} for the Magellanic systems, and \citealt{ibata2020} and \citealt{ramos2022} for Sgr). Our dwarf inventory is therefore all other systems out to Leo T (heliocentric distance of $\sim$450 kpc), including a handful of recently discovered dwarfs. Newly discovered systems we include are: Bo\"otes~5 (\citealt{smith2022}, \citealt{cerny2022}), DELVE~J0155-6815 (\citealt{cerny2021_delve2}), Eridanus~4 (\citealt{cerny2021}), Leo~Minor and Virgo~2 (\citealt{cerny2022}), and Pegasus~4 (\citealt{cerny2023}). In Figure \ref{fig:galactic_plot}, we show the positions of all the candidate and confirmed dwarfs studied in this work in Galactic coordinates, and we list their names and positions in Table \ref{tab:dSphs_all}.

For each dwarf, we use the measured structural properties, distances, radial velocities, and metallicities as provided in the updated tables of \cite{mcconnachie2012}, with some exceptions. For the newly added systems, we use parameters as reported in the relevant detection papers. We also updated parameters for Bo\"otes~3 and Grus 1, both of which have recent structural measurements/updates (\citealt{moskowitz2020} and \citealt{cantu2021}, respectively). The parameters used for the newly added and updated systems are reported in Table \ref{tab:updates} in the Appendix.

While all dwarfs have measured distances and half-light radii, not all have measured ellipticities, position angles, radial velocities (RVs) and/or metallicities. We use all the aforementioned information for these systems, when available, to create the model likelihoods (discussed in Section \ref{sect:methods}) which represent the dwarf galaxy as a whole. For all dwarf parameters used in this work (structural parameters, positions, distance moduli, RVs, and metallicities) with the exception of updates/new dwarfs listed in Table \ref{tab:updates}, we refer the reader to Table 1 of \citetalias{McVenn2020}. 

Our analysis is based on the data available in Gaia early Data Release 3 (eDR3; \citealt{gaia_edr3_2021}). Specifically, we use stellar sources with all five astrometric solutions (positions, parallaxes, and proper motions) and photometry (G, B$_{P}$, and R$_{P}$ magnitudes). For most dwarfs, we select all such sources within 2$^{\circ}$ of the center of the dwarf, which corresponds to on-sky distances well beyond 10 half-light radii for any dwarf. However, for larger systems (Antlia 2, Bo\"otes~3, Carina, Crater 2, Fornax, Sculptor, Sextans 1, and Ursa Minor), we use a larger selection radius of 4 $-$ 9$^{\circ}$ to better capture any distant members. 

To select stars with reliable photometry and astrometry in Gaia eDR3, we require that they pass the same criteria discussed in \citet[hereafter \citetalias{McVenn2020_updated}]{McVenn2020_updated}. Specifically, the sources must have high-quality astrometry ({\tt ruwe} $<$ 1.3), as given by \citet{lindegren2018} and \citet{lindegren2021_astrometricsolution}. We de-redden the photometry using the dust maps in \cite{schlegel1998} and relevant extinction coefficients (see Equation 1 of \citealt{gaia_collab_2018_HRD}). We then correct the parallaxes of the stellar sources by the global zero-point of $-$0.017 mas (\citealt{lindegren2021}). As part of the membership selection algorithm, we filter the parallaxes to only select stars in the field that are at broadly consistent distances to the dwarf (i.e., stars whose 3-$\sigma$ parallax uncertainty places them within the 3-$\sigma$ range of the dwarf, inferred from the distance modulus). 


Following \citetalias{McVenn2020} and \citetalias{McVenn2020_updated}, we do not \textit{require} stellar radial velocities (RVs) to identify members.  However, we later use these data to validate the method and estimate contamination fractions (Section \ref{sect:RV_purity}) when RVs are available. The radial velocities used in this work are from:

\begin{enumerate}
    \item the APOGEE Survey (\citealt{APOGEE_majewski}) using the 17th data release of the Sloan Digital Sky Survey (SDSS; \citealt{blanton2017, APOGEE_DR17_abdurrouf2022}). The APOGEE instruments \citep{wilson2019} are located on the 2.5-m SDSS and 2.5-m DuPont telescopes \citep{bowen_vaughan1973, gunn2006} providing a dual-hemisphere coverage of Milky Way dwarf galaxies, whose targeting is described in \cite{zasowski2013, zasowski2017}, \cite{beaton2021}, and \cite{santana2021}. In this work we use the radial velocities from the APOGEE survey which are measured as described in \citet{nidever2015} and Holtzman et al. (in prep). These exist for a limited subset of the dwarfs. The number of APOGEE stars within the Gaia field of each dwarf are reported in the sixth column of Table \ref{tab:dSphs_all}; 
    \item a compilation of surveys targeting individual dwarf galaxies. The number of stars per Gaia field from these surveys are reported in the seventh column of Table \ref{tab:dSphs_all}.
\end{enumerate}

\noindent In practice, for stars with more than one RV measurement, we take the weighted mean and weighted standard deviation of the multiple RV measurements.

\begin{table*}
    \centering
    \renewcommand{\arraystretch}{1.1}
    \begin{tabular}{c|cccc|ccc}
    Galaxy & R.A. ($^{\circ}$) & Dec ($^{\circ}$) & $\ell$ ($^{\circ}$) & b ($^{\circ}$) & No. APOGEE stars &  No. Survey stars & Spec. Follow-up References \\
    \hline
        Antlia 2 (Ant2) &  143.887 & -36.767 &  264.895 &  11.248 &        1374 &     221 & \citet{torrealba2019} \\
        Aquarius 2 (Aq2) &  338.481 &  -9.328 &   55.107 & -53.009 &         616 &      11 & \citet{torrealba2016} \\
        Bootes 1 (Boo1) &  210.025 &  14.500 &  358.082 &  69.624 &         237 &     282 & \citet{munoz2006} \\
        & & & & & & & \citet{martin2007} \\
        & & & & & & & \citet{norris2008, norris2010_boo1_seg1} \\
        & & & & & & & \citet{koposov2011} \\
        & & & & & & & \citet{waller2023} \\
        Bootes 2 (Boo2) &  209.500 &  12.850 &  353.694 &  68.871 &         103 &       8 & \citet{koch2009} \\
        & & & & & & & \citet{ji2016_boo2} \\
        
        Bootes 3 (Boo3) &  209.300 &  26.800 &   35.405 &  75.353 &        1483 &    20 & \citet{carlin2018} \\
        *Bootes 4 (Boo4) &  233.689 &  43.726 &   70.682 &  53.305 &         524 &    $-$ & $-$ \\
        *Bootes 5 (Boo5) &  213.911 &  32.912 &   55.661 &  70.916 &        $-$ &    $-$ & $-$ \\
        Canes Venatici 1 (CVn1) &  202.015 &  33.556 &   74.305 &  79.822 &         217 &     145 & \citet{simon_geha2007} \\
        Canes Venatici 2 (CVn2) &  194.292 &  34.321 &  113.577 &  82.703 &         146 &      42 & \citet{martin2007} \\
        & & & & & & & \citet{simon_geha2007} \\
          Carina (Car) &  100.403 & -50.966 &  260.112 & -22.223 &         260 &    1855 & \citet{walker2009} \\
        Carina 2 (Car2) &  114.107 & -57.999 &  269.982 & -17.140 &          62 &     283 & \citet{li2018_car_UFDs} \\
        Carina 3 (Car3) &  114.630 & -57.900 &  270.006 & -16.846 &          32 &     283 & \citet{li2018_car_UFDs} \\
        *Centaurus 1 (Cen1) &  189.585 & -40.902 &  300.265 &  21.902 &        $-$ &    $-$ & $-$ \\
         *Cetus 2 (Cet2) &   19.470 & -17.420 &  156.465 & -78.531 &        $-$ &    $-$ & $-$ \\
         *Cetus 3 (Cet3) &   31.331 &  -4.270 &  163.810 & -61.133 &         116 &    $-$ & $-$ \\
        Columba 1 (Col1) &   82.860 & -28.030 &  231.621 & -28.879 &        $-$ &      49 & \citet{fritz2019} \\
        Coma Berenices (CBer) &  186.746 &  23.904 &  241.889 &  83.611 &         155 &      47 & \citet{simon_geha2007} \\
        & & & & & & & \citet{waller2023} \\
        Crater 2 (Cra2) &  177.310 & -18.413 &  282.908 &  42.028 &         621 &     404 & \citet{caldwell2017} \\
        & & & & & & & \citet{fu2019} \\
        DELVE J0155-6815 (DELVE2) &  28.772  & -68.253  & 294.236 & -47.786  &   $-$ & $-$ & $-$ \\
        *DESJ0225+0304 (DES) &   36.427 &   3.069 &  163.582 & -52.201 &          25 &    $-$ & $-$ \\
           Draco (Dra) &  260.052 &  57.915 &   86.367 &  34.722 &        1246 &    1420 & \citet{walker2009} \\
         Draco 2 (Dra2) &  238.198 &  64.565 &   98.294 &  42.880 &          43 &      45 & \citet{martin2016} \\
         & & & & & & & \citet{longeard2018} \\
        Eridanus 2 (Eri2) &   56.088 & -43.533 &  249.781 & -51.646 &        $-$ &      44 & \citet{li2017} \\
        & & & & & & & \citet{zoutendijk2020} \\
        *Eridanus 3 (Eri3) &   35.690 & -52.284 &  274.960 & -59.599 &        $-$ &    $-$ & $-$ \\
        *Eridanus 4 (Eri4) &   76.438 &  -9.515 &  209.499 & -27.772 &        $-$ &    $-$ & $-$ \\
          Fornax (Fnx) &   39.997 & -34.449 &  237.103 & -65.652 &         263 &    2486 & \citet{walker2009} \\
        Grus 1 (Gru1) &  344.177 & -50.163 &  338.679 & -58.245 &        $-$ &      79 & \citet{walker2016} \\
          Grus 2 (Gru2) &  331.020 & -46.440 &  351.143 & -51.939 &         234 &     256 & \citet{simon2020} \\
        Hercules (Her) &  247.758 &  12.792 &   28.728 &  36.872 &          99 &      81 &  \citet{simon_geha2007} \\
        & & & & & & & \citet{aden2009} \\
        & & & & & & & \citet{deason2012} \\
        & & & & & & & \citet{gregory2020} \\
        Horologium 1 (Hor1) &   43.882 & -54.119 &  271.388 & -54.733 &        $-$ &      19 & \citet{nagasawa2018} \\
        Horologium 2 (Hor2) &   49.134 & -50.018 &  262.472 & -54.137 &        $-$ &      92 & \citet{fritz2019} \\
         Hydra 2 (Hya2) &  185.425 & -31.985 &  295.617 &  30.464 &        $-$ &      19 & \citet{kirby2015_hydra_pisc} \\
        Hydrus 1 (Hyi1) &   37.389 & -79.309 &  297.416 & -36.746 &        $-$ &     139 & \citet{koposov2018} \\
         *Indus 1 (Ind1) &  317.205 & -51.166 &  347.156 & -42.071 &        $-$ &    $-$ & $-$ \\
           Leo 1 (Leo1) &  152.117 &  12.306 &  225.986 &  49.112 &         223 &     382 & \citet{mateo2008} \\
           Leo 2 (Leo2) &  168.370 &  22.152 &  220.168 &  67.231 &         275 &     220 & \citet{spencer2017} \\
           Leo 4 (Leo4) &  173.238 &  -0.533 &  265.442 &  56.515 &         239 &      27 & \citet{simon_geha2007} \\
           Leo 5 (Leo5) &  172.790 &   2.220 &  261.862 &  58.537 &         246 &     122 & \citet{walker2009_leo5} \\
        *Leo Minor (LeoMi) &  164.261 &  28.875 &  202.232 &  64.750 &        $-$ &    $-$ & $-$ \\
            Leo T (LeoT) &  143.722 &  17.051 &  214.853 &  43.660 &        $-$ &      42 & \citet{simon_geha2007} \\
        Pegasus 3 (Peg3) &  336.094 &   5.420 &   69.852 & -41.813 &        $-$ &    $-$ & $-$ \\
        Pegasus 4 (Peg4) &  328.539 &  26.620 &   80.797 & -21.403 &        $-$ &    24 & \citet{cerny2023}  \\
         Phoenix (Phx) &   27.776 & -44.445 &  272.160 & -68.949 &        $-$ &     118 & \citet{kacharov2017} \\
        Phoenix 2 (Phx2) &  354.997 & -54.406 &  323.687 & -59.749 &        $-$ &      75 & \citet{fritz2019} \\
   \hline
    \end{tabular}
    \caption{The Milky Way dwarf satellites explored in this work (1; marked with * if no systemic RV is reported for the dwarf) and their equatorial positions in the sky (2, 3). Galactic coordinates are included in (4, 5). We highlight the number of stars in a given dwarf field which also contain RV information from APOGEE DR17 (6), as well as the number of stars with RVs available in focused dwarf studies (7). Relevant references for (7) are listed in the final column (8).}
    \label{tab:dSphs_all}
\end{table*}

\begin{table*}
    \centering
    \renewcommand{\arraystretch}{1.1}
    \begin{tabular}{c|cccc|ccc}
    Galaxy & R.A. ($^{\circ}$) & Dec ($^{\circ}$) & $\ell$ ($^{\circ}$) & b ($^{\circ}$) & No. APOGEE stars &  No. Survey stars & Spec. Follow-up References \\
    \hline
    *Pictor 2 (Pic2) &  101.180 & -59.897 &  269.633 & -24.052 &        1150 &    $-$ & $-$ \\
    *Pictoris 1 (Pic1) &   70.948 & -50.283 &  257.299 & -40.645 &        $-$ &    $-$ & $-$ \\
    Pisces 2 (Pis2) &  344.629 &   5.952 &   79.206 & -47.106 &          78 &       7 & \citet{kirby2015_hydra_pisc} \\
     Reticulum 2 (Ret2) &   53.925 & -54.049 &  266.296 & -49.736 &        $-$ &      59 & \citet{koposov2015} \\
     & & & & & & & \citet{simon2015} \\
     & & & & & & & \citet{walker2015} \\
     Reticulum 3 (Ret3) &   56.360 & -60.450 &  273.878 & -45.648 &        $-$ &      45 & \citet{fritz2019} \\
   Sagittarius 2 (Sag2) &  298.169 & -22.068 &   18.936 & -22.898 &        $-$ &     121 & \citet{longeard2020, longeard2021} \\
       Sculptor (Scl) &   15.039 & -33.709 &  287.535 & -83.157 &        248 &    1462 & \citet{tolstoy2004} \\
       & & & & & & & \citet{walker2009} \\
         Segue 1 (Seg1) &  151.767 &  16.082 &  220.478 &  50.426 &          21 &     327 & \citet{simon2011} \\
         & & & & & & & \citet{norris2010_boo1_seg1} \\
         & & & & & & & \citet{geha2009} \\
         Segue 2 (Seg2) &   34.817 &  20.175 &  149.433 & -38.135 &        $-$ &     219 & \citet{belokurov2009} \\
         & & & & & & & \citet{kirby2013} \\
       Sextans 1 (Sex1) &  153.262 &  -1.615 &  243.498 &  42.272 &        1517 &     914 & \citet{walker2009} \\
    Triangulum 2 (Tri2) &   33.322 &  36.178 &  140.899 & -23.822 &         435 &      28 & \citet{kirby2015} \\
    & & & & & & & \citet{martin2016_tri2} \\
    & & & & & & & \citet{kirby2017} \\
        Tucana 2 (Tuc2) &  342.980 & -58.569 &  328.086 & -52.325 &        $-$ &      95 & \citet{walker2016} \\
        & & & & & & & \citet{chiti2021} \\
        Tucana 3 (Tuc3) &  359.150 & -59.600 &  315.376 & -56.184 &        $-$ &     675 & \citet{simon2017} \\
        & & & & & & & \citet{li2018} \\
        Tucana 4 (Tuc4) &    0.730 & -60.850 &  313.287 & -55.292 &        $-$ &    81 & \citet{simon2020} \\
        Tucana 5 (Tuc5) &  354.350 & -63.270 &  316.310 & -51.892 &        $-$ &    29 & \citet{simon2020} \\
     UrsaMajor 1 (UMa1) &  158.720 &  51.920 &  159.431 &  54.414 &         100 &     106 & \citet{kleyna2005} \\
     & & & & & & & \citet{martin2007} \\
     & & & & & & & \citet{simon_geha2007} \\
     UrsaMajor 2 (UMa2) &  132.875 &  63.130 &  152.464 &  37.443 &          86 &    42 & \citet{martin2007}  \\
     & & & & & & & \citet{simon_geha2007} \\
       Ursa Minor (UMi) &  227.285 &  67.222 &  104.965 &  44.801 &        273 &     958 & \citet{spencer2018} \\
         *Virgo 1 (Vir1) &  180.038 &   0.681 &  275.852 &  60.827 &         330 &    $-$ & $-$ \\
         *Virgo 2 (Vir2) &  225.059 &   5.909 &    4.067 &  52.754 &        $-$ &    $-$ & $-$ \\
       Willman 1 (Wil1) &  162.337 &  51.050 &  158.577 &  56.780 &         375 &  97 & \citet{martin2007} \\
       & & & & & & & \citet{willman2011} \\
    \bottomrule
    \hline
    \end{tabular}
    \contcaption{from previous page.}
\end{table*}

\section{Methods}
\label{sect:methods}

Our methodology is an extension of \citetalias{McVenn2020} (inspired by \citealt{pace_li2019}; see other similar implementations in \citealt{pace2022}, \citealt{qi2022}, and \citealt{battaglia2022}), to which we direct the reader for specifics in the approach of the analysis. In summary, we assume every star in a given field is a member of one of two populations: (a) the satellite, or (b) the MW foreground. A star’s total likelihood is therefore given as:

\begin{equation}
 \mathcal{L} = f_{sat} \mathcal{L}_{sat} + (1 - f_{sat}) \mathcal{L}_{MW}
 \label{eq:likelihoods}
\end{equation}

\noindent where $\mathcal{L}_{MW}$ and $\mathcal{L}_{sat}$ are the likelihoods of the MW and dwarf models respectively, and $f_{sat}$ represents the fraction of stars in the dataset that are members of the dwarf. In our Bayesian framework, the two likelihoods in Equation \ref{eq:likelihoods} are based on three main components: (i)  the stellar color and apparent magnitude ($\mathcal{L}_{CMD}$); (ii) the radial distance of the star from the dwarf's center ($\mathcal{L}_{S}$); (iii) the stellar proper motions ($\mathcal{L}_{PM}$), such that:

\begin{equation}
 \mathcal{L}_{sat} = \mathcal{L}_{CMD} \mathcal{L}_{PM} \mathcal{L}_{S}
\end{equation}

\noindent The three likelihood functions $\mathcal{L}_{S}$, $\mathcal{L}_{CMD}$, and $\mathcal{L}_{PM}$ are constructed for both the satellite and MW foreground populations. The final probability of a star being a member of a dwarf is therefore:

\begin{equation}
 P_{sat} = \frac{f_{sat} \mathcal{L}_{sat}}{f_{sat} \mathcal{L}_{sat} + (1 - f_{sat}) \mathcal{L}_{MW}}
\end{equation}

For each aforementioned likelihood function, we map each as a 2-D “look-up map”. The method to create the look-up maps for the MW foreground models are the same as in \citetalias{McVenn2020}. However, our methodology for the dwarf model varies from the original implementation. In the following subsections, we highlight the differences between the \citetalias{McVenn2020} paper and our approach. All dwarf parameters that we use in the satellite model are adopted from Table 1 in \citetalias{McVenn2020}, with the exceptions of the new and updated systems described in Section \ref{sect:data} and listed in Table \ref{tab:updates}.



\subsection{Inclusion of the Horizontal Branch to $\mathcal{L}_{CMD}$}
\label{sect:HB}

\begin{figure*}
 \centering
 \includegraphics[width=0.9\textwidth]{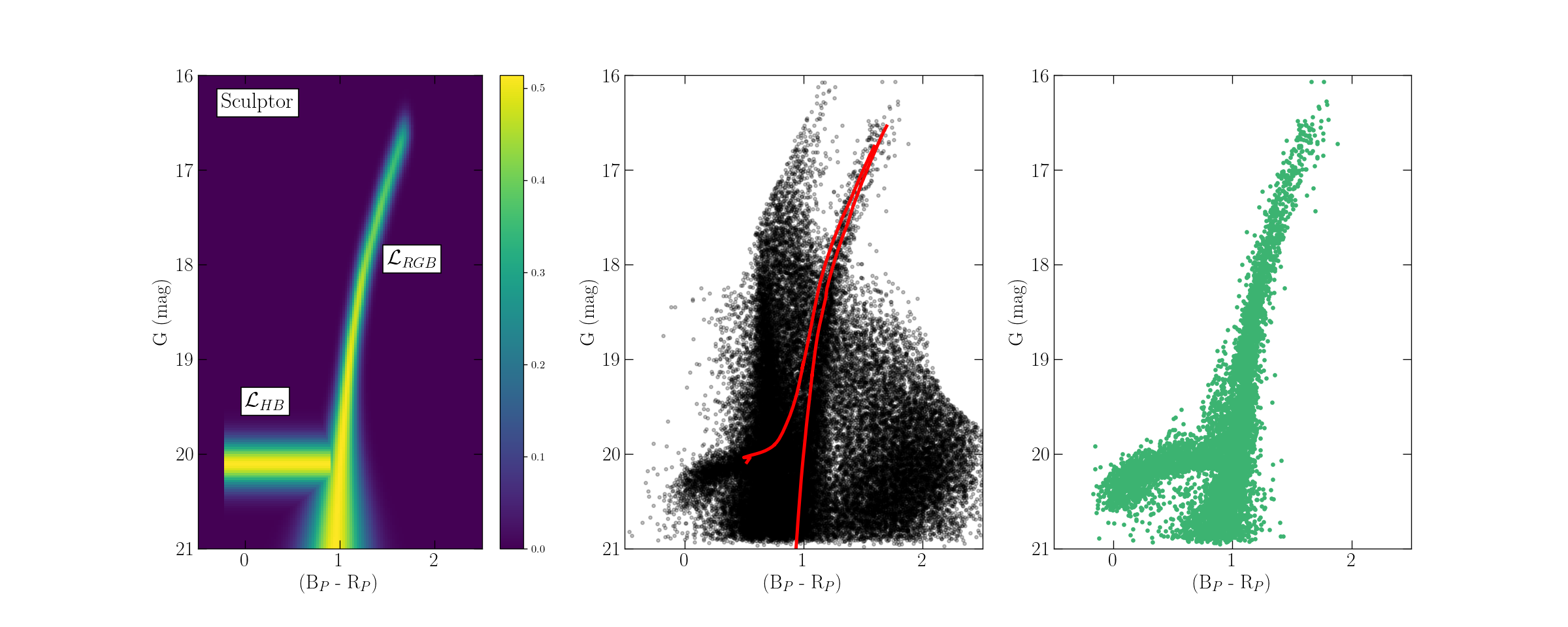}
 \caption{An example of the CMD likelihood ($\mathcal{L}_{CMD}$) calculated for the Sculptor dwarf galaxy. The left panel shows the CMD likelihood map, consisting of two models: one for the horizontal branch ($\mathcal{L}_{HB}$) and another for the red giant branch ($\mathcal{L}_{RGB}$), as described in the text. The middle panel is the CMD of Sculptor field stars, with the corresponding PADOVA isochrone overlain in red. The RGB of this model approximates the data well; however it is clear that the HB does not extend as blueward, and misses many Sculptor HB stars. Stars in the right panel are those with $P_{sat}$ $\geq$ 20\%, using our CMD likelihood map in the left panel, which clearly captures the full extent of the HB and exhibits little contamination.}
 \label{fig:scul_CMD}
\end{figure*}

The implementation for the CMD satellite model is a likelihood function consisting of two separate likelihood maps: one for the red giant branch (RGB) and one for the horizontal branch (HB). The creation of the RGB likelihood map is performed in the same manner as in \citetalias{McVenn2020}. We additionally include a model for the HB, which was not available in \citetalias{McVenn2020} or \citetalias{McVenn2020_updated}. Our HB implementation is simplified compared to the RGB, since we had found that the synthetic populations that fit the RGB well do not describe the HB population to a sufficient accuracy. In particular, the synthetic HB does not generally extend as blue as necessary to match the data, nor does it reproduce the observed HB shape\footnote{In addition to synthetic isochrones not reproducing the HB's shape appropriately, there is also the “second parameter problem” (as seen in globular clusters; e.g., \citealt{moehler2003}) that dictates how stars are distributed along the HB. These differences in similar metallicity systems are not well captured in synthetic stellar populations; nonetheless, these observed effects need to be accounted for.}. We highlight this in the central panel of Figure \ref{fig:scul_CMD}, which shows the positions on the CMD for stars in the field of Sculptor and the PADOVA isochrone (\citealt{girardi2002}; converted to Gaia photometric bands using \citealt{weiler2018} corrections) that best matches the RGB.

For the HB likelihood function, we choose to represent the HB as a constant value in G-magnitude, ranging between the blue end set by (B$_{P}$ $-$ R$_{P}$) = $-$0.25, and a red end extending to the RGB isochrone. This constant magnitude corresponds to the mean magnitude of an old (12 Gyr) and very metal-poor ([Fe/H] = $-$2.19) isochrone that is shifted by the dwarf’s distance modulus. 

In a similar manner to the implementation of the RGB, the likelihood function of the HB is essentially a Gaussian whose sigma is two factors added in quadrature. The first is an inherently assumed width of the stellar population; for the HB, we assume an approximate width of 0.1 mag in Gaia G. The second is the mean errors in (B$_{P}$ $-$ R$_{P}$), to account for the errors in magnitude.



After the RGB and HB models are created, they are merged to create one likelihood map, such that each pixel in $\mathcal{L}_{CMD}$ is the maximum likelihood from either HB or RGB model. Then, the map is normalize such that the integral of the entire likelihood function ($\mathcal{L}_{CMD}$ = $\mathcal{L}_{RGB}$ + $\mathcal{L}_{HB}$) is equal to 1. 

The left panel of Figure \ref{fig:scul_CMD} shows an example of the constructed CMD likelihood for our example dwarf, Sculptor. On the right, we show the Sculptor stars whose probabilities obtained in the algorithm correspond to likely members ($P_{sat}$ $\geq$ 20\%) in green. The HB model, though certainly simplistic, effectively identifies probable HB members of the dwarf. 

\subsection{Spatial likelihood Modifications to Identify Distant Members}
\label{sect:spatial}


For the dwarf spatial likelihood model, the original approach in \citetalias{McVenn2020} is to create a look-up map, by co-adding many realizations of a 2-D exponential elliptical profile to approximate the dwarf's shape. Each realization is created by randomly sampling the structural parameters of the dwarf (position angle, ellipticity, and half-light radius) using Gaussian uncertainty distributions, or assuming a circular profile shape if not all structural information is available.

In contrast, our approach assumes that there are two spatial components to every dwarf. The first ($\mathcal{L}_{S, inner}$) represents the main body, and corresponds to an exponential stellar density profile accounting for the structural parameters of the dwarf from the literature similar to the methods of \citetalias{McVenn2020}. Here, we expand this expression to include a secondary profile ($\mathcal{L}_{S, outer}$) which represents a putative outer, extended component. We choose to model this as an additional exponential profile with two unknown parameters. The first is the normalization ($B$) and the second is its scale radius ($r_{s}$). The overall spatial likelihood is therefore given by:

\begin{equation}
 \mathcal{L}_{S} \propto \exp[\frac{-r}{r_e}] + B \exp[\frac{-r}{r_s}]
 \label{eq:Ls}
\end{equation}

\noindent where $r$ is the distance of the star from the dwarf's center (accounting for position angle and ellipticity), $r_{e}$ is the exponential scale radius of the dwarf (derived from the literature value of the half-light radius, $r_{h}$, such that $r_{h} \simeq$~1.68$r_{e}$), $B$ is the normalization of the outer component to the inner component (such that 0~$\le B <$~1), and $r_{s}$ is the exponential scale radius of the putative outer component (such that $r_{s} > r_{e}$). If the preferred value for a dwarf is $B = 0$, this implies that no outer component is necessary in order to explain the data. In addition to requiring $r_{s} > r_{e}$, we set a weak prior on $r_{s}$ such that it must be smaller than the radius of the field. 


Our implementation of the spatial likelihood allows us to search over every dwarf to determine which dwarfs prefer an outer profile to exist in order to explain the data (i.e., $B > 0$). However, we do not want to make strong assumptions about the shape of such an extra component, should it exist. Tidal tails can be very long and extended in specific directions (i.e., along the dwarf's orbit); alternatively, there may be an “excess” of stars at all position angles around the dwarf due to an extended halo. To approach the problem agnostically, we adopt two simple assumptions and run the algorithm using both in order to check for consistency. The first is that the projected shape (i.e., ellipticity) of the outer profile is the same as the main body of the dwarf; the second is that the outer component is spherical. For the first, $r_{s, ell}$~=~$\sqrt{\xi_{rot}^2 + (\frac{\eta_{rot}}{1 - ell})^2}$ where ($\xi_{rot}$,~$\eta_{rot}$) are the tangent plane positions ($\xi$, $\eta$) of a star, whose axes have been rotated by the position angle. For the circular outer profile, $r_{s, circ}$~=~$\sqrt{\xi^2 + \eta^2}$.

In contrast to \citetalias{McVenn2020}, the spatial likelihood function for the dwarf now contains unknown parameters. This means we cannot easily construct look-up maps for the spatial likelihood as was done previously, but instead we need to calculate the spatial likelihood function direct from Equation \ref{eq:Ls} for each realization. While this allows the easy addition of a second spatial component, it does not allow the inclusion of the uncertainties in the measured parameters for the inner profile. We comment further on the influence of the spatial uncertainties in Section~\ref{sect:limits}.

\begin{figure*}
 \centering
 \includegraphics[width=0.75\textwidth]{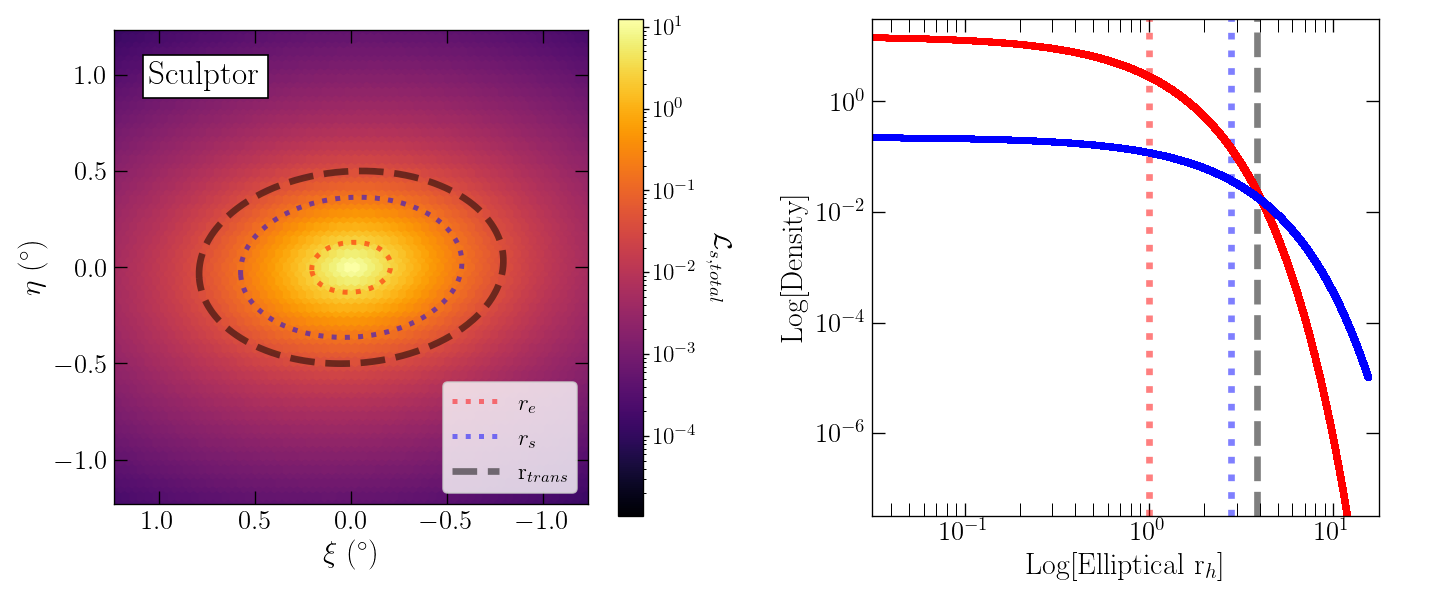}
 \caption{An example of the spatial likelihood ($\mathcal{L}_{S}$) for the Sculptor dwarf galaxy when assuming an elliptical outer profile. The 2-D spatial likelihood in the tangent plane is calculated using the values ($B$ and $r_{s}$) solved for by the algorithm (see values in Table \ref{tab:outer_profile_dwarfs}). The inner half-light exponential radius ($r_{e}$, red dotted), the outer scale radius ($r_{s}$, blue dotted), and the transition radius designating the approximate radii where the fraction of starlight is dominated by the secondary outer profile ($r_{trans}$, black dashed) are shown for comparison in both panels. The right plot shows the 1-D spatial likelihood for the inner (red line) and outer (blue line) spatial likelihood profiles.}
 \label{fig:spatial_likelihood}
\end{figure*}

In Figure \ref{fig:spatial_likelihood}, we show an example of the spatial likelihood implementation for the Sculptor dwarf galaxy using an elliptical outer profile. The left panel shows a 2-D tangent plane map color-coded by the total spatial likelihood ($\mathcal{L}_{S}$) and the right represents the likelihood in 1-D as a function of elliptical half-light radius. The relevant scale radii of the two components are shown as dotted lines, where the scale radius of the inner profile (r$_{e}$) is shown in red and the best-fit scale radius of the outer profile (r$_{s}$) is shown in blue for comparison. In the 1-D panel on the right, we show the relative densities for the inner (red) and outer (blue) components, where the outer profile is scaled by the algorithm's estimate for $B$. The associated scale radii are shown as vertical dotted lines in this panel. For dwarfs where the algorithm identifies an outer profile (i.e., $B$ is non-zero), we can also estimate where the outer profile begins to dominate over the inner profile. We refer to this boundary as the “transition radius” ($r_{trans}$), and for the Sculptor dwarf we highlight this boundary as a black dashed line in Figure \ref{fig:spatial_likelihood}. Stars beyond this boundary will primarily belong to the outer component, and may have a particularly interesting physical origin.

To confirm internal consistency when using our 2-component $\mathcal{L}_{S}$ model to the previous work in \citetalias{McVenn2020} and \citetalias{McVenn2020_updated}, we additionally compare our results to what we refer to as the “1-component” model. In practice, this means that the $\mathcal{L}_{S}$ model is fixed per dwarf, as was done in the preceding works (i.e., errors in structural parameters are accounted for, and only one exponential is used to describe the dwarf's density). The difference here is our addition of the HB from Section \ref{sect:HB} to the CMD model. We find it necessary to compare both the 1-component and 2-component models for internal consistency, and discuss this further in the following subsection.

\subsection{Proper Motion Estimates and Added Prior}

In the original implementation of \citetalias{McVenn2020} and \citetalias{McVenn2020_updated}, the spatial likelihood model per dwarf was fixed, and so only three unknowns were to be solved for. These parameters being: the dwarf’s systemic PMs in both directions, and the fraction of stars in the satellite. With the addition of an outer component whose unknowns describe the structure of the outer profile, our 2-component model now has five total variables to be estimated. We maintain the same requirement as the previous implementations to solve for the systemic PM of the dwarf, as it provides an additional method of verification in our methods.

For our 1-component runs, we first confirm that the addition of the HB in $\mathcal{L}_{CMD}$ does not change the results of the proper motion estimate. In the Appendix Figure \ref{fig:pm_comparisons}, we compare the relative measurements and 1-$\sigma$ errors of our estimates in the 1-component model (black) to the reported PMs in \citetalias{McVenn2020_updated} (grey dashed) and \citet[blue square]{battaglia2022}. For the more recently discovered systems mentioned in Section \ref{sect:data}, we also plot the reported PM estimates from each dwarf’s respective detection papers (pink triangle). In nearly all systems with previous measurements, we find consistency to past works within 1- to 2-$\sigma$. 

In both versions of the 2-component runs (elliptical and circular outer profiles), we found that 51 out of the 64 satellites also have consistent PM estimates to \citetalias{McVenn2020_updated}. The 13 systems that fail the 2-component model do not produce consistent PM estimates due to an inability to converge in the spatial parameters $B$ and $r_{s}$. We found that these particular systems have two properties in common: firstly, these dwarfs are the smallest UFDs such that the 1-component model only identifies $\lesssim$5 member stars within 6r$_{h}$. And secondly, these small systems have substantially higher MW contamination fractions within the same area, such that the fraction of members within 6r$_{h}$ is $\lesssim$20\%. This means that the MW contamination is significantly higher than the signal of the satellite within those radii (a MW fraction of $\gtrsim$80\%). Figure \ref{fig:bad_systems} in the Appendix shows this trend, with the systems that perform poorly highlighted in red text.

Given this issue, we decided to adopt a loose PM prior to better constrain $\mathcal{L}_{PM}$ in the case of the 2-component spatial model, such that the proper motion estimate must be within 5-$\sigma$ of the PM of the dwarf derived in the single component case (accounting also for systematic errors; \citealt{lindegren2021_astrometricsolution}). Even with the addition of the PM prior, we find that these same 13 systems are unable to converge on any preferred values for the secondary component. These systems are: Boo4, Cet3, DES, Hor2, Ind1, LeoMi, Peg3, Pic2, Pisc2, Ret3, Tuc5, Vir1, and Vir2. It is unsurprising that the algorithm fails in these cases. We are essentially attempting to fit 5 parameters to less than 10 stars, and the algorithm is unable to resolve the dwarf’s signal against the presence of the MW foreground. For the remainder of this work, we remove these systems from any further analysis.


\subsection{A Note on Reported Parameters}
\label{sect:note_params}

Using the three likelihood functions ($\mathcal{L}_{CMD}$, $\mathcal{L}_{S}$, and $\mathcal{L}_{PM}$) for both the MW foreground and satellite models, we search over parameter space for the combination that maximize the total likelihood for a given dwarf. The estimated parameters are the following 5 variables: $\mu_{\alpha*}$, $\mu_{\delta}$, $B$, $r_{s}$, and $f_{sat}$. These parameters are determined using the \textsc{python} package \textsc{emcee} (\citealt{foreman-mackey2013}). We apply the algorithm to all Gaia fields obtained for the dwarfs listed in Table \ref{tab:dSphs_all}.

Typically, the parameters for the outer profile have asymmetric posterior distributions. Therefore, we opt to report the values corresponding to the mode of the posterior distribution. The mode is determined by creating a histogram of the posterior using very fine, equally spaced bins. 


The uncertainties on this estimate are then reported as the high and low bounds of the highest density interval (HDI). The HDI is defined as the shortest interval in a posterior distribution that contains a specified credible mass. Consequently, the mode is always contained within the credible region of the PDF. We elect to use a credible mass of 68\%, which is essentially the Bayesian equivalent to $\overline{x}$~$\pm$~1$\sigma$, or the 68\% confidence interval in frequentist statistics. 



\section{Results and Validation}
\label{sect:results}

\begin{table*}
\renewcommand{\arraystretch}{1.5}
    \centering
    \begin{tabular}{c|ccccc|ccccc}
         &  & & Elliptical & &  & & & Circular & \\
         \hline
         Galaxy & B & $r_{s, ell}$  & $r_{trans, ell}$ & & Ellipticity & B & $r_{s, circ}$ &  $r_{trans, circ}$ &  & Ellipticity \\
         & & ($^{\circ}$) &  ($^{\circ}$) & (pc) &  & & ($^{\circ}$) &  ($^{\circ}$) &  (pc) &  \\
         \hline
         
         Boo1 & $0.245_{-0.35}^{+0.151}$ & $0.206_{-0.036}^{+0.021}$  & 0.34 & 400 & 0.25 &  $0.086_{-0.098}^{+0.058}$ & $0.217_{-0.055}^{+0.029}$  &  0.57 & 660 & 0.41  \\

         Boo3 & $0.015_{-0.025}^{+0.006}$ & $1.901_{-1.841}^{+0.554}$ & 1.66 & 1350 & 0.33 & $0.014_{-0.021}^{+0.006}$ & $1.805_{-1.53}^{+0.585}$ & 1.71 & 1390 & 0.38 \\
         
         Dra2 & $0.019_{-0.148}^{+0.018}$ & $0.069_{-0.05}^{+0.016}$  & 0.21 & 80 & 0.23 & $0.026_{-0.198}^{+0.023}$ & $0.063_{-0.031}^{+0.018}$  &  0.21 & 80 & 0.36 \\
         
         
         Gru2 & $0.004_{-0.002}^{+0.002}$ & $0.088_{-2.912}^{+1.704}$  & 1.02 & 940 & 0.0 & $-$  & $-$ &  $-$ & $-$ & $-$  \\
         
         Scl & $0.015_{-0.004}^{+0.004}$ & $0.317_{-0.022}^{+0.022}$  & 0.84 & 1260 & 0.37 & $0.013_{-0.003}^{+0.003}$  & $0.356_{-0.048}^{+0.021}$  &  1.06 & 1520 & 0.52 \\
         
         Seg1 & $0.174_{-0.562}^{+0.082}$ & $0.108_{-0.022}^{+0.029}$  & 0.11 & 40 & 0.34 &  $0.165_{-0.507}^{+0.109}$ & $0.081_{-0.029}^{+0.013}$  &  0.14 & 60 & 0.50 \\
         
         Tuc2 & $0.009_{-0.219}^{+0.007}$ & $0.284_{-0.187}^{+0.087}$ & 0.70 & 700 & 0.39 &  $0.037_{-0.201}^{+0.035}$ &  $0.244_{-0.104}^{+0.07}$ &  0.54 & 540 & 0.51 \\
         
         Tuc3 & $0.005_{-0.004}^{+0.002}$ & $0.889_{-0.954}^{+0.369}$ & 0.34 & 150 & 0.0 & $-$  & $-$ &  $-$ & $-$ & $-$ \\
         
         UMi & $0.015_{-0.004}^{+0.005}$ & $0.619_{-0.063}^{+0.056}$ & 1.00 & 1320 &  0.55 &  $0.022_{-0.006}^{+0.00. }$  &  $0.356_{-0.048}^{+0.021}$ &  1.29 & 1710 & 0.70 \\
         \hline
    \end{tabular}
    \caption{Dwarfs with secondary outer profiles, as determined in both the elliptical and circular runs. The transition boundary ($r_{trans}$, see text) is defined as the approximate boundary where the outer profile dominates the total fraction of starlight. We provide parameters describing this boundary by the semi-major axis and ellipticity of the best-fit ellipse.}
    \label{tab:outer_profile_dwarfs}
\end{table*}

The best-fit estimates for the spatial parameters ($B$ and $r_{s}$) allow us to identify a handful of dwarfs whose outer profile solutions are non-zero. That is, in the $\sim$60 dwarfs we explore in this work, we find a total of 9 which are suggested to host extended stellar substructure. In both of the 2-component runs, we find that the inventory of dwarfs where B is non-zero is the same between both assumed outer profile shapes. These dwarfs are: Bo\"otes 1, Bo\"otes 3, Draco 2, Grus 2, Sculptor, Tucana 2, Tucana 3, and Ursa Minor. The preferred parameter values for their outer stellar component in both the elliptical and circular runs are given in Table \ref{tab:outer_profile_dwarfs}.

A natural consequence of the existence of an outer spatial profile is the transition at which the secondary component dominates the fraction of starlight. Largely, the location of this transition depends on the outer profile parameters, $B$ and $r_{s}$. The location of this transition can be estimated empirically using the spatial likelihood function, and can then be used as a modest proxy to highlight stars that may be considered to be outskirt members. We reference this feature as the “transition boundary” ($r_{trans}$). In Table \ref{tab:outer_profile_dwarfs}, we report the angular and physical semi-major axis of this transition, as well as the ellipticity of the transition boundary ellipse. In all systems, the position angle of this boundary is the same as that assumed for the dwarf itself (as in Table 1 of \citetalias{McVenn2020}).

Prior to discussing the dwarfs which appear to host outer profiles in detail, we first explore the differences in the probability of membership between the two outer component shape assumptions in the following subsection. We then conduct an analysis to validate our methods using stars with RVs, when available. 

\subsection{Elliptical and Circular Outer Profiles}

Our main rationale of this work is to determine, with high fidelity, which dwarf galaxy systems host outskirt stellar substructures. However, we also do not know $\textit{a priori}$ what the properties of the putative outer component may be. With respect to the projected shape of this component, our strategy has been to adopt two different shapes (one the same shape as the inner regions, i.e. elliptical, while the other is circular).

As reported in Table \ref{tab:outer_profile_dwarfs}, we observe that the set of dwarfs the algorithm identifies as having two components is the same for both outer profile runs. The conclusion that these dwarfs have more stars in the outskirts than expected from a simple exponential, is apparently robust to our outer profile shape assumption.

\begin{figure*}
    \centering
    \includegraphics[width=0.9\textwidth]{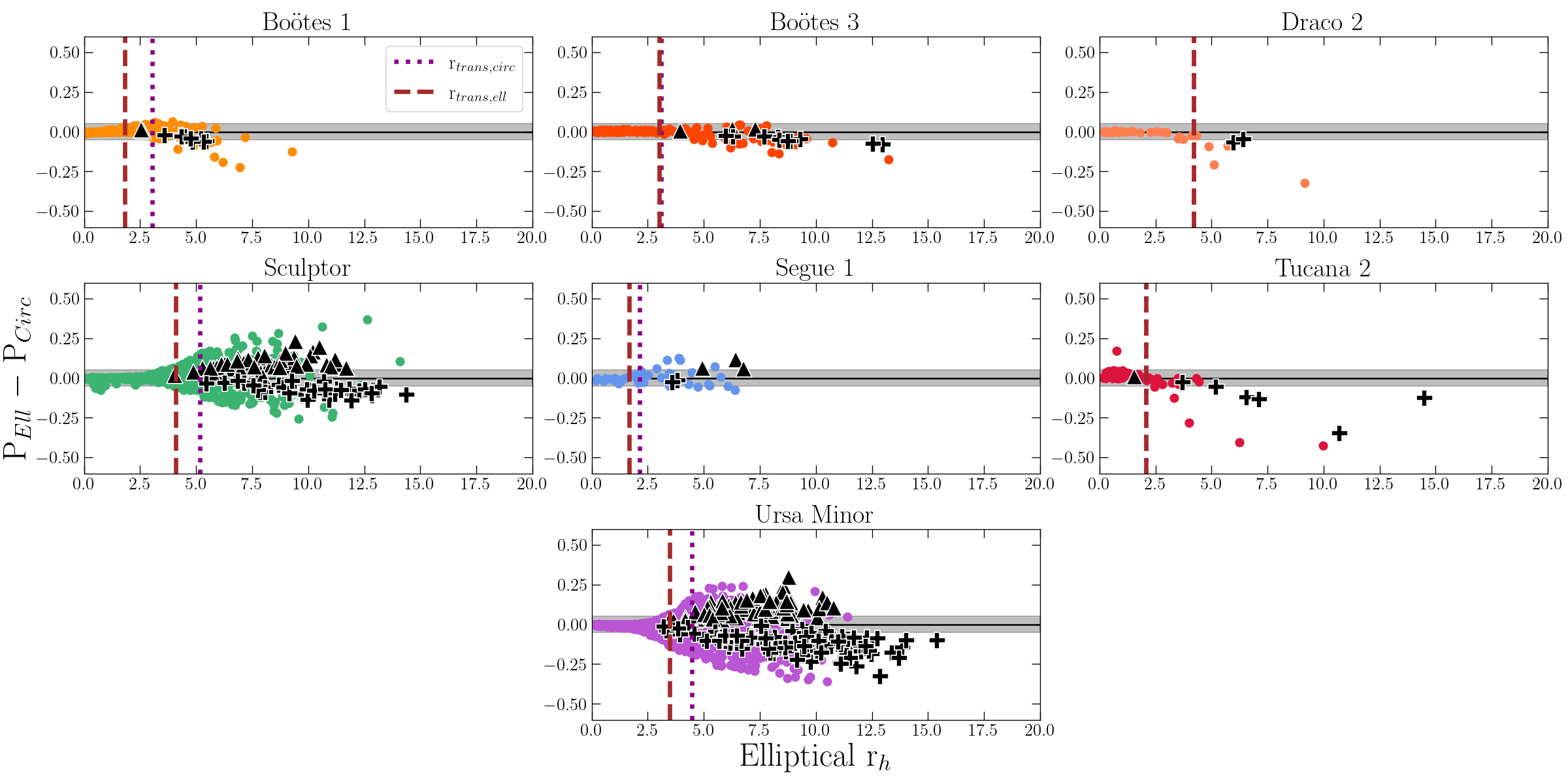}
    \caption{Differences between calculated probabilities in the elliptical and circular 2-component runs, for dwarfs with detectable outer profiles. In all panels, three subsets of the data are shown: stars which appear only in the elliptical dataset are shown as black triangles ($P_{ell} > 20$\%), stars which are only in the circular dataset are shown as black \textbf{+} icons ($P_{circ} > 20$\%), and stars in common between both profile runs are highlighted as colored points (both $P_{ell}$ and $P_{circ} > 20$\%). Over 90\% of the data shown here have minimal changes in probabilities ($\pm$5\%) between elliptical and circular models (highlighted by the grey region in each panel). The vertical magenta dotted and brown dashed lines are the transition radii solutions for the circular and elliptical profiles, respectively. Gru2 and Tuc3 are not included in the figure, since their structural parameters are assumed to be circular and there is no difference in probabilities between either 2-component run.}
    \label{fig:p1_p2}
\end{figure*}

In Figure \ref{fig:p1_p2}, we show the differences in probabilities ($P_{Ell} - P_{Circ}$) between these two datasets as a function of $r_{h}$. For clarity, stars with $P_{sat}$~$\geq$~20\% in at least one of these datasets are plotted. Those with $P_{sat}$ above this threshold in $\textit{both}$ datasets are shown as filled circles. For stars whose probability is $\geq$20\% in one set but not the other, we plot them as black triangles (elliptical set only) and black \textbf{+} icons (circular set only). 


Figure \ref{fig:p1_p2} shows that, for stars interior to the transition boundaries, the choice of an elliptical or circular outer profiles does not significantly change the probabilities. To emphasize this, each panel also shows a grey highlighted region, corresponding a change in probability of $\pm$5\% between the two runs. We find that $>$90\% of the overall data presented in this figure are found within these bounds.  

\begin{figure*}

    \centering
     \includegraphics[width=0.9\textwidth]{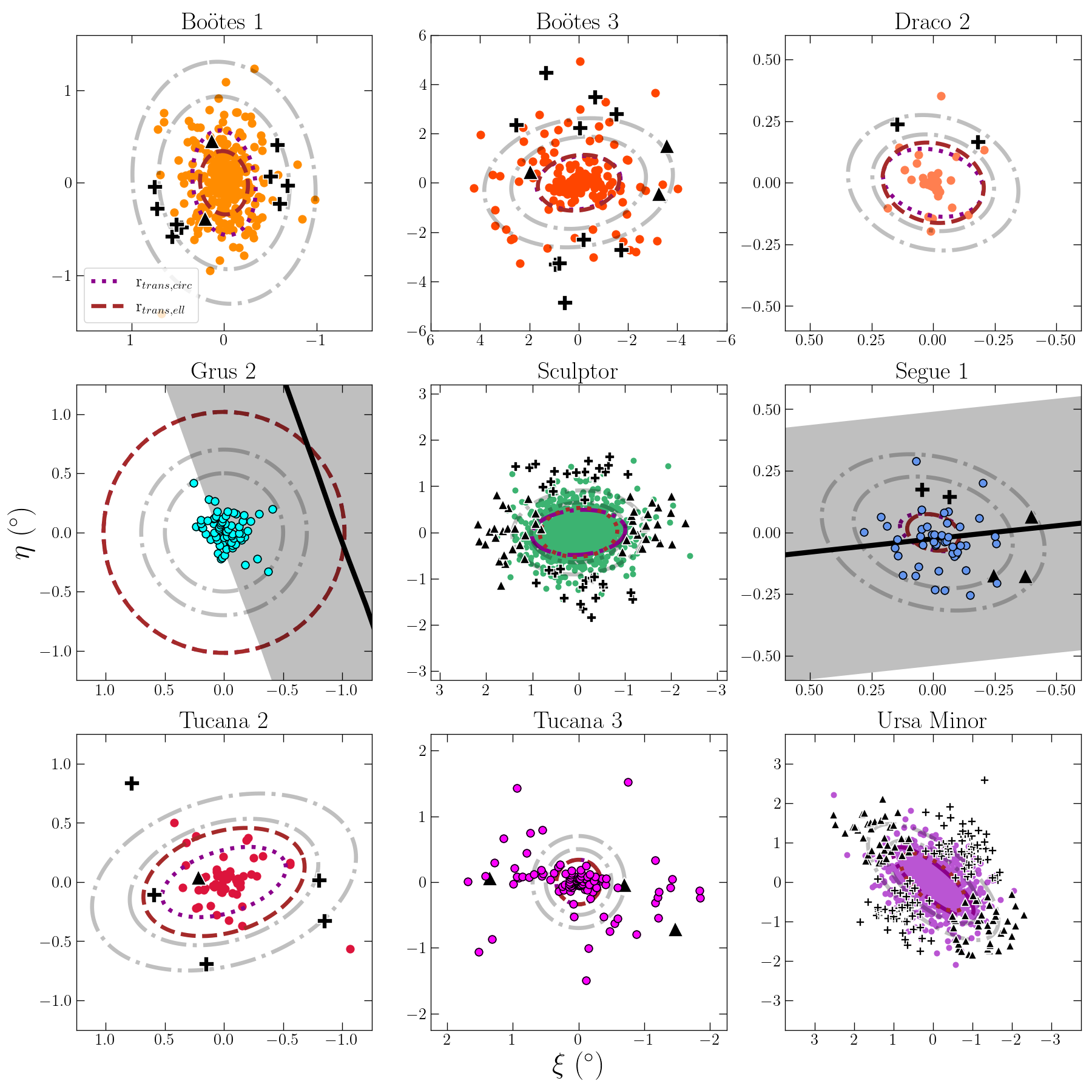}
     \caption{Tangent plane plots for all dwarfs with identified outer profiles. The relevant radii representing the transition boundary solutions ($r_{trans, circ}$ and $r_{trans, ell}$), and the three subsets of data (elliptical only, circular only, and stars in common) are represented the same as in Figure \ref{fig:p1_p2}. We additionally plot in grey the 5 and 7$r_{h}$ ellipses for each system. For direct comparison to the incidence of the Orphan/Chenab stream to Gru2, and the 300~km~s$^{-1}$ stream to Seg1, we include the approximate stream position (black line; from the \textsc{python} package \textsc{galstreams};  \citealt{mateu2023}) and highlight the approximate width ($\sigma$~=~1$^{\circ}$ for Orphan/Chenab, and 0.4$^{\circ}$ for the 300~km~s$^{-1}$ stream) in grey.}
     \label{fig:tangent_outer_profiles}
    
\end{figure*}

As seen at larger radii, the greatest probability differences occur at farther radial distances. We find that these differences are a result of the outer profile shape: for example, when examining the majority of the data which appear in only one set versus the other (all black icons), the positive y-axis consists only of the “elliptical only” set, while the negative y-axis are those of the “circular only” set. The differences seen in Figure \ref{fig:p1_p2}, whereby some stars are more likely to be members in one realisation than the other, is a result of the stars’ angle from the major axis; this concept is further corroborated when examining the tangent plane positions of these same stars in Figure \ref{fig:tangent_outer_profiles}. As shown in the figure, stars in these mutually exclusive sets are located along the dwarf's major axis (“elliptical only”) and minor axis (“circular only”). As a final confirmation, we further examined the residuals of the tangent plane spatial likelihoods ($\mathcal{L}_{S, Ell}$~$-$~$\mathcal{L}_{S, Circ}$) and found the major and minor axes are enhanced in their respective model. 


At this time, it is not obvious which model best fits these outer profile dwarfs as the structure will no doubt be linked to the stars’ physical origin. However, our approach in this study is not to define the shape of dwarf galaxy outskirts explicitly, but instead is to identify whichever stars plausibly belong to these systems (of course, particularly those at large radii). Given our overall motivations, we remain agnostic about which shape should be utilized, and instead assume that any star with a reasonably high probability should be considered a member. For this reason, we decide to take the maximum probability ($P_{max}$) per star, between all three runs (the single and the two 2-component versions $-$ i.e, elliptical and circular outer profiles) to determine the possibility of membership. As evidenced by the data presented here, member stars may be overlooked when assuming one model over the other. In using the maximum probability of the 3 runs, we ensure all plausible stars are accounted for. We anticipate spectroscopic follow-up will not only be necessary to confirm that a star is a member of its nearby satellite, but also to confirm their origin (accreted or tidal), and in turn derive the dwarf's stellar halo shape.

\subsection{Examination of Purity Fractions and Contamination}
\label{sect:RV_purity}

In the Bayesian framework presented here, a typical method to classify data as a member of one population or the other is to select a decision boundary for the probabilities. Generally, a reasonable boundary is 50\%; in our case, $>$50\% implies a dwarf member, and $<$50\% is of the MW population. In this work, we instead choose to adopt a boundary that is inferred from the most likely members, by confirming their membership via radial velocity information (when available), to better interpret a reasonable limit. This approach also allows us to implement a consistency check, by comparing the samples of true MW foreground and true spectroscopically confirmed members to their associated probability (i.e., $P_{max}$).

We do this by first calculating what we refer to as the “purity fraction”. For each star in a given field with spectroscopic data, we determine whether its measured RV is consistent with being a member of the dwarf. A star is considered a true member if its individual RV is within 3-$\sigma$ of the satellite’s RV, accounting for the dwarf’s velocity dispersion, the star’s individual RV error, and the dwarf’s systemic RV error in the calculation. RV, errors, and dispersion of the dwarf are taken directly from Table 1 of \citetalias{McVenn2020} or from the updated systems listed in Table \ref{tab:updates}. The purity fraction is defined as the fraction of stars per probability interval that have RVs consistent with membership of the dwarf, specifically in each probability bin from 0 to 1 (0\% to 100\%). We note that $\sim$3 out of every 100 stars may have an RV more than 3-$\sigma$ from the satellite, but still be a genuine member, although we will end up classifying it as a non-member based on its RV.

\begin{figure*}
 \centering
 \includegraphics[width=0.9\textwidth]{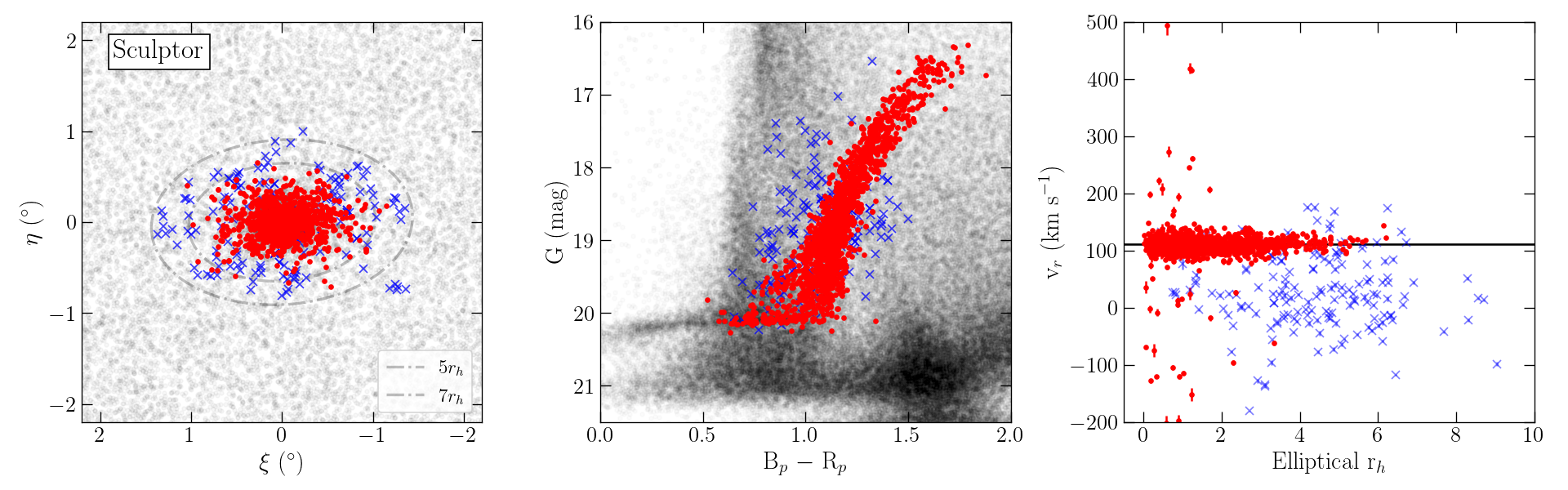}
 \caption{An example of the radial velocity purity test, for the Sculptor dwarf. Grey points in the tangent plane (left) and CMD (middle) represent all field stars analyzed in this work. Stars with radial velocity information available are shown as colored points in all three panels; red indicates stars which we consider as dwarf members ($P_{max}$ $>$ 10\%) are marked as points, while foreground MW contamination ($P_{max}$ $<$ 10\%) are shown as faint x's. We show these radial velocities as a function of distance in the right panel, where Sculptor's systemic radial velocity is represented as the black horizontal line for comparison.}
 \label{fig:scul_pure}
\end{figure*}

\begin{table}
    \centering
    \renewcommand{\arraystretch}{1.3}
    \begin{tabular}{c|ccc}
    Probability bin & No. RV stars & No. Consistent & Fractional Purity \\
    \hline
        0.0 $\leq$ $P_{max}$ $<$ 0.1 &  166   &   35  & 0.21 $\pm$ 0.08 \\
        0.1 $\leq$ $P_{max}$ $<$ 0.2 &  2     &   2   & 1.00 $\pm$ 0.71 \\
        0.2 $\leq$ $P_{max}$ $<$ 0.3 &  2     &   1   & 0.50 $\pm$ 0.70 \\
        0.3 $\leq$ $P_{max}$ $<$ 0.4 &  1     &   1   & 1.00 $\pm$ 1.0 \\
        0.4 $\leq$ $P_{max}$ $<$ 0.5 &  N/A   &   N/A & N/A \\
        0.5 $\leq$ $P_{max}$ $<$ 0.6 &  4     &   4   & 1.00 $\pm$ 0.50 \\
        0.6 $\leq$ $P_{max}$ $<$ 0.7 &  5     &   5   & 1.00 $\pm$ 0.45 \\
        0.7 $\leq$ $P_{max}$ $<$ 0.8 &  9     &   8   & 0.89 $\pm$ 0.33 \\
        0.8 $\leq$ $P_{max}$ $<$ 0.9 &  18    &   17  & 0.94 $\pm$ 0.24 \\
        0.9 $\leq$ $P_{max}$ $\leq$ 1.0 &  1781  & 1731  & 0.97 $\pm$ 0.02 \\
    \hline 
    \end{tabular}
    \caption{Purity fraction estimates for the Sculptor dwarf, as described in the text. Each probability bin is in increments of approximately 10\%. Note that there are no stars in our sample in the 0.4 $\leq$ $P_{max}$ $<$ 0.5 bin (i.e., no data with measured RVs are assigned probabilities in this range).}
    \label{tab:purity_frac_scul}
\end{table}

We highlight an example of this validation test as applied to the Sculptor dwarf galaxy in Figure \ref{fig:scul_pure}. The RVs are taken as the weighted mean of APOGEE DR17 and individual survey measurements, if either exist for a given dwarf (listed in Table \ref{tab:dSphs_all}). For Sculptor, we plot all stars in the field from Gaia eDR3 with no known spectroscopic data (at the time of this publication) in grey in the tangent plane (left) and CMD (center) panels. Colored symbols show stars with RV data separated by probability of membership to Sculptor; stars with $P_{max}$~$>$~10\% (dwarf members) are shown as red points, while those with $P_{max}$~$<$~10\% (MW foreground) are marked as blue \textbf{x} icons. The right panel shows individual star velocities as a function of elliptical half-light radii (in units of $r_{h}$), where the dwarf’s systemic RV is highlighted as the black horizontal line. 

The overwhelming fraction of stars flagged as members by this probability division (red points in Figure~\ref{fig:scul_pure}) exhibit RVs that are consistent with the dwarf satellite. This remains true, and is particularly apparent, for stars found out to large distances where the MW foreground population dominates the total number of stars in any radial annulus. We find this result to be particularly encouraging, as our kinematic model for the dwarf is independent of RV considerations. 

We now estimate the level of contamination in our Sculptor sample by determining the purity fractions. Per probability bin (in steps of 0.1, or 10\%), we determine the number of RV stars which are consistent to dwarf membership, divided by the total number (N) of RV stars in that bin. Errors for the purity fraction are reported as $\frac{\sqrt{N}}{N}$ to account for Poisson statistics. Sculptor’s purity fraction (and relative number of stars to obtain this statistic) are reported in Table \ref{tab:purity_frac_scul}. 

Largely, the fractional purity for Sculptor across all $P_{max}$ bins $>$10\% is high (though, small number statistics in multiple bins suggest rather large errors), and a significant number of RV members are successfully identified with the highest probabilities (1731 member stars out of 1781, in the $>$90\% bin). In the  $>$90\% bin, we note that there are 50 or so stars which are not members based on their RVs, yet their probabilities in the algorithm are high. We note these $\sim$50 stars are all located in the innermost regions of Sculptor (i.e., $<$2 r$_{h}$; shown as the red points in the RV panel of Figure ~\ref{fig:scul_pure}). Though they indeed have inconsistent RVs to Sculptor, we note that these particular stars are all a) towards the center of the dwarf, b) have consistent proper motions to Sculptor, and c) overlap the same region of the CMD as other Sculptor stars. Given that our model would therefore indicate that these stars are indeed members based on these three criteria, it is anticipated that they would be ascribed high probabilities ($>$95\%, in fact). As evidenced by the purity fraction estimates in Table \ref{tab:purity_frac_scul}, these data represent the 3\% of stars which contaminate the largest probability bin. We expect this type of contamination, and at close elliptical distances we do expect to have a larger $\textit{number}$ of stars with inconsistent RVs (though the $\textit{fraction}$ of contamination remains low).

We repeat this process for every dwarf which has a measured systemic RV. After obtaining the relative numbers of true members and the total per probability bin, we then choose to express the purity as a function of probability by marginalizing the data over all satellites. This “purity curve” (as it is hereafter titled) is essential to determining an appropriate decision boundary for dwarf membership based on the probability assigned, with the idea in mind to retain the most complete member sample out to large radii.

\begin{figure*}
 \centering
 \includegraphics[width=0.9\textwidth]{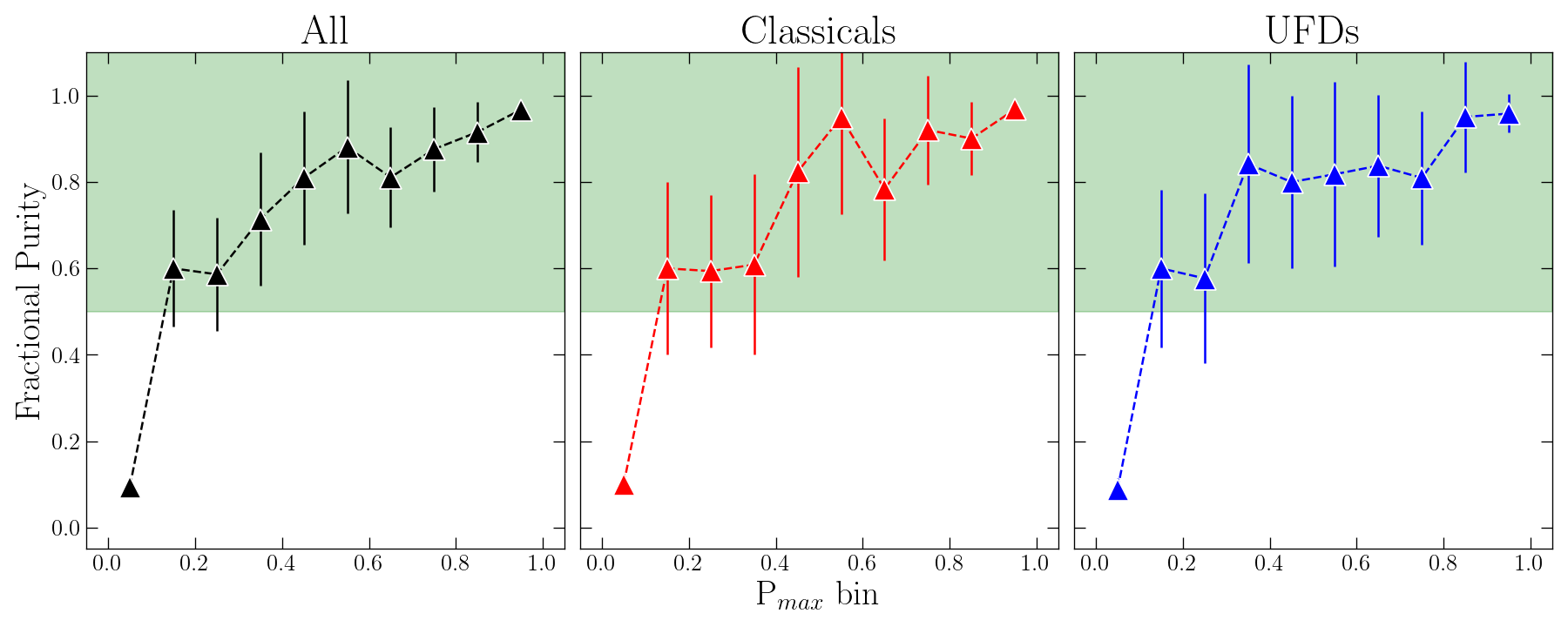}
 \caption{The fractional purity curves provided by the radial velocities of stars observed each dwarf, compiled together for both of the 2-component runs. These curves have been compiled for (a) all dwarfs, (b) classicals only, and (c) ultra-faints only. Green shading highlights the fractional purity regimes where statistically, at least 1 in 2 stars is correctly identified as a member ($>$50\% fractional purity).}
 \label{fig:frac_pure}
\end{figure*}

Figure \ref{fig:frac_pure} shows, for all stars with a radial velocity, the fraction of stars that are RV confirmed members as a function of $P_{max}$. The first panel shows this for all dwarfs (with measured systemic RVs), the middle panel for dwarfs with $M_V \lesssim -$8 mag, and the right panel for dwarfs with $M_V \gtrsim -$8 mag.

In the idealized case, we might expect that the probability of membership is synonymous with the purity fraction, such that the correlation between purity and probability is one-to-one. However, all panels of Figure \ref{fig:frac_pure} show that the $P_{max}$ estimated from the algorithm is in fact conservative. For example, we observe that $\sim$60\% of stars with $P_{max}$ = 20\% are members (whereas, we would expect only 20\% to be members). We find that the traditional limit of $P_{max}$ = 50\% actually results in membership fractions of $>$85\%. 

We also stress the unequal distribution of the number of stars in each $P_{max}$ bin, resulting in vastly different errors between the intermediate bins versus the first and last ones (corresponding to the $<$10\% and $>$90\%, respectively; also seen in Table \ref{tab:purity_frac_scul}). This observation continues across all 3 dwarf galaxy groupings. It is first obvious that most stars in the field are not members of dwarfs, and so the lowest bin therefore contains a large number of stars (thereby resulting in small errors). Secondly, most spectroscopic member stars are correctly flagged in the algorithm as members with high $P_{max}$ (such that the purity fraction is 0.97 for $P_{max} >$ 0.9). For comparison, the raw number of RV stars whose probabilities are intermediate (i.e., 0.1~$\leq$~$P_{max} <$~0.9) is $\sim$620 stars, while the first and last bins contain $\sim$3360 and $\sim$6670, respectively. Given the corresponding purity fractions in the first and last bins (0.09 and 0.97, respectively), we conclude that we appropriately identify the majority of MW and dwarf member stars in a given field, with very little contamination at high $P_{max}$.

\section{Discussion}
\label{sect:discussion}

In this section, we conduct a detailed review of each dwarf for which an outer profile has been detected via our methodology. For each dwarf, we search the literature for three main findings: (1) what is the most distant stellar member observed to date, and how far does each work trace each system; (2) do any previous works observe a break or transition in the stellar density profile, and if so, how do their results directly compare to ours; and (3) what evidence (if any) is there in the literature to argue for/against the dwarf’s disruption via tidal influence imparted by the MW? 

\begin{figure*}
    \centering
    \includegraphics[width=0.9\textwidth]{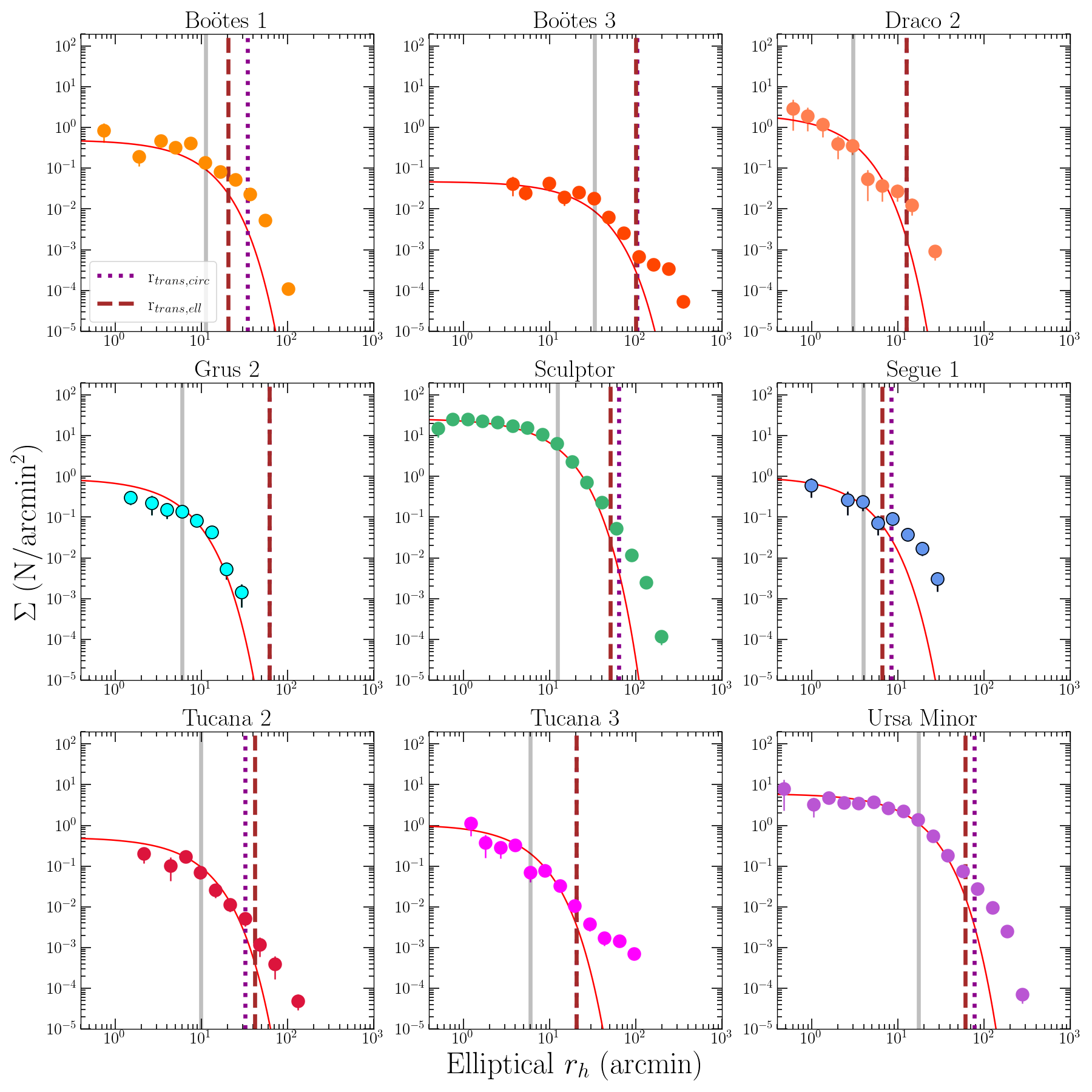}
    \caption{Stellar density profiles for all dwarfs with identified outer profiles. Each plot is made with stars whose $P_{max} \geq 10\%$. Vertical lines represent important radii: grey is each dwarf's half-light radius, and magenta dotted and brown dashed represent the transition radii ($r_{trans, circ}$ and $r_{trans, ell}$, respectively). For comparison, we additionally plot the 1-component exponential function in red to highlight the excess of dwarf members we detect using the 2-component model.}
    \label{fig:profiles}
\end{figure*}

Throughout this discussion, we note whether previous works find evidence for a bump or break in the stellar density profiles of our 9 systems, and compare their findings to our estimated transition boundaries from Table \ref{tab:outer_profile_dwarfs}. For easy comparison, we construct surface number density profiles (constructed by taking candidate members stars whose $P_{max} \geq$ 10\%) presented in Figure \ref{fig:profiles}. To highlight the benefit of the 2-component model and the detectability of radially distant stars, we also include an exponential function in each panel (red) for reference.

\begin{figure*}
    \centering
    \includegraphics[width=0.8\textwidth]{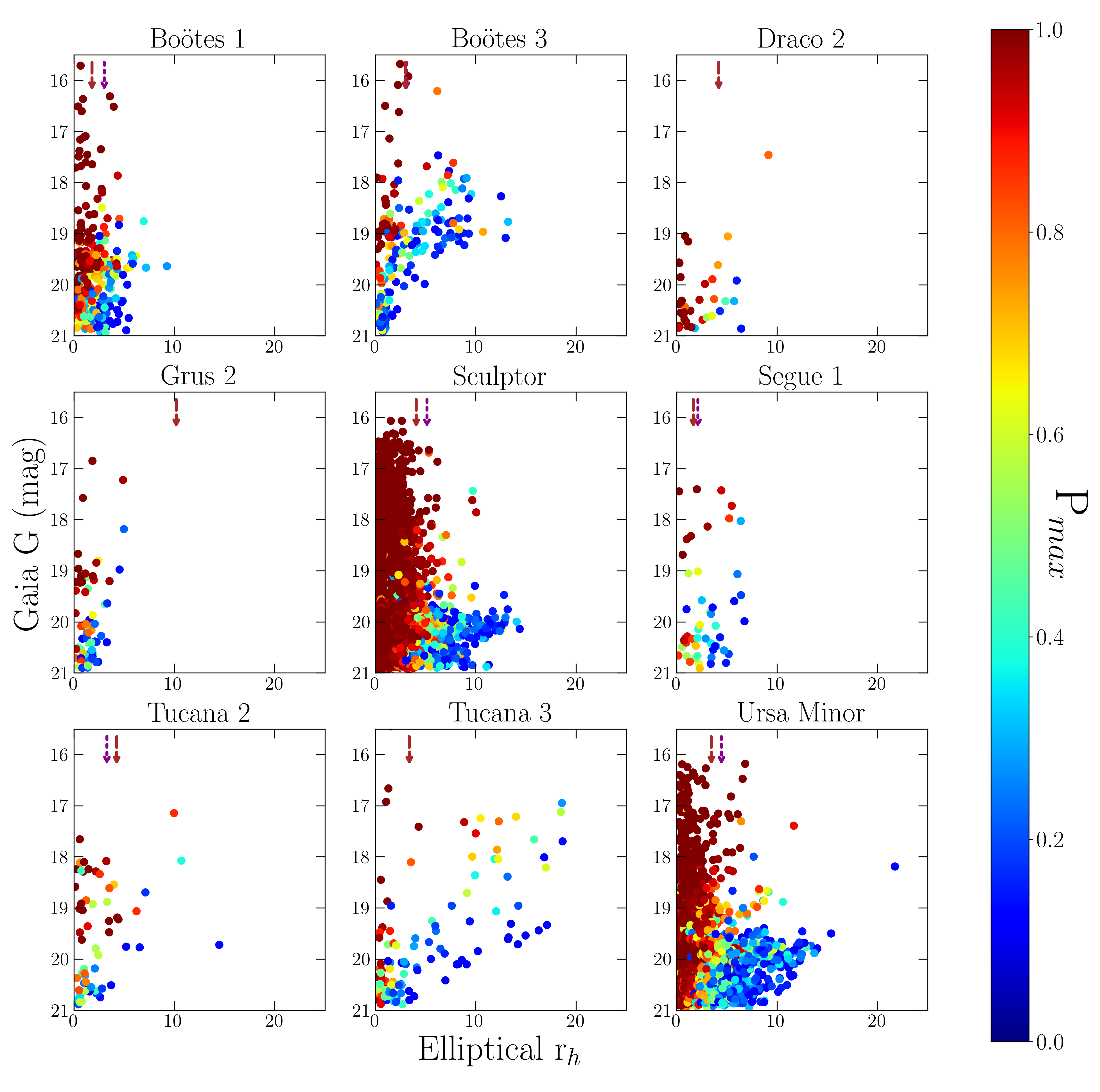}
    \caption{Magnitude versus distance for dwarf member stars with $P_{max} >~$10\%. Arrows in each panel represent $r_{trans}$ to highlight the outskirts, and are color-coded the same as Figure \ref{fig:tangent_outer_profiles}.}
    \label{fig:candidates}
\end{figure*}

Additionally, we report the number “outskirt” candidate members; these are defined as stars whose radial distances place them farther than the transition boundary in Table \ref{tab:outer_profile_dwarfs}, and are brighter than Gaia G $\leq$ 19.5 mag. The spread of members radially and in magnitude are shown in Figure \ref{fig:candidates}, where each point is color-coded by $P_{max}$. For each dwarf, we also highlight the transition boundary ($r_{trans}$) as arrows to show the approximate “outskirt” region.

The following section contains common themes and questions for each dwarf. Complementing figures for this section are Figure \ref{fig:tangent_outer_profiles}, Figure \ref{fig:profiles}, and Figure \ref{fig:candidates}.


\subsection{Bo\"otes 1}

Boo1 is one of the most luminous ($M_V$ $\approx$ $-$6.3 mag; $L_V$ $\approx$ 2.9~$\times$~10$^{4}$~L$_{\odot}$; \citealt{munoz2018}) and relatively nearby (66 kpc; \citealt{dall'ora2006}) UFDs. With a physical half-light radius of $\sim$200 pc (10$\arcmin$, or 0.17$^{\circ}$; \citealt{longeard2022}), it is comparable in $r_{h}$ to other MW classical systems. Boo1 has been the subject of many recent studies, resulting in a significant number of spectroscopic members ($>$70 stars; see compilation in \citealt{waller2023}). While its size and chemical trends [$\alpha$/Fe]-plane suggest it is a more massive UFD, \citet{frebel2016} determined Boo1’s abundances of neutron-capture elements are much lower than observed in MW halo stars, a trait unique to UFDs (\citealt{frebel2015}). 

Of these spectroscopic data in the literature, the most distant, confirmed members of Boo1 have been traced out to $\sim$4$r_{h}$ (\citealt{longeard2022}; \citealt{waller2023}). \citet{filion2021} also obtained photometric candidate members by probing the nearby blue horizontal branch (BHB) population, and found these BHBs occupy an extended outer envelope of the dwarf. For comparison, we detect 61 stars on the HB, and find the spatial distribution of these stars largely resembles that of Boo1 as presented in \citet{filion2021} (see their Figure 6). The farthest BHB in their study was located at a radial distance of $\sim$10$r_{h}$, though this star does not have consistent proper motions to the dwarf and does not appear as a member in this work. We do confirm that all 29 spectroscopic members from \citet{longeard2022} are correctly identified as members, and with high probabilities of membership ($P_{max} >$~50\%). For comparison to previous works, the most distant member identified via our algorithm is at an elliptical radius of $\sim$9$r_{h}$ from the center of Boo1, and does show consistent PMs to other spectroscopically confirmed members of \citet{longeard2022}. 

As reported in Table \ref{tab:outer_profile_dwarfs}, we find two possible solutions for the transition boundary in the elliptical ($r_{trans, ell}$) and circular ($r_{trans, circ}$) 2-component models. These values correspond to $r_{trans, ell}$ = 0.34$^{\circ}$ ($\sim$1.8$r_{h}$) and $r_{trans, circ}$ = 0.57$^{\circ}$ ($\sim$3$r_{h}$), respectively. Multiple works have previously reported transitions in Boo1's stellar components. \citet{koposov2011} probed stars in the inner half-light radius and detected two separate kinematically “hot” and “cold” samples. The authors suggest the nature of these two samples are clearly segregated by different velocity dispersions and metallicities in each population. In a more extended spectroscopic study, \citet{longeard2022} claim there is evidence of a break in the velocity dispersion and metallicity profiles. This break is located at $\sim$2$r_{h}$, which corresponds to our solution for $r_{trans, ell}$. As evidenced by the stellar density plot in Figure \ref{fig:profiles}, we also observe that our stellar density profiles break from the exponential at approximately this same position. 

The tidal influence of Boo1 has been suspected in multiple previous works, such as \citet{roderick2016} who observed a distinct overdensity near Boo1’s estimated tidal radius (reported as 0.54$^{\circ}$ and approximated via a King density profile). We note here that this particular report also coincides with our solution for $r_{trans, circ}$. However, recent works in \citet{battaglia2022} and \citet{pace2022} have concluded that the orbit of Boo1 has as pericenter of $\sim$35 kpc, and N-body simulations from \citet{read2006} have found that tidal shocking and stripping may be negligible for systems whose pericenters are $>$35 kpc. It is therefore questionable if tides are afflicting the morphology of Boo1. 

At present, no work has confirmed if Boo1 is actively disrupting. Considering that (1) the spectroscopic data in \citet{longeard2022} shows a moderate velocity gradient, (2) studies argue for a relatively small pericenter, and (3) the dwarf’s elongation aligns with the direction of orbit (\citealt{pace2022}), a natural conclusion is that tides could have affected the distribution of the outermost stars in Boo1.

However, there is some evidence favoring the alternative model (an extended stellar halo induced by a merger), or even a combination of the two scenarios. Firstly, \citet{longeard2022} report a metallicity gradient for Boo1, which has only ever been observed in one other UFD (Tucana 2, see below). \citet{pace2022} also argue that MW dwarfs with appreciably low central densities within the half-light radius, compared to the average density of the MW at the dwarf’s pericenter, can be used as a proxy to identify likely disrupting dwarfs. The authors report that Boo1's density ratio is larger than that of known disrupting systems (Ant2, Boo3, Tuc3, and Sagittarius), placing it more in accordance with other typical MW satellites. And finally, abundances from \citet{waller2023} showed that the outer stars (selected via the algorithm and shown to have consistent RVs to Boo1) have the same chemical properties as inner ones. This result may indicate that stars from the inner regions of Boo1 have been stripped to the outskirts, but additionally it cannot rule out a scenario where these stars were contributed from a chemically similar dwarf. 

It is unclear at present if the outskirt members of Boo1 are due to the effect of tides, accreted members, or both. To conclude that Boo1’s halo is more consistent with one scenario over the other would require ruling out the possibility of an extended halo from a contributing accretion, via evidence of a separate lower mass galaxy (e.g. a lower [$\alpha$/Fe] knee). Therefore, more spectroscopic observations with detailed abundances will be necessary to conclude the dynamics in this intriguing UFD. For future campaigns of outer profile stars, we identify 34 outskirt candidates whose magnitudes are G $\leq$ 19.5 mag and radial distance is greater than $r_{trans}$. Only 9 of these stars have reported RVs in the literature.

\subsection{Bo\"otes 3}



Boo3 was first identified in SDSS by \citet{grillmair2009}; yet since its discovery, it has remained relatively unstudied. Compared to other UFDs, it is very typical in absolute magnitude ($M_{V}$ $\approx$ $-$5.75 mag; \citealt{correnti2009}) and luminosity ($L_{V}$ $\approx$ 1.74 $\times$ 10$^{4}$ L$_{\odot}$). However, it is a particularly diffuse and large ($r_{h}$ = 33.03$\arcmin$ = 0.56$^{\circ}$; \citealt{moskowitz2020}) system. At a distance of $\sim$46 kpc (\citealt{carlin2018}), its physical half-light radius ($\sim$450 pc) is relatively extensive for a UFD. Boo3's faint luminosity and diffuse nature in part explains the lack of substantial literature on this dwarf. 


Since its discovery, Boo3 has been argued to be the progenitor (or at a minimum, associated with) the Styx stellar stream (\citealt{grillmair2009, carlin2009, carlin2018}). A number of results support this conclusion: firstly, Styx overlaps Boo3 in the matched filter map of \citet{grillmair2009} at a distance corresponding to $\sim$45 kpc (i.e., the same distance of Boo3; \citealt{grillmair2009}). And secondly, Boo3 appears misshapen, double-lobed, and very diffuse (\citealt{grillmair2009}). These morphological characteristics suggest that the system is not in dynamic equilibrium and is being actively disrupted. 

Currently, there are no publications that report any positions or kinematics of resolved stars in Styx. Only the distance to the stream and its length on the sky have been estimated, while orbit properties of Boo3 have been established and used for comparison to the stream's path. \citet{grillmair2009} first attempted to constrain the orbit of Styx (via an orbit-fitting method applied to the on-sky path), and was able to give first estimates for the PMs of Styx stars at the position of Boo3. This requires the assumption of either prograde or retrograde motion for the stream, but recent proper motion estimates of Boo3 in \citet{pace2022} and \citet{battaglia2022} would be consistent with Styx if the stream is in a retrograde orbit, providing further evidence that Boo3 may be the progenitor of Styx.

The updated orbit of Boo3 provided by \citet{pace2022} and \citet{battaglia2022} also strongly supports a disrupted dwarf. Both studies find Boo3's orbit to be highly eccentric ($\epsilon$ $\approx$ 0.85 and 0.95, respectively) and to have a small pericenter distance (7 $-$ 9 kpc). It is important to note that only two MW dwarfs are known to have pericentric orbits of $<$10~kpc. Aside from Boo3, Tuc3’s pericenter is also extremely close to Galactic Center, and is the only UFD known to be tidally disrupting (\citealt{li2018}). Boo3 allegedly recently passed pericenter ($\sim$140 Myrs ago; \citealt{carlin2018}) and, according to \citealt{pace2022}, does have a relatively low density ratio (on the order of Sagittarius and Tuc3), which they argue is in favor of a tidal disruption scenario.

At such a moderate distance, it is surprising that no resolved 
Styx members have yet been identified and kinematically associated to Boo3, though the evidence appears to favor that Boo3 is likely a disrupting dwarf. The most distant spectroscopic member of Boo3 to date is located at 2.56$r_{h}$ (\citealt{carlin2009, carlin2018}), which we find falls within our estimated transition boundaries of $r_{trans, ell}$~=~3.02$r_{h}$ and $r_{trans, circ}$~=~3.11$r_{h}$. We successfully identify all spectroscopic members reported in \citet{carlin2018}, save for 2 stars whose updated proper motions in Gaia eDR3 suggest they are not consistent with membership to Boo3. Though the globular clusters M3 and NGC 5466 (and its stream; see \citealt{jensen2021}) are close to Boo3 in projection, we find no contamination from these systems as their proper motions and CMDs are unequivocally distinct from Boo3 stars. 

At present, the outskirts of Boo3 remain largely unexplored. In this work, we report 65 candidate stars (with no known spectroscopic follow-up) whose distances are greater than $r_{trans}$ and have magnitudes $\leq$19.5 mag.

\subsection{Draco 2}

Dra2 is a very small and nearby collection of stars whose estimated distance and brightness is $\sim$21 kpc and $L_{V}$ $\approx$ 180 L$_{\odot}$ ($M_{V}$ $\approx$ $-$0.8 mag; \citealt{longeard2018}) respectively. With a half-light radius of a mere 3$\arcmin$ (0.05$^{\circ}$, or $\sim$20 pc) and a sufficiently low (marginally resolved) velocity dispersion ($\lesssim$6 km s$^{-1}$ at the 95\% confidence interval), it has been a subject of controversy as to whether it is a true, dynamically cold UFD, or alternatively, a star cluster.

Though Dra2 is very nearby, spectroscopic follow-up of this intriguing system has been fairly limited due to a lack of bright (g $<$ 19 mag) stars. First discovered in the Pan-STARRS 1 (PS1) 3$\pi$ survey (\citealt{chambers2016}) by \citet{laevens2015}, only two works have published further results on the system. \citet{martin2016} and \citet{longeard2018} obtained a total of 14 spectroscopic members reported for Dra2. Of these stars bright enough to be included in Gaia eDR3, we report that we successfully identify all 12 spectroscopic members successfully via our 2-component algorithm. 

The true nature of Dra2 is still largely speculated. \citet{longeard2018} suggest that Dra2 seemingly conforms with other UFD trends in its absolute magnitude, metallicity, and larger size than most globular clusters, but ultimately it lies on the boundary between either object. Given the small number statistics, \citet{longeard2018} are unable to deduce a metallicity dispersion for Dra2 which would yield further evidence towards a dwarf scenario. In addition, the marginally resolved velocity dispersion is very cold for a UFD, but is still comparable to other known dwarf systems like Boo1 (2.4~km~s$^{-1}$; \citealt{koposov2011}) and Seg1 (3.7~km~s$^{-1}$; \citealt{simon2011}). 

If Dra2 is indeed a cluster, the approximate Jacobi tidal radius is only 10 pc (according to \citealt{longeard2018} and assuming the mass is solely stellar), suggesting that the stars we find at many half-light radii would be members that have migrated to significant distances as a result of tides. However, as shown in Figure \ref{fig:tangent_outer_profiles}, our sample of outer members are largely perpendicular to Dra2’s elongation (also perpendicular to the direction of Dra2’s orbit). This particular situation is contrary to the expected distribution of tidal tails, and therefore not likely to be initiated from a tidal stream. 

However, multiple studies argue that Dra2 could be in the midst of tidal disruption. Namely, \citet{pace2022} and \citet{longeard2018} conclude that it is currently at pericenter ($\sim$21 kpc), or close enough to experience tidal disturbance from the MW. Without a resolved velocity dispersion, \citet{pace2022} note that the central density ratio argument (as suggested for Boo1) cannot be utilized for Dra2. However, they note that if the velocity dispersion is sufficiently low ($\sim$1 km s$^{-1}$), Dra2's morphology is likely caused by tidal disruption. 

Currently, the most distant spectroscopic member of Dra2 is located at 2.6$r_{h}$ ($\sim$8$\arcmin$; \citealt{longeard2018}). In our work presented here, we identify members up to 9.15$r_{h}$, and contrary to past works, we identify an additional, bright (G-band = 17.5 mag) RGB star at this large elliptical distance. This particular star was not identified in our 1-component runs, suggesting that the spatial likelihood without the free parameters of an outer profile was too restrictive to obtain a complete sample. Follow-up of this individual star may lead to a better understanding of Dra2, as it is located perpendicular to the system's orbit (as in the stellar features of Tuc2, see below). Further, if this most distant star is bound to the system, it could be argued that Dra2 (under the assumption of a dwarf galaxy nature) could have a sizeable DM halo.




Regardless, Dra2 lacks a substantial sample of bright enough RGBs for spectroscopic follow-up (save the one most radially distant example), as many would-be targets in the outskirts lie at G-band $>$ 19 mag. In this work, we only find 4 stars brighter than G $\leq$ 19.5 mag, and only 2 of these are beyond the estimated $r_{trans}$ = 0.21$^{\circ}$ (4.2$r_{h}$, or 80 pc).

\subsection{Grus 2}

Gru2 was first discovered in the Dark Energy Survey (DES; \citealt{drlica-wagner2015}). It is a moderately distant system ($\sim$55 kpc; \citealt{martinez-vazquez2019}), whose luminosity is consistent to other UFDs ($L_{V}$ $\approx$ 3150 L$_{\odot}$). Though a resolved velocity dispersion has not yet been determined for this system, it is classified as a dwarf galaxy, mainly based on its absolute magnitude ($M_{V}$ $\approx$ $-$3.9 mag), metallicity ([Fe/H] = $-$2.51; \citealt{simon2020}) and relatively large physical size ($r_{h}$ = 5.9$\arcmin$ = 0.1$^{\circ}$, or $\sim$95 pc).

Gru2 is located behind a more metal-rich and diffuse substructure, known as the Orphan/Chenab (OC) stream (\citealt{grillmair_dionatos2006}; \citealt{belokurov2007}; \citealt{shipp2018}). \citet{koposov2019} first reported on these incident structures, finding the proper motions were similar (particularly at stream longitude position of Gru2, or $\phi_{1}$ = 66.1$^{\circ}$) and suggesting that these structures may be linked. However, there is strong evidence to suggest that Gru2 is not the stream’s progenitor, as it differs in heliocentric distances by $\sim$10 kpc (\citealt{martinez-vazquez2019}) and it’s heliocentric RV is offset by $\sim$90~km~s$^{-1}$ (and Gru2's Galactocentric RV v$_{GSR, Gru2}$ $\approx$ $-$131~km~s$^{-1}$, while v$_{GSR, OC}$ $\approx$ $-$200~km~s$^{-1}$; \citealt{erkal2019}; \citealt{koposov2019}). 

As we show in Figure \ref{fig:tangent_outer_profiles}, the OC stream clearly overlaps Gru2 (grey highlight), but additionally the calculated transition boundary reaches the stream’s center. Given that our method relies on proper motions, but not parallaxes/distances or RVs, it is very likely that the OC stream affects our determination of an outer profile for Gru2. It can also be noted that Gru2 is the only dwarf with no stars beyond our calculated $r_{trans}$. 

Though Gru2's outer profile detection may be compromised by an unrelated structure in the field, we note that the algorithm identifies only member stars out to 5$r_{h}$ (all of which are also members in our 1-component model). In comparison, previous work in \citet{simon2020} estimated a tidal radius for Gru2 to be roughly 300 pc, or $\sim$3$r_{h}$. According to the central density to MW pericenter density ratio calculated in \citet{pace2022}, Gru2 is potentially tidally disrupting. The spectroscopic members traced in \citealt{simon2020} extend out to 2$r_{h}$, and so it is possible that some radially distant stars are evidence of this disruption. However, given the OC stream’s width and overlap with these stars, it is certainly possible that the outer stars we identify are indeed members of the OC stream and not Gru2. 

In dynamical models constructed in \citet{koposov2022}, the authors claim that, while Gru2 itself is not the progenitor of the OC stream, its orbit coincides with recent interactions with the OC progenitor. These close passages suggest that Gru2 may have been accreted with the OC progenitor, which is more similar in mass to a classical dwarf system. Determining the influence of tides on Grus2 is complicated by the interplay with not only the OC stream, but additionally the LMC. Orbit projections in \citet{battaglia2022} suggest that, while Gru2 was not accreted with the LMC, it has been interacting/potentially captured by the Magellanic group within the past 200 Myrs. This is further corroborated by results in \citet{santos-santos2021} who reported that Gru2 is a potential associate of the LMC. Given the relatively close pericenters (24 $-$ 31 kpc) determined by \citet{pace2022} and \citet{battaglia2022}, and the dwarf’s low central density, tides may have been important in shaping this system. Suffice to say, the complicated interplay of Gru2, the LMC, and the OC stream make for intriguing future work.

\subsection{Sculptor}

Of the many MW satellites, Scl was the first dwarf spheroidal (dSph) discovered (\citealt{shapley1938}), thereby earning its title as a classical dwarf. It is the fourth brightest dSph at $M_{V}$ $\approx$ $-$10.82 mag ($L_{V}$ $\approx$ 1.8 $\times$ 10$^{6}$ L$_{\odot}$; \citealt{munoz2018}), preceded only by Fornax, Leo1, and Sagittarius. It is a typical-sized system with a physical half-light radius of $\sim$310 pc ($r_{h}$ = 12.33$\arcmin$ = 0.21$^{\circ}$) and a large velocity dispersion ($\sigma_{RV}$ $\sim$ 9.2 km~s$^{-1}$; \citealt{walker2009}) indicative of the presence of dark matter. 

Given the brightness and angular size of this system, Scl’s resolved stellar populations have long been studied, yielding a wealth of information and detail into this particular system. Multiple studies have explored the chemo-dynamics of Scl and concur that it consists largely of old ($>$10 Gyrs) metal-poor ([Fe/H] = $-$1.8 dex; \citealt{tolstoy2023}) stars with well-mixed enrichment (e.g., \citealt{hill2019}). Star formation in Scl is proposed to have had one epoch which ceased 8 $-$ 10 Gyrs ago and lasted $\lesssim$1.4 Gyrs (\citealt{bettinelli2019}); \citealt{delosreyes2022}). Much like Boo1 and UMi (see below), Scl’s innermost regions ($<$2$r_{h}$) are composed of a metal-rich and kinematically cold population that transitions to a metal-poor, kinematically hot population (e.g., \citealt{tolstoy2004}; \citealt{battaglia2008}). The transition from the inner to the outer population occurs at $\sim$0.2$^{\circ}$ ($\sim$0.95$r_{h}$), and is potentially a result of outside-in star formation. 

Regarding kinematic details, little evidence has been found to suggest that Scl is in the process of losing stellar mass via tides. For example, both \citet{pace2022} and \citet{battaglia2022} find large pericentric distances for Scl, in the range of 45 $-$ 65 kpc, depending largely on the assumed MW mass. Additionally, \citet{tolstoy2023} recently consolidated archive spectra for Scl RGBs out to the nominal tidal radius ($\sim$6.2$r_{h}$ via a King density profile; \citealt{irwin_hatzidimitriou1995}) and did not resolve an RV gradient (see also \citealt{martinez-garcia2022}). At a heliocentric distance of 86 kpc (\citealt{munoz2018}), MW tidal influence is not purportedly large enough to strongly distort Scl’s morphology. Recent simulations from \citet{iorio2019} show that the stellar component of Scl is not obviously affected by tides, but up to 60\% of the DM halo may have been stripped by present-day. 

In comparison, recent work by \citet{sestito2023_scul} utilizes our candidate members for Scl and obtained spectroscopic follow-up for the 2 most radially distant (out to 10$r_{h}$) and relatively bright stars in our sample. The authors confirm that these distant stars are \textit{bona fide} members of Scl based on their metallicities and RVs. As this finding expands the radial stellar distribution of Scl out to $\sim$3 kpc, the authors also explored the stellar density profile and its derivative $-$ $\Gamma$($r_{ell}$) $-$ using the full sample of candidates in this present work and compared them to a tidal disruption simulation from \citet{penarrubia2008}. With our candidate members, \citet{sestito2023_scul} report a sharp deviation from an exponential profile in the stellar density in Scl (also shown in this work in Figure \ref{fig:profiles}), located at about 25$\arcmin$. According to the tidal model of \citet{penarrubia2008}, tides can be responsible for an outer excess over the initial exponential profile, also called at a “kink” radius. This is followed by a region where $\Gamma$($r_{ell}$) approaches a power-law with a slope of $-$4, before flattening back to $\sim$0. The second change in $\Gamma$($r_{ell}$) occurs where the local crossing time of stars in the dwarf exceeds the time since the last pericentric passage, and is referred to as a “break”. Remarkably, the \citet{penarrubia2008} tidal model matches well with our observed Scl data. The conclusion of work done in \citet{sestito2023_scul} is that the extended profile in Scl is indeed indicative of tidal disturbance. 



The “kink” in $\Gamma$($r_{ell}$) is located at $\sim$25$\arcmin$ ($\sim$2$r_{h}$), corresponding with a profile excess also seen in the density profiles of \citet{irwin_hatzidimitriou1995} and Figure \ref{fig:profiles}. In our work, we find the transition boundary solutions are $r_{trans, ell}$ = 0.84$^{\circ}$ ($\sim$4$r_{h}$) and $r_{trans, circ}$ = 1.06$^{\circ}$ ($\sim$5.1$r_{h}$). These are much larger than the “kink” radius from \citet{sestito2023_scul}, yet are less than the nominal tidal radius from \citet[and less than the “break” radius from \citealt{sestito2023_scul} which was found to be $\sim$80$\arcmin$ = 6.5$r_{h}$]{irwin_hatzidimitriou1995}. Seemingly, our proposed values do not correspond to these estimates. 


\begin{figure}
    \centering
    \includegraphics[width=0.75\columnwidth]{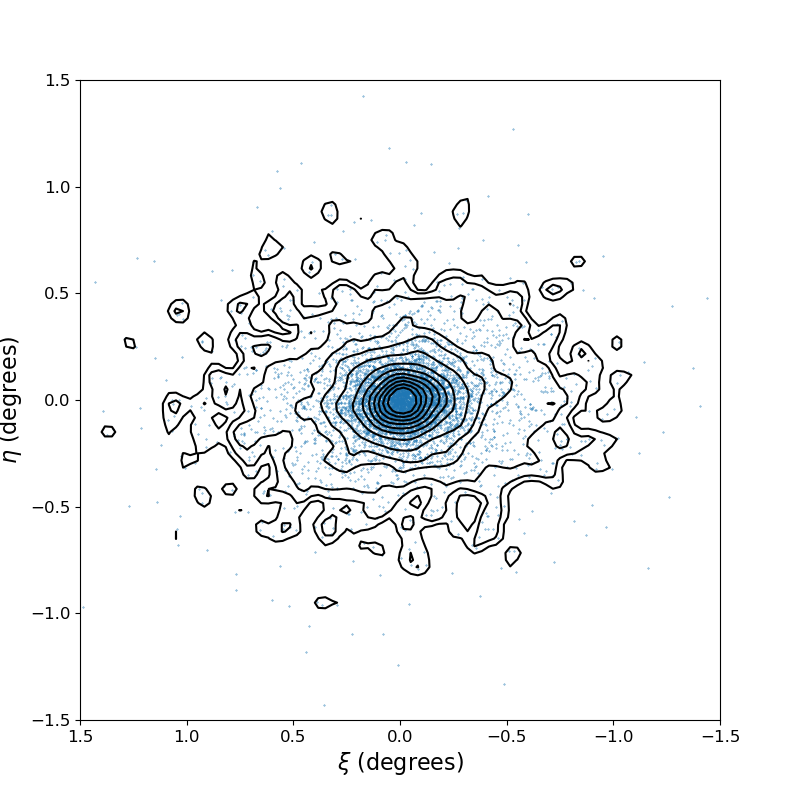}
    \caption{Isophote contour map of Sculptor using candidate members with $P_{max} \geq$~10\%. The contours are made with 2$\arcmin$x2$\arcmin$ pixels, smoothed with a Gaussian kernel ($\sigma$ = 2$\arcmin$). Contour levels are set to [0.0025, 0.005, 0.01, 0.03, 0.05, 0.1, 0.2, 0.3, 0.5, 0.7, 0.8] of the maximum counts. The central surface brightness corresponds to 29.8 mag/arcsec$^{2}$, while the outermost contour corresponds to 23.3 mag/arcsec$^{2}$.}
    \label{fig:scul_contour}
\end{figure}

Given the number of candidate members identified in this work, we also are able to produce an isophote contour map for Scl to highlight any morphological distortions in the outskirts (Figure \ref{fig:scul_contour}). The contours are made with 2$\arcmin$x2$\arcmin$ pixels, smoothed with a Gaussian kernel ($\sigma$ = 2$\arcmin$). Contour levels are set to [0.0025, 0.005, 0.01, 0.03, 0.05, 0.1, 0.2, 0.3, 0.5, 0.7, 0.8] of the maximum counts. The central surface brightness corresponds to 29.8 mag/arcsec$^{2}$, while the outermost contour corresponds to 23.3 mag/arcsec$^{2}$. In agreement with past works, we find that Scl's morphology does not appear to be strongly distorted, but is rather nicely elliptical (particularly for the innermost contours). 

Prior to \citet{sestito2023_scul}, the most distant spectroscopic member was 6$r_{h}$, which we now expand out to 10.1$r_{h}$. Not including those already observed by \citet{sestito2023_scul}, we find 48 outskirt ($>r_{trans}$) candidates for spectroscopic follow-up (G $\leq$ 19.5 mag).

\subsection{Segue 1}

At a heliocentric distance of 23 kpc (\citealt{belokurov2007_cats}), Seg1 is one of the nearest dwarf galaxies. Its size is relatively compact ($r_{h}$~=~3.95$\arcmin$~=~0.07$^{\circ}$, or $\sim$30 pc; \citealt{munoz2018}) and similar in luminosity to other MW UFDs ($M_{V}$ $\approx$ $-$1.5 mag; $L_{V}$ $\approx$ 340 L$_{\odot}$; \citealt{martin2008}). In the discovery paper of \citet{belokurov2007_cats}, this system was proposed as a globular cluster containing very few stars. However, spectroscopic follow-up conducted by \citet{geha2009} and \citet{martinez2011} suggested that Seg1's internal kinematics are consistent with a dwarf galaxy. The former determined that Seg1 has an exceptionally large dark matter content (M/L~$\approx$~1340~$-$~2440~M$_{\odot}$/L$_{\odot}$) while the latter claimed this system may be one of the highest DM density systems in the Local Group.

As the focus of many individual spectroscopic studies, Seg1 has consistently proven to be a particularly unique system. In addition to its high M/L ratio, Seg1 is one of the most metal-poor dwarfs in the MW ([Fe/H] = $-$2.7; \citealt{frebel2014}; \citealt{norris2010_boo1_seg1}), comparable to other UFDs like Tuc2, Hor1, and Boo2 (\citealt{simon2019}). Intriguingly, \citet{frebel2014} observed a large metallicity dispersion in this system ($\sigma_{[Fe/H]}$ = 0.95), yet all stars were found to have high $\alpha$-enhancement regardless of metallicity (see also \citealt{vargas2013}). Given the high [$\alpha$/Fe] ratios across all measured stars in this study, the authors deduced that Seg1 lacks enrichment from Type Ia supernovae (SNe) which pollute the dwarf’s ISM with iron over time, and that its star formation must have only lasted for a few hundred Myrs. Enriched only by the core-collapse of massive stars, Seg1's exceptionally distinctive chemical trends suggest it is a candidate “first galaxy” (\citealt{webster2016}), or one that experienced a short burst of star formation, and has remained since unchanged. 

Seg1 is located in a particularly advantageous position for spectroscopic study. Its close proximity, in addition to its high Galactic latitude (+50$^{\circ}$) means that a) stars down to $\sim$1 mag below the main sequence turn-off are accessible for spectroscopic follow-up (\citealt{frebel2014}), and b) contaminating MW foreground stars can be effectively removed. As such, previous work by \citet{simon2011} obtained a fairly complete sample of radial velocity measurements for 98.2\% of candidate member stars within 10$\arcmin$ (or $\sim$2.5$r_{h}$) of Seg1. In an independent study, \citet{norris2010_boo1_seg1} targeted stars out to $\sim$9$r_{h}$ in the $\textit{g}$-band range of [17, 21.4] mag. Additionally, Seg1’s field was partially covered by APOGEE DR17. The relatively large number of stars with radial velocities in and around the dwarf (348 stars in total; see Table \ref{tab:dSphs_all}) alongside the high completeness achieved in previous studies for the inner regions of Seg1 provides us with a more complete dataset to test the robustness of our algorithm detection. 

Using the samples from \citet{norris2010_boo1_seg1} and \citet{simon2011}, we compare our probabilities estimated via the algorithm to their velocity-defined members. For the \citet{simon2011} sample, of which 30 stars are bright enough to be in Gaia eDR3 and the most distant member is located at 2.9$r_{h}$, we obtain consistent membership probabilities for all stars. For the \citet{norris2010_boo1_seg1} sample, we find similar consistency such that all but 2 of their spectroscopic members are also found as members in the algorithm. These 2 remaining cases were found in this work to have inconsistent PMs to the rest of Seg1, and thus have been flagged as non-members.  


Thus far, the most radially distant spectroscopic member of this system is 5.25$r_{h}$. For comparison, our most distant member (with no present spectroscopic follow-up) is located at 6.8$r_{h}$. We determine the radius of the transition boundary is  $r_{trans, ell}$ = 0.11$^{\circ}$ ($\sim$1.57$r_{h}$) and $r_{trans, circ}$ = 0.14$^{\circ}$ ($\sim$2$r_{h}$) in the elliptical and circular cases, respectively. For this well-studied system, we only identify 2 candidates beyond $r_{trans}$ with G $\leq$ 19.5. These are located at $r_{ell}$ = 5.5 and 6.3$r_{h}$, and G = 17.7 and 18.0 mag respectively. 

Though Seg1 has been extensively studied, it is at present unclear if it is currently undergoing tidal disruption. A photometric study conducted by \citet{niederste-osthold2009} claimed to have found a 1$^{\circ}$ extension of stars they deemed “tidal debris”. However, a more detailed analysis in \citet{geha2009} reports no kinematic tracers nor any other clear evidence of tidal disruption from their sample. \citet{geha2009} also determine a Jacobi radii of 220 pc (or 7.4$_{h}$) where stars in Seg1 are likely to be unbound. We confirm in our work that no member stars (in Gaia eDR3) are beyond this radius. 

Searching for tidal signatures in Seg1 is further complicated by the multitude of stellar streams intersecting in the field of the dwarf. These are: Slidr, Sagittarius, Gaia-10, Orphan/Chenab, and the 300 km~s$^{-1}$ stream (\citealt{mateu2023}; see Figure \ref{fig:tangent_outer_profiles}). In this less than fortunate circumstance, we  do find sources of contamination from these substructures. 6 stars with measured RVs are flagged as Seg1 members, but their origin is clearly the 300 km~s$^{-1}$ stream (at 18 kpc in heliocentric distances, and approximately the same proper motions; \citealt{frebel2013}; \citealt{fu2018}) while 8 more originate from a lower RV structure (likely Sagittarius, see \citealt{geha2009}). However, we reiterate that the algorithm performs well at minimizing the number of false negatives, though there is contamination of false positives from other substructures (especially when they have similar proper motions to the dwarf) than would be ideal. Given the detailed spectroscopic follow-up thus far of the system, it is indeed impressive that only a few interlopers from these stellar streams managed to contaminate our sample. 


\subsection{Tucana 2}

Upon first glance, Tuc2’s characteristics are fairly typial compared to other UFDs. Its intrinsic brightness ($M_{V}$ $\approx$ $-$3.9 mag; $L_{V}$ $\approx$ 3150 L$_{\odot}$) and physical size ($r_{h}$ = 9.83$\arcmin$ = 0.16$^{\circ}$, or $\sim$160 pc; \citealt{bechtol2015}) places it squarely in the locus of other MW satellites in the size-luminosity relation (see Figure 2 of \citealt{simon2019} and Figure 3 of \citealt{bechtol2015}). Velocity dispersion measurements also support Tuc2’s classification as a dwarf galaxy ($\sigma$~$\sim$~3.8~km~s$^{-1}$; \citealt{chiti2023}), and its mean metallicity falls well below the proposed “metallicity floor” for most MW globular clusters (\citealt[2010 version]{harris1996}; \citealt{wan2020} and references therein). A somewhat unique characteristic is its low metallicity, suggesting Tuc2 is one of the most metal-poor systems in the MW ([Fe/H]~=~$-$2.71 dex; \citealt{ji2016}; \citealt{chiti2018}), on average.

Despite these parallels to other UFDs, Tuc2 has been shown to be a particularly interesting dwarf. In 2021, \citeauthor{chiti2021} searched the outskirts of Tuc2 for spectroscopic members and reported two compelling results. Firstly, they detected the most radially distant Tuc2 member, located at 1.1~kpc (10.71$r_{h}$ in elliptical radii). This observation was (at the time) the most distant star in a UFD, preceded only by spectroscopic members of other dwarfs up to 4$r_{h}$ (\citealt{frebel2016}; \citealt{norris2010_boo1_seg1}). Secondly, the authors identified 7 RGBs in Tuc2’s outskirts, located at distances $>$2$r_{h}$. The authors suggested these stars were members of an extended stellar component which originated from the dwarf. As to the physical motivation for these stars to be displaced to such large radii, \citet{chiti2021, chiti2023} argue that their location conflicts with tidal stripping. Instead, \citet{chiti2021, chiti2023} suggest that this structure is a result of an early galactic merger, whose remnant accreted stars remain in the outskirts of Tuc2 (corroborated by the outside-in formation scenario in dwarf-dwarf mergers; \citealt{benitez-llambay2016}). This initial discovery is significant to our understanding of dynamics of stars in low-mass systems, particularly as we do not yet have a census of dwarfs whose outskirts show evidence for this type of scenario.

\citet{chiti2021, chiti2023} propose three main possibilities for the extended substructure in Tuc2. One hypothesis is that Tuc2’s structure could be the result of bursty feedback or energetic SNe. By comparing the innermost stellar chemistries to outskirt members, \citet{chiti2023} found that the metallicities and chemical abundances are similar to other UFDs. While UFDs in their formation may have experienced early bursty feedback (e.g., \citealt{wheeler2019}), the conclusion that Tuc2 seems to have had similar enrichment compared to other UFD systems suggests that the energies of past SNe would make the extension in Tuc2 a much more common feature than currently observed. In agreement with this conclusion, the authors find that the chemical abundance trends favor a low mass progenitor, and therefore does not suggest particularly energetic SNe that could act to displace these stars.

The second (and most common conclusion) is a tidal scenario. However, the authors' observations find no conclusive evidence for tides. When modelling its tidal disruption, \citet{chiti2021} determined that Tuc2’s tidal debris would be observed following the direction of orbit, as is typical for many other dwarf and globular cluster streams. Even when including the time-varying potential of the LMC in their orbits, they find that Tuc2’s observed stellar extension is indeed located perpendicular to the proposed direction of tidal debris. The authors further confirmed in later high-resolution spectroscopic studies that there is no clear RV gradient in Tuc2 that would be indicative of disruption (\citealt{chiti2023}). 

Also contrary to tides is Tuc2's current distance (58 kpc; \citealt{bechtol2015}), but most importantly, its orbit places its pericenter between $\sim$35 kpc (\citealt{battaglia2022}) and $\sim$45 kpc (\citealt{pace2022}), depending on the model for MW mass (both including the LMC). In either case, this pericenter is not particularly small, and indeed \citet{pace2022} do not find a particularly large density ratio for this system that would suggest it is actively tidally disrupting. 

The lack of evidence for both the previous scenarios (i.e., tidal origins or expulsion due to energetic SNe) led \citet{chiti2021, chiti2023} to this final hypothesis: that Tuc2’s morphology and steep metallicity gradient could be the remnants of an ancient merger. Simulations of a UFD (similar to Tuc2) in \citet{tarumi2021} showed that the early (first $\sim$100s Myrs) merging of two galaxies can produce extended features in the direction of the merger, which could explain the perpendicular stellar extension in Tuc2. These simulations also showed that accreted stars from the merger can be deposited in the outskirts of the central dwarf, as the central dwarf is dynamically heated which forms an extended stellar distribution. Similarly, simulations by \citet{benitez-llambay2016} find that mergers in dwarfs can induce metallicity (and age) gradients in the surviving dwarf galaxy. Arguably, this could be why \citet{chiti2023} observed a metal-poor inner component ([Fe/H] = $-$2.71 dex) yet an even lower metallicity outer component ($>$2$r_{h}$, [Fe/H] = $-$3.02 dex) in Tuc2.

The nullification of the previous two hypothesis presents a strong case for the dwarf-dwarf merger remnant scenario, though direct observational evidence is clearly difficult to ascertain. Nonetheless, in our methods presented here we successfully identify all spectroscopic members in the literature as members of Tuc2. We find only one member star with a particularly low probability (7\%); however, this star is evidently an AGB, and considering we do not actively model AGB in our methods, the CMD likelihood is very low for this star. 

Given the expansive work conducted by previous studies, very few stars in our sample are candidates for follow-up. The most radially distant star we identify is the same as \citet{chiti2021} at a distance of $r_{ell}$ = 10.71$r_{h}$. Considering that the transition boundaries are $r_{trans, ell}$ = 0.70$^{\circ}$ ($\sim$4.38$r_{h}$) and $r_{trans, circ}$ = 0.54$^{\circ}$ ($\sim$3.38$r_{h}$), there are only 4 candidates in the outer profile regime which have not yet been observed. We note that 3 of these stars are HBs.

\subsection{Tucana 3}


Tuc3 is a fairly small ($r_{h}$ = 6$\arcmin$ = 0.1$^{\circ}$, or $\sim$45 pc) and faint ($L_{V}$~$\approx$~800~L$_{\odot}$; $M_{V}$ $\approx$ $-$2.4 mag; \citealt{drlica-wagner2015}) collection of stars located near the LMC. Discovered in DES alongside Gru2, the first observation of the dwarf was reported simultaneous to the $\sim$4$^{\circ}$-long tidal tails emanating from the dwarf's center. Besides the aforementioned Dra2 and Seg1 dwarf galaxies, Tuc3 is also one of the closest dwarfs ($\sim$25 kpc). Its proximity has enabled multiple high-resolution spectroscopic studies such that its stellar stream and chemical trends have been studied in detail (e.g., \citealt{simon2017}; \citealt{hansen2017}; \citealt{li2018}; \citealt{marshall2019}). 

Though the presence of tidal tails is a strong indication of tidal disruption, the velocity dispersion of Tuc3 is smaller than typically associated with dwarf galaxies (an upper limit reported as $\sigma_{RV}$ $\sim$ 0.1 km~s$^{-1}$ in \citealt{simon2017}). However, \citet{simon2017} argued that the low velocity dispersion could have resulted from the removal of stars which then acts to decrease the velocity dispersion in the system. Indeed, \citet{marshall2019} observed a metallicity dispersion in Tuc3 $-$ a result in favor of a dwarf galaxy origin since the potential wells in globular clusters are not typically large enough to retain the gas ejected from SNe. At present, its dwarf nature is not ultimately confirmed but certainly likely. 


From these spectroscopic studies, it has also been found that Tuc3 is one of only two $\textit{r}$-process enhanced UFDs in the MW (\citealt{hansen2017}). \citet{marshall2019} showed that this enhancement must have occurred before Tuc3’s most recent pericentric passage, as these same enrichment features are observed in the core and most distant stars in the tails. 

We identify most spectroscopic members; however, we find 7 members with substantially low probabilities that are not included in our member lists. Given their proper motions and position on the CMD, they indeed appear to be consistent with membership to Tuc3. We note that these 7 stars are the most radially distant of the sample, resulting in low spatial probabilities. This result is a limit of using a centrally concentrated outer component, rather than having a prior that would be more consistent to a longer stream structure. Nonetheless, we find that the most distant spectroscopic member within our probability limit is at 14.25$r_{h}$, though the actual most distant spectroscopic member to date is located at 16.63$r_{h}$. 

The reported transition boundary for Tuc3 is $r_{trans}$ = 0.34$^{\circ}$ = 3.4$r_{h}$. 
Though we know Tuc3 to be tidally disrupting, it is unclear if this boundary is physically motivated. Close inspection of the stellar density profile in Figure 8 of \citet{drlica-wagner2015} shows a break at this radius, which at minimum appears to be a transition from the Tuc3 core to the tidal tails. We note however that a break can also be seen in our stellar density profiles in Figure \ref{fig:profiles}. Given our transition boundary, we identify 18 stars (5 of which are HBs) in the outskirts with reasonable magnitudes that have not been already spectroscopically observed.

\citet{pace2022} and \citet{battaglia2022} updated kinematic information for Tuc3 and both reported very small pericenter distances ($\sim$1 $-$ 4 kpc), a very radial orbit, and high eccentricities ($\approx$ 0.85 $-$ 0.95). \citet{pace2022} additionally note that Tuc3 is a system whose central density ratio is indicative of likely tidal influence. Interestingly, Tuc3 has also been proposed to be intertwined with the LMC system, as \citet{erkal2018} reported that Tuc3 passed within tens of kpcs to the LMC some $\sim$75 Myrs ago.


\subsection{Ursa Minor}


\begin{figure}
    \centering
    \includegraphics[width=0.75\columnwidth]{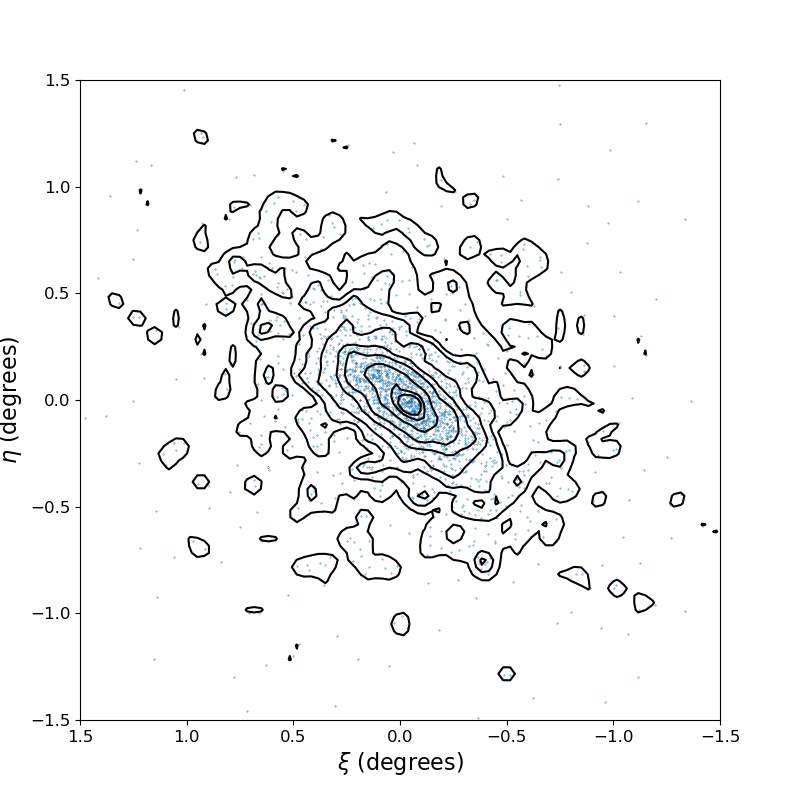}
    \caption{Isophote contour map of Ursa Minor using candidate members with $P_{max} \geq$~10\%. The contours are made with 2$\arcmin$x2$\arcmin$ pixels, smoothed with a Gaussian kernel ($\sigma$ = 2$\arcmin$). Contour levels are set to [0.01, 0.03, 0.05, 0.1, 0.2, 0.3, 0.5, 0.7, 0.8] of the maximum counts. The central surface brightness corresponds to 30.8 mag/arcsec$^{2}$, while the outermost contour corresponds to 25.8 mag/arcsec$^{2}$ (100$\times$ fainter than the central surface brightness).}
    \label{fig:umi_contour}
\end{figure}

The final system where we find evidence for an outer profile is the classical system, UMi. UMi resides at a relatively large heliocentric distance ($\sim$76 kpc; \citealt{bellazini2002}), yet it was one of the first discovered MW satellites (\citealt{wilson1955}), owing largely to its luminosity, $M_{V}$~$\approx$~$-$9.03~mag ($L_{V}$ $\approx$ 3.6 $\times$ 10$^{5}$ L$_{\odot}$; \citealt{munoz2018}). It is a sufficiently large system ($r_{h}$ = 17.32$\arcmin$ = 0.28$^{\circ}$ or 380 pc; \citealt{munoz2018}) with a large velocity dispersion (8.6~km~s$^{-1}$; \citealt{spencer2018}) similar to other classical dwarfs.

It has been speculated in many studies that tides may affect UMi; this was concluded from (i) morphological asymmetries, (ii) a relatively large ellipticity ($\epsilon$ = 0.55; \citealt{munoz2018}), and (iii) stellar clumps that are offset from the dwarf's centroid, and appear aligned with UMi's orbital direction (e.g., \citealt{olszewski_aaronson1985}; \citealt{irwin_hatzidimitriou1995}). To better constrain the stellar density of UMi out to its nominal tidal radius, \citet{palma2003} conducted a large-area ($\sim$3$^{\circ}$ radius) photometric survey of UMi, and confirmed two main results: the first being that UMi’s isocontours appear S-shaped rather than elliptical (see also \citealt{irwin_hatzidimitriou1995}), and the second was that UMi has two main peaks in stellar densities which are offset from the dwarf’s center. Though the photometric data do not confirm if the most distant stars are bound or unbound to UMi, the authors propose their presence and UMi’s morphology are likely as a result of tides. 

Multiple properties of UMi that would be further indicators of tidal influence remain speculated. For example, UMi’s most recent orbit estimates seemingly do not suggest a substantially close pericenter such that tides can strongly affect the dwarf (see simulations in \citealt{read2006}). Pericenters estimated from \citet{pace2022} and \citet{battaglia2022} range from 35 $-$ 55 kpc, depending on the potential used. Secondly, UMi’s Jacobi tidal radius\footnote{We note that the tidal radius assumes dynamical equilibrium. In the case of UMi, it has been suggested that the dwarf recently passed apocenter (see \citealt{sestito2023_umi} and Figure 6 in \citealt{martinez-garcia2022}) and may be reasonably relaxed such that the tidal radius could be useful.} estimates range from $\sim$76$\arcmin$ (1.3$^{\circ}$, or $\sim$500 pc; \citealt{palma2003,piatek2005}) up to 4.4$^{\circ}$ (or 6 kpc; \citealt{pace2020}). We further note that \citet{pace2022} do not find a substantially low central density ratio such that UMi is a tidal candidate, and though a velocity gradient in UMi has been detected in multiple studies (e.g., \citealt{hargreaves1994}), the trend is identified along the minor axis, suggesting a modest internal rotation and not necessarily tides.

Spectroscopic study of UMi’s outskirts is already in progress. \citet{sestito2023_umi} obtained spectroscopic follow-up for 5 of the most distant and bright stellar candidates identified in this work, ranging from distances of 5.1 to 11.7$r_{h}$, confirming that all targets are indeed UMi members. According to previous studies, the most distant member was located at 6.8$r_{h}$; with this work, we extend the outskirts of UMi out from $\sim$2.6 kpc out to $\sim$4.4 kpc. We find virtually no contamination in this sample, as we find \textit{only} 4 contaminants in the RV sample (i.e., they show inconsistent RVs but are flagged as members in the algorithm). We note the contamination rate for UMi is only 4 stars out of the total 436 that have RV measurements.


In this work, we determined the transition boundaries in UMi are equal to $r_{trans, ell}$ = 1$^{\circ}$ (or 3.46$r_{h}$) and $r_{trans, circ}$ = 1.29$^{\circ}$ (or 4.47$r_{h}$). We note the circular solution coincides not only with a previous estimate for the tidal radius (from \citealt{piatek2005} and \citealt{palma2003} at 1.3$^{\circ}$) but also to a break in the outskirts of the stellar density profile in \citet{palma2003}, located at $\sim$80$\arcmin$ (see their Figure 14, and our stellar density profile in Figure \ref{fig:profiles}). In comparison, \citet{sestito2023_umi} report the “kink” radius for UMi is $\sim$30$\arcmin$ (or 1.7$r_{h}$) while the “break” is $\sim$225$\arcmin$ (or 13$r_{h}$).

In the same manner as Scl, we also created an isophote contour map for UMi using our candidate members from this work. Figure \ref{fig:umi_contour} shows contour levels from [0.01, 0.03, 0.05, 0.1, 0.2, 0.3, 0.5, 0.7, 0.8] of the maximum counts, where the central surface brightness corresponds to 30.8 mag/arcsec$^{2}$ and the outermost contour corresponds to 25.8 mag/arcsec$^{2}$. Particularly in the central contours, there is some evidence of distortion (i.e., less clearly elliptical) moreso than in Figure \ref{fig:scul_contour} for Scl.

Aside from the 5 candidates in \citet{sestito2023_umi}, we find that there are 29 candidate outskirt($>r_{trans}$) members with no current spectroscopic follow-up.


\subsection{Additional Considerations}
\label{sect:limits}

The algorithm has proven to be robust in identifying (i) candidate members of MW dwarf galaxies and (ii) systems that exhibit evidence of outer substructure, in agreement with much of the literature. In this section, we emphasize three main considerations:

\begin{itemize}
    \item The 9 systems identified in this work may not be the only systems in our sample with two components, but they are the only systems where the second component is identifiable using this technique with Gaia data.
    \item There are many systems with stars at relatively large radii ($>$5$r_{h}$) that do not require two components to explain the overall stellar distribution.
    \item We have determined that the detection of an outer profile in our 9 systems is robust against uncertainties in the structural parameters (ellipticity, position angle, and half-light radius). We have confirmed that all systems use the most recent updates in structural parameters.
\end{itemize}

To address the first of these points, we note that our detection of an outer profile in each dwarf is limited by the density and physical extent of the structure. For example, it will be more difficult to detect a secondary component around small systems that contain only a handful of stars. As such, we cannot say there exist \textit{only} 9 systems with extended stellar features. We anticipate the total number of dwarfs with outer profiles will likely increase with upcoming Gaia data releases and future large-sky surveys like LSST.

The second consideration relates to the fact that stars can exist at large radius from the host galaxy without requiring the galaxy to have a secondary component. For example, recent follow-up by \citet{roederer2023} identified 5 member stars in Sextans at distances between 3.5 $-$ 10$r_{h}$. In comparison, our 1-component model of Sextans does similarly identify distant candidate members, but only out to 6.8$r_{h}$. This is also the case in Fornax for which we find candidate members out to $\sim$7$r_{h}$ (2.1$^{\circ}$), or the same radial distance as stars reported in \citet{yang2022}. In these cases, a secondary component is not necessary to find member out to beyond 5$r_{h}$, and the existence of members at these radii is consistent with these systems having a single structural component.

And finally, we consider the limitations of the outer component as it relates to the assumed structural parameters. As discussed in Section \ref{sect:spatial}, a consequence of the 2-component spatial model is that we cannot easily create a likelihood map that accounts for uncertainties in the dwarf’s spatial parameters. A particular case arose with Gru1, as recent structural parameters for this system have been significantly changed. Specifically, Gru1’s new $r_{h}$ is in fact 2.5 times larger, and the position angle is now rotated into the next quadrant (i.e., it now is oriented West of North instead of East of North) resulting in a vastly different orientation and size in the likelihood model. In using the previous values for Gru1, we found that the algorithm determined that Gru1 had an outer profile, albeit the transition boundary was only slightly larger than the new measurement for Gru1’s half-light radius. Additionally, the scale of the outer profile ($B$) was found to be much larger than in any of the other detections.

Determining the stability of the outer profile is an important outcome of this work, and so we thoroughly examined our list of outer profile systems to confirm that our detections are not hindered by large structural uncertainties. In doing so, we were able to verify that:

\begin{enumerate}
    \item The uncertainties in position angle, ellipticity, and half-light radius are all sufficiently smaller than the changed percent error in Gru1 (or for position angle, a small enough change such that the orientation is not in the opposing quadrant); 
    \item If updated structural information exists, we confirmed that using these new values do not produce a different membership list for these dwarfs, even if the individual probabilities may be different; and
    \item The updated values do not invalidate the detection of an outer profile. 
\end{enumerate}

\noindent Of the dwarfs where we detected outer profiles, the only systems with recently updated spatial parameters were Boo1 and Gru1. We found no change in Boo1’s results when using the updated parameters in \citet{longeard2022}.


\section{Conclusions \& Summary}
\label{sect:summary}

In this work, we present an updated algorithm for the detection of dwarf galaxy stellar members using Gaia eDR3 photometry and astrometry. Our application differs from previous works in that we allow for the detection of an outer component in the stellar distribution (if there is one), enabling us to develop a  census of dwarfs which likely host extended structure. We applied the algorithm to all known MW dwarf galaxy satellites (and dwarf candidates). Of these, we detect outer profiles in a total of 9 systems; these are: Boo1, Boo3, Dra2, Gru2, Seg1, Scl, Tuc2, Tuc3, and UMi.

We have shown that the presence of an extended profile in our model does successfully acquire additional members in the outskirts of these dwarfs. Many of these stars would have likely been missed without this change, as evidenced by the excess of stars seen in the stellar density profiles of Figure \ref{fig:profiles}. The inclusion of these distant members indicates a divergence from the fiducial 1-component exponential function and at distances $>r_{trans}$. The only exception is Gru2, which we discuss is limited by the presence of the OC stream that overlaps the dwarf in projection (and inconveniently has the same proper motions as Gru2).

For systems with an extended profile, we note that the addition of a secondary component does not change the vast majority of stellar members in the interior ($<r_{trans}$). Primarily, the change is evident in the outskirts where we observe that the 1-component case restricts the detectability of distant members. Using our membership lists for Scl and UMi, \citet{sestito2023_scul, sestito2023_umi} showed that the 1- and 2-component data produce exceptionally similar inner stellar density profiles, save for the evident truncation of stars in more distant radial bins. As the 1-component model has been shown to limit these detections, we report that our adjustments merely allow for the inclusion of additional stars at large radii, allowing us to probe further into the outskirts of these satellites. 


Of the systems that we identify as having extended structure, a handful in the literature have already been reported as possibly disrupted systems (Boo1, Boo3, Tuc3, Scl, UMi; see references for each in Section \ref{sect:discussion}). Encouragingly, we also find evidence of an extended feature in Tuc2. We therefore conclude that the algorithm performs well at locating systems with extended haloes and tidal disturbances. We again note however that we remain agnostic regarding the origin of these outer features, and that it will be up to future follow-up campaigns to make this conclusion. 

An additional concept we report here is the boundary at which the spatial likelihood transitions from the inner exponential to the outer, which may prove to be an interesting feature. This feature corresponds to known breaks/bumps in the stellar density profiles of other works as explored in Section \ref{sect:discussion}. As Figure \ref{fig:profiles} shows, the transition boundary nicely corresponds with the divergence of the stellar density from a singular exponential. In some clear cases (like Boo3 and Tuc3), we note the substantial excess of stars beyond this boundary appear as a stellar density bump. We argue it may be a useful limit to examine the central dwarf members compared to the stellar halo outskirt stars, for exploration in future works.

We also showed in this work that the algorithm performs exceptionally well at removing MW foreground contamination from our samples. By comparing our membership probabilities to the RV-confirmed dwarf members for which we have data, we confirmed that we obtain reasonable purity down to $P_{max} \geq~$10\% and determined that a purity of $>$85\% can be obtained for stars with $P_{max} \geq~$50\%. However, in fields where the MW foreground contamination dominates significantly and the system in question is particularly small, we find that the algorithm is unable to effectively converge on spatial parameters for the outer profile (see Section \ref{sect:note_params}). 

This algorithm proves to be very efficient at detecting true members of dwarf galaxies, even those at large radial distances. Already, our source list of individual resolved members has proven useful for detailed study of these interesting dwarfs (Boo1, UMa1, and ComaBer in \citealt{waller2023}; Boo5 in \citealt{smith2022}; Ret2 in \citealt{hayes2023}; Scl in \citealt{sestito2023_scul}; and UMi in \citealt{sestito2023_umi}). With the availability of deeper photometric campaigns (e.g., LSST) paired with the impressive precision and increasing accuracy of Gaia kinematics in each new data release, we anticipate the detectability of dwarf stellar haloes will be exceptionally promising. We look forward to disentangling the origins of these extended features, with the advent of the Gemini High-resolution Optical SpecTrograph (GHOST; \citealt{pazder2016}) in the near future.


\section*{Acknowledgements}

AWM and JJ acknowledge support of a Natural Sciences and Engineering Research Council of Canada Discovery Grant RGPIN2018-03853. FS thanks the Dr. Margaret "Marmie" Perkins Hess postdoctoral fellowship for funding his work at the University of Victoria.

The work detailed above was conducted at the University of Victoria in Victoria, British Columbia, as well as in the Township of Esquimalt in Greater Victoria. We acknowledge with respect the Lekwungen peoples on whose unceded traditional territory the university stands, and the Songhees, Esquimalt and WS\'{A}NE\'{C} peoples whose historical relationships with the land continue to this day.

This work has made use of data from the European Space Agency (ESA) mission $\textit{Gaia}$ (\url{https://www.cosmos.esa.int/gaia}), processed by the $\textit{Gaia}$ Data Processing and Analysis Consortium (DPAC, \url{https://www.cosmos.esa.int/web/gaia/dpac/consortium}). Funding for the DPAC has been provided by national institutions, in particular the institutions participating in the Gaia Multilateral Agreement.

Funding for the Sloan Digital Sky 
Survey IV has been provided by the 
Alfred P. Sloan Foundation, the U.S. 
Department of Energy Office of 
Science, and the Participating 
Institutions. 

SDSS-IV acknowledges support and 
resources from the Center for High 
Performance Computing  at the 
University of Utah. The SDSS 
website is www.sdss4.org.

SDSS-IV is managed by the 
Astrophysical Research Consortium 
for the Participating Institutions 
of the SDSS Collaboration including 
the Brazilian Participation Group, 
the Carnegie Institution for Science, 
Carnegie Mellon University, Center for 
Astrophysics | Harvard \& 
Smithsonian, the Chilean Participation 
Group, the French Participation Group, 
Instituto de Astrof\'isica de 
Canarias, The Johns Hopkins 
University, Kavli Institute for the 
Physics and Mathematics of the 
Universe (IPMU) / University of 
Tokyo, the Korean Participation Group, 
Lawrence Berkeley National Laboratory, 
Leibniz Institut f\"ur Astrophysik 
Potsdam (AIP),  Max-Planck-Institut 
f\"ur Astronomie (MPIA Heidelberg), 
Max-Planck-Institut f\"ur 
Astrophysik (MPA Garching), 
Max-Planck-Institut f\"ur 
Extraterrestrische Physik (MPE), 
National Astronomical Observatories of 
China, New Mexico State University, 
New York University, University of 
Notre Dame, Observat\'ario 
Nacional / MCTI, The Ohio State 
University, Pennsylvania State 
University, Shanghai 
Astronomical Observatory, United 
Kingdom Participation Group, 
Universidad Nacional Aut\'onoma 
de M\'exico, University of Arizona, 
University of Colorado Boulder, 
University of Oxford, University of 
Portsmouth, University of Utah, 
University of Virginia, University 
of Washington, University of 
Wisconsin, Vanderbilt University, 
and Yale University.

\section*{Data Availability}


The data underlying this article is from Gaia DR3. The membership catalogues derived as part of this work will be shared upon reasonable request, understanding that some restrictions may apply due to ongoing observational follow-up studies.








\bibliographystyle{mnras}
\bibliography{MAIN.bbl}

\begin{thebibliography}{}
\makeatletter
\relax
\def\mn@urlcharsother{\let\do\@makeother \do\$\do\&\do\#\do\^\do\_\do\%\do\~}
\def\mn@doi{\begingroup\mn@urlcharsother \@ifnextchar [ {\mn@doi@}
  {\mn@doi@[]}}
\def\mn@doi@[#1]#2{\def\@tempa{#1}\ifx\@tempa\@empty \href
  {http://dx.doi.org/#2} {doi:#2}\else \href {http://dx.doi.org/#2} {#1}\fi
  \endgroup}
\def\mn@eprint#1#2{\mn@eprint@#1:#2::\@nil}
\def\mn@eprint@arXiv#1{\href {http://arxiv.org/abs/#1} {{\tt arXiv:#1}}}
\def\mn@eprint@dblp#1{\href {http://dblp.uni-trier.de/rec/bibtex/#1.xml}
  {dblp:#1}}
\def\mn@eprint@#1:#2:#3:#4\@nil{\def\@tempa {#1}\def\@tempb {#2}\def\@tempc
  {#3}\ifx \@tempc \@empty \let \@tempc \@tempb \let \@tempb \@tempa \fi \ifx
  \@tempb \@empty \def\@tempb {arXiv}\fi \@ifundefined
  {mn@eprint@\@tempb}{\@tempb:\@tempc}{\expandafter \expandafter \csname
  mn@eprint@\@tempb\endcsname \expandafter{\@tempc}}}

\bibitem[\protect\citeauthoryear{{Abdurro'uf} et~al.,}{{Abdurro'uf}
  et~al.}{2022}]{APOGEE_DR17_abdurrouf2022}
{Abdurro'uf} et~al., 2022, \mn@doi [\apjs] {10.3847/1538-4365/ac4414}, \href
  {https://ui.adsabs.harvard.edu/abs/2022ApJS..259...35A} {259, 35}

\bibitem[\protect\citeauthoryear{{Ad{\'e}n} et~al.,}{{Ad{\'e}n}
  et~al.}{2009}]{aden2009}
{Ad{\'e}n} D.,  et~al., 2009, \mn@doi [\aap] {10.1051/0004-6361/200912718},
  \href {https://ui.adsabs.harvard.edu/abs/2009A&A...506.1147A} {506, 1147}

\bibitem[\protect\citeauthoryear{{Battaglia}, {Helmi}, {Tolstoy}, {Irwin},
  {Hill}  \& {Jablonka}}{{Battaglia} et~al.}{2008}]{battaglia2008}
{Battaglia} G.,  {Helmi} A.,  {Tolstoy} E.,  {Irwin} M.,  {Hill} V.,
  {Jablonka} P.,  2008, \mn@doi [\apjl] {10.1086/590179}, \href
  {https://ui.adsabs.harvard.edu/abs/2008ApJ...681L..13B} {681, L13}

\bibitem[\protect\citeauthoryear{{Battaglia}, {Taibi}, {Thomas}  \&
  {Fritz}}{{Battaglia} et~al.}{2022}]{battaglia2022}
{Battaglia} G.,  {Taibi} S.,  {Thomas} G.~F.,   {Fritz} T.~K.,  2022, \mn@doi
  [\aap] {10.1051/0004-6361/202141528}, \href
  {https://ui.adsabs.harvard.edu/abs/2022A&A...657A..54B} {657, A54}

\bibitem[\protect\citeauthoryear{{Beaton} et~al.,}{{Beaton}
  et~al.}{2021}]{beaton2021}
{Beaton} R.~L.,  et~al., 2021, \mn@doi [\aj] {10.3847/1538-3881/ac260c}, \href
  {https://ui.adsabs.harvard.edu/abs/2021AJ....162..302B} {162, 302}

\bibitem[\protect\citeauthoryear{{Bechtol} et~al.,}{{Bechtol}
  et~al.}{2015}]{bechtol2015}
{Bechtol} K.,  et~al., 2015, \mn@doi [\apj] {10.1088/0004-637X/807/1/50}, \href
  {https://ui.adsabs.harvard.edu/abs/2015ApJ...807...50B} {807, 50}

\bibitem[\protect\citeauthoryear{{Bellazzini}, {Ferraro}, {Origlia}, {Pancino},
  {Monaco}  \& {Oliva}}{{Bellazzini} et~al.}{2002}]{bellazini2002}
{Bellazzini} M.,  {Ferraro} F.~R.,  {Origlia} L.,  {Pancino} E.,  {Monaco} L.,
   {Oliva} E.,  2002, \mn@doi [\aj] {10.1086/344794}, \href
  {https://ui.adsabs.harvard.edu/abs/2002AJ....124.3222B} {124, 3222}

\bibitem[\protect\citeauthoryear{{Belokurov} et~al.,}{{Belokurov}
  et~al.}{2007a}]{belokurov2007_cats}
{Belokurov} V.,  et~al., 2007a, \mn@doi [\apj] {10.1086/509718}, \href
  {https://ui.adsabs.harvard.edu/abs/2007ApJ...654..897B} {654, 897}

\bibitem[\protect\citeauthoryear{{Belokurov} et~al.,}{{Belokurov}
  et~al.}{2007b}]{belokurov2007}
{Belokurov} V.,  et~al., 2007b, \mn@doi [\apj] {10.1086/511302}, \href
  {https://ui.adsabs.harvard.edu/abs/2007ApJ...658..337B} {658, 337}

\bibitem[\protect\citeauthoryear{{Belokurov} et~al.,}{{Belokurov}
  et~al.}{2009}]{belokurov2009}
{Belokurov} V.,  et~al., 2009, \mn@doi [\mnras]
  {10.1111/j.1365-2966.2009.15106.x}, \href
  {https://ui.adsabs.harvard.edu/abs/2009MNRAS.397.1748B} {397, 1748}

\bibitem[\protect\citeauthoryear{{Ben{\'\i}tez-Llambay}, {Navarro}, {Abadi},
  {Gottl{\"o}ber}, {Yepes}, {Hoffman}  \& {Steinmetz}}{{Ben{\'\i}tez-Llambay}
  et~al.}{2016}]{benitez-llambay2016}
{Ben{\'\i}tez-Llambay} A.,  {Navarro} J.~F.,  {Abadi} M.~G.,  {Gottl{\"o}ber}
  S.,  {Yepes} G.,  {Hoffman} Y.,   {Steinmetz} M.,  2016, \mn@doi [\mnras]
  {10.1093/mnras/stv2722}, \href
  {https://ui.adsabs.harvard.edu/abs/2016MNRAS.456.1185B} {456, 1185}

\bibitem[\protect\citeauthoryear{{Bettinelli}, {Hidalgo}, {Cassisi},
  {Aparicio}, {Piotto}, {Valdes}  \& {Walker}}{{Bettinelli}
  et~al.}{2019}]{bettinelli2019}
{Bettinelli} M.,  {Hidalgo} S.~L.,  {Cassisi} S.,  {Aparicio} A.,  {Piotto} G.,
   {Valdes} F.,   {Walker} A.~R.,  2019, \mn@doi [\mnras]
  {10.1093/mnras/stz1679}, \href
  {https://ui.adsabs.harvard.edu/abs/2019MNRAS.487.5862B} {487, 5862}

\bibitem[\protect\citeauthoryear{{Blanton} et~al.,}{{Blanton}
  et~al.}{2017}]{blanton2017}
{Blanton} M.~R.,  et~al., 2017, \mn@doi [\aj] {10.3847/1538-3881/aa7567}, \href
  {https://ui.adsabs.harvard.edu/abs/2017AJ....154...28B} {154, 28}

\bibitem[\protect\citeauthoryear{{Bowen} \& {Vaughan}}{{Bowen} \&
  {Vaughan}}{1973}]{bowen_vaughan1973}
{Bowen} I.~S.,  {Vaughan} A.~H. J.,  1973, \mn@doi [\ao]
  {10.1364/AO.12.001430}, \href
  {https://ui.adsabs.harvard.edu/abs/1973ApOpt..12.1430B} {12, 1430}

\bibitem[\protect\citeauthoryear{{Caldwell} et~al.,}{{Caldwell}
  et~al.}{2017}]{caldwell2017}
{Caldwell} N.,  et~al., 2017, \mn@doi [\apj] {10.3847/1538-4357/aa688e}, \href
  {https://ui.adsabs.harvard.edu/abs/2017ApJ...839...20C} {839, 20}

\bibitem[\protect\citeauthoryear{{Cantu} et~al.,}{{Cantu}
  et~al.}{2021}]{cantu2021}
{Cantu} S.~A.,  et~al., 2021, \mn@doi [\apj] {10.3847/1538-4357/ac0443}, \href
  {https://ui.adsabs.harvard.edu/abs/2021ApJ...916...81C} {916, 81}

\bibitem[\protect\citeauthoryear{{Carlin} \& {Sand}}{{Carlin} \&
  {Sand}}{2018}]{carlin2018}
{Carlin} J.~L.,  {Sand} D.~J.,  2018, \mn@doi [\apj]
  {10.3847/1538-4357/aad8c1}, \href
  {https://ui.adsabs.harvard.edu/abs/2018ApJ...865....7C} {865, 7}

\bibitem[\protect\citeauthoryear{{Carlin}, {Grillmair}, {Mu{\~n}oz}, {Nidever}
  \& {Majewski}}{{Carlin} et~al.}{2009}]{carlin2009}
{Carlin} J.~L.,  {Grillmair} C.~J.,  {Mu{\~n}oz} R.~R.,  {Nidever} D.~L.,
  {Majewski} S.~R.,  2009, \mn@doi [\apjl] {10.1088/0004-637X/702/1/L9}, \href
  {https://ui.adsabs.harvard.edu/abs/2009ApJ...702L...9C} {702, L9}

\bibitem[\protect\citeauthoryear{{Cerny} et~al.,}{{Cerny}
  et~al.}{2021a}]{cerny2021_delve2}
{Cerny} W.,  et~al., 2021a, \mn@doi [\apj] {10.3847/1538-4357/abe1af}, \href
  {https://ui.adsabs.harvard.edu/abs/2021ApJ...910...18C} {910, 18}

\bibitem[\protect\citeauthoryear{{Cerny} et~al.,}{{Cerny}
  et~al.}{2021b}]{cerny2021}
{Cerny} W.,  et~al., 2021b, \mn@doi [\apjl] {10.3847/2041-8213/ac2d9a}, \href
  {https://ui.adsabs.harvard.edu/abs/2021ApJ...920L..44C} {920, L44}

\bibitem[\protect\citeauthoryear{{Cerny} et~al.,}{{Cerny}
  et~al.}{2022}]{cerny2022}
{Cerny} W.,  et~al., 2022, arXiv e-prints, \href
  {https://ui.adsabs.harvard.edu/abs/2022arXiv220912422C} {p. arXiv:2209.12422}

\bibitem[\protect\citeauthoryear{{Cerny} et~al.,}{{Cerny}
  et~al.}{2023}]{cerny2023}
{Cerny} W.,  et~al., 2023, \mn@doi [\apj] {10.3847/1538-4357/aca1c3}, \href
  {https://ui.adsabs.harvard.edu/abs/2023ApJ...942..111C} {942, 111}

\bibitem[\protect\citeauthoryear{{Chambers} et~al.,}{{Chambers}
  et~al.}{2016}]{chambers2016}
{Chambers} K.~C.,  et~al., 2016, arXiv e-prints, \href
  {https://ui.adsabs.harvard.edu/abs/2016arXiv161205560C} {p. arXiv:1612.05560}

\bibitem[\protect\citeauthoryear{{Chiti}, {Frebel}, {Ji}, {Jerjen}, {Kim}  \&
  {Norris}}{{Chiti} et~al.}{2018}]{chiti2018}
{Chiti} A.,  {Frebel} A.,  {Ji} A.~P.,  {Jerjen} H.,  {Kim} D.,   {Norris}
  J.~E.,  2018, \mn@doi [\apj] {10.3847/1538-4357/aab4fc}, \href
  {https://ui.adsabs.harvard.edu/abs/2018ApJ...857...74C} {857, 74}

\bibitem[\protect\citeauthoryear{{Chiti} et~al.,}{{Chiti}
  et~al.}{2021}]{chiti2021}
{Chiti} A.,  et~al., 2021, \mn@doi [Nature Astronomy]
  {10.1038/s41550-020-01285-w}, \href
  {https://ui.adsabs.harvard.edu/abs/2021NatAs...5..392C} {5, 392}

\bibitem[\protect\citeauthoryear{{Chiti} et~al.,}{{Chiti}
  et~al.}{2023}]{chiti2023}
{Chiti} A.,  et~al., 2023, \mn@doi [\aj] {10.3847/1538-3881/aca416}, \href
  {https://ui.adsabs.harvard.edu/abs/2023AJ....165...55C} {165, 55}

\bibitem[\protect\citeauthoryear{{Correnti}, {Bellazzini}  \&
  {Ferraro}}{{Correnti} et~al.}{2009}]{correnti2009}
{Correnti} M.,  {Bellazzini} M.,   {Ferraro} F.~R.,  2009, \mn@doi [\mnras]
  {10.1111/j.1745-3933.2009.00677.x}, \href
  {https://ui.adsabs.harvard.edu/abs/2009MNRAS.397L..26C} {397, L26}

\bibitem[\protect\citeauthoryear{{Dall'Ora} et~al.,}{{Dall'Ora}
  et~al.}{2006}]{dall'ora2006}
{Dall'Ora} M.,  et~al., 2006, \mn@doi [\apjl] {10.1086/510665}, \href
  {https://ui.adsabs.harvard.edu/abs/2006ApJ...653L.109D} {653, L109}

\bibitem[\protect\citeauthoryear{{Deason}, {Belokurov}, {Evans}, {Watkins}  \&
  {Fellhauer}}{{Deason} et~al.}{2012}]{deason2012}
{Deason} A.~J.,  {Belokurov} V.,  {Evans} N.~W.,  {Watkins} L.~L.,
  {Fellhauer} M.,  2012, \mn@doi [\mnras] {10.1111/j.1745-3933.2012.01314.x},
  \href {https://ui.adsabs.harvard.edu/abs/2012MNRAS.425L.101D} {425, L101}

\bibitem[\protect\citeauthoryear{{Deason}, {Bose}, {Fattahi}, {Amorisco},
  {Hellwing}  \& {Frenk}}{{Deason} et~al.}{2022}]{deason2022}
{Deason} A.~J.,  {Bose} S.,  {Fattahi} A.,  {Amorisco} N.~C.,  {Hellwing} W.,
  {Frenk} C.~S.,  2022, \mn@doi [\mnras] {10.1093/mnras/stab3524}, \href
  {https://ui.adsabs.harvard.edu/abs/2022MNRAS.511.4044D} {511, 4044}

\bibitem[\protect\citeauthoryear{{Drlica-Wagner} et~al.,}{{Drlica-Wagner}
  et~al.}{2015}]{drlica-wagner2015}
{Drlica-Wagner} A.,  et~al., 2015, \mn@doi [\apj]
  {10.1088/0004-637X/813/2/109}, \href
  {https://ui.adsabs.harvard.edu/abs/2015ApJ...813..109D} {813, 109}

\bibitem[\protect\citeauthoryear{{Erkal} et~al.,}{{Erkal}
  et~al.}{2018}]{erkal2018}
{Erkal} D.,  et~al., 2018, \mn@doi [\mnras] {10.1093/mnras/sty2518}, \href
  {https://ui.adsabs.harvard.edu/abs/2018MNRAS.481.3148E} {481, 3148}

\bibitem[\protect\citeauthoryear{{Erkal} et~al.,}{{Erkal}
  et~al.}{2019}]{erkal2019}
{Erkal} D.,  et~al., 2019, \mn@doi [\mnras] {10.1093/mnras/stz1371}, \href
  {https://ui.adsabs.harvard.edu/abs/2019MNRAS.487.2685E} {487, 2685}

\bibitem[\protect\citeauthoryear{{Filion} \& {Wyse}}{{Filion} \&
  {Wyse}}{2021}]{filion2021}
{Filion} C.,  {Wyse} R. F.~G.,  2021, \mn@doi [\apj]
  {10.3847/1538-4357/ac2df1}, \href
  {https://ui.adsabs.harvard.edu/abs/2021ApJ...923..218F} {923, 218}

\bibitem[\protect\citeauthoryear{{Foreman-Mackey}, {Hogg}, {Lang}  \&
  {Goodman}}{{Foreman-Mackey} et~al.}{2013}]{foreman-mackey2013}
{Foreman-Mackey} D.,  {Hogg} D.~W.,  {Lang} D.,   {Goodman} J.,  2013, \mn@doi
  [\pasp] {10.1086/670067}, \href
  {https://ui.adsabs.harvard.edu/abs/2013PASP..125..306F} {125, 306}

\bibitem[\protect\citeauthoryear{{Frebel} \& {Norris}}{{Frebel} \&
  {Norris}}{2015}]{frebel2015}
{Frebel} A.,  {Norris} J.~E.,  2015, \mn@doi [\araa]
  {10.1146/annurev-astro-082214-122423}, \href
  {https://ui.adsabs.harvard.edu/abs/2015ARA&A..53..631F} {53, 631}

\bibitem[\protect\citeauthoryear{{Frebel}, {Lunnan}, {Casey}, {Norris}, {Wyse}
  \& {Gilmore}}{{Frebel} et~al.}{2013}]{frebel2013}
{Frebel} A.,  {Lunnan} R.,  {Casey} A.~R.,  {Norris} J.~E.,  {Wyse} R. F.~G.,
  {Gilmore} G.,  2013, \mn@doi [\apj] {10.1088/0004-637X/771/1/39}, \href
  {https://ui.adsabs.harvard.edu/abs/2013ApJ...771...39F} {771, 39}

\bibitem[\protect\citeauthoryear{{Frebel}, {Simon}  \& {Kirby}}{{Frebel}
  et~al.}{2014}]{frebel2014}
{Frebel} A.,  {Simon} J.~D.,   {Kirby} E.~N.,  2014, \mn@doi [\apj]
  {10.1088/0004-637X/786/1/74}, \href
  {https://ui.adsabs.harvard.edu/abs/2014ApJ...786...74F} {786, 74}

\bibitem[\protect\citeauthoryear{{Frebel}, {Norris}, {Gilmore}  \&
  {Wyse}}{{Frebel} et~al.}{2016}]{frebel2016}
{Frebel} A.,  {Norris} J.~E.,  {Gilmore} G.,   {Wyse} R. F.~G.,  2016, \mn@doi
  [\apj] {10.3847/0004-637X/826/2/110}, \href
  {https://ui.adsabs.harvard.edu/abs/2016ApJ...826..110F} {826, 110}

\bibitem[\protect\citeauthoryear{{Frenk}, {White}, {Davis}  \&
  {Efstathiou}}{{Frenk} et~al.}{1988}]{frenk1998}
{Frenk} C.~S.,  {White} S. D.~M.,  {Davis} M.,   {Efstathiou} G.,  1988,
  \mn@doi [\apj] {10.1086/166213}, \href
  {https://ui.adsabs.harvard.edu/abs/1988ApJ...327..507F} {327, 507}

\bibitem[\protect\citeauthoryear{{Fritz}, {Carrera}, {Battaglia}  \&
  {Taibi}}{{Fritz} et~al.}{2019}]{fritz2019}
{Fritz} T.~K.,  {Carrera} R.,  {Battaglia} G.,   {Taibi} S.,  2019, \mn@doi
  [\aap] {10.1051/0004-6361/201833458}, \href
  {https://ui.adsabs.harvard.edu/abs/2019A&A...623A.129F} {623, A129}

\bibitem[\protect\citeauthoryear{{Fu} et~al.,}{{Fu} et~al.}{2018}]{fu2018}
{Fu} S.~W.,  et~al., 2018, \mn@doi [\apj] {10.3847/1538-4357/aad9f9}, \href
  {https://ui.adsabs.harvard.edu/abs/2018ApJ...866...42F} {866, 42}

\bibitem[\protect\citeauthoryear{{Fu}, {Simon}  \& {Alarc{\'o}n Jara}}{{Fu}
  et~al.}{2019}]{fu2019}
{Fu} S.~W.,  {Simon} J.~D.,   {Alarc{\'o}n Jara} A.~G.,  2019, \mn@doi [\apj]
  {10.3847/1538-4357/ab3658}, \href
  {https://ui.adsabs.harvard.edu/abs/2019ApJ...883...11F} {883, 11}

\bibitem[\protect\citeauthoryear{{Gaia Collaboration} et~al.,}{{Gaia
  Collaboration} et~al.}{2016}]{gaia_collab_2016}
{Gaia Collaboration} et~al., 2016, \mn@doi [\aap]
  {10.1051/0004-6361/201629272}, \href
  {https://ui.adsabs.harvard.edu/abs/2016A&A...595A...1G} {595, A1}

\bibitem[\protect\citeauthoryear{{Gaia Collaboration} et~al.,}{{Gaia
  Collaboration} et~al.}{2018a}]{gaia_collab_2018_HRD}
{Gaia Collaboration} et~al., 2018a, \mn@doi [\aap]
  {10.1051/0004-6361/201832843}, \href
  {https://ui.adsabs.harvard.edu/abs/2018A&A...616A..10G} {616, A10}

\bibitem[\protect\citeauthoryear{{Gaia Collaboration} et~al.,}{{Gaia
  Collaboration} et~al.}{2018b}]{gaia_collab_2018_kinematics}
{Gaia Collaboration} et~al., 2018b, \mn@doi [\aap]
  {10.1051/0004-6361/201832698}, \href
  {https://ui.adsabs.harvard.edu/abs/2018A&A...616A..12G} {616, A12}

\bibitem[\protect\citeauthoryear{{Gaia Collaboration} et~al.,}{{Gaia
  Collaboration} et~al.}{2021}]{gaia_edr3_2021}
{Gaia Collaboration} et~al., 2021, \mn@doi [\aap]
  {10.1051/0004-6361/202039657}, \href
  {https://ui.adsabs.harvard.edu/abs/2021A&A...649A...1G} {649, A1}

\bibitem[\protect\citeauthoryear{{Gaia Collaboration} et~al.,}{{Gaia
  Collaboration} et~al.}{2023}]{gaia_collab_2023}
{Gaia Collaboration} et~al., 2023, \mn@doi [\aap]
  {10.1051/0004-6361/202243940}, \href
  {https://ui.adsabs.harvard.edu/abs/2023A&A...674A...1G} {674, A1}

\bibitem[\protect\citeauthoryear{{Geha}, {Willman}, {Simon}, {Strigari},
  {Kirby}, {Law}  \& {Strader}}{{Geha} et~al.}{2009}]{geha2009}
{Geha} M.,  {Willman} B.,  {Simon} J.~D.,  {Strigari} L.~E.,  {Kirby} E.~N.,
  {Law} D.~R.,   {Strader} J.,  2009, \mn@doi [\apj]
  {10.1088/0004-637X/692/2/1464}, \href
  {https://ui.adsabs.harvard.edu/abs/2009ApJ...692.1464G} {692, 1464}

\bibitem[\protect\citeauthoryear{{Girardi}, {Bertelli}, {Bressan}, {Chiosi},
  {Groenewegen}, {Marigo}, {Salasnich}  \& {Weiss}}{{Girardi}
  et~al.}{2002}]{girardi2002}
{Girardi} L.,  {Bertelli} G.,  {Bressan} A.,  {Chiosi} C.,  {Groenewegen}
  M.~A.~T.,  {Marigo} P.,  {Salasnich} B.,   {Weiss} A.,  2002, \mn@doi [\aap]
  {10.1051/0004-6361:20020612}, \href
  {https://ui.adsabs.harvard.edu/abs/2002A&A...391..195G} {391, 195}

\bibitem[\protect\citeauthoryear{{Goater} et~al.,}{{Goater}
  et~al.}{2024}]{goater2024}
{Goater} A.,  et~al., 2024, \mn@doi [\mnras] {10.1093/mnras/stad3354}, \href
  {https://ui.adsabs.harvard.edu/abs/2024MNRAS.527.2403G} {527, 2403}

\bibitem[\protect\citeauthoryear{{Gregory} et~al.,}{{Gregory}
  et~al.}{2020}]{gregory2020}
{Gregory} A.~L.,  et~al., 2020, \mn@doi [\mnras] {10.1093/mnras/staa1553},
  \href {https://ui.adsabs.harvard.edu/abs/2020MNRAS.496.1092G} {496, 1092}

\bibitem[\protect\citeauthoryear{{Grillmair}}{{Grillmair}}{2009}]{grillmair2009}
{Grillmair} C.~J.,  2009, \mn@doi [\apj] {10.1088/0004-637X/693/2/1118}, \href
  {https://ui.adsabs.harvard.edu/abs/2009ApJ...693.1118G} {693, 1118}

\bibitem[\protect\citeauthoryear{{Grillmair} \& {Carlin}}{{Grillmair} \&
  {Carlin}}{2016}]{carlin2016}
{Grillmair} C.~J.,  {Carlin} J.~L.,  2016, in {Newberg} H.~J.,  {Carlin} J.~L.,
   eds,  Astrophysics and Space Science Library Vol. 420, Tidal Streams in the
  Local Group and Beyond. p.~87 (\mn@eprint {arXiv} {1603.08936}),
  \mn@doi{10.1007/978-3-319-19336-6_4}

\bibitem[\protect\citeauthoryear{{Grillmair} \& {Dionatos}}{{Grillmair} \&
  {Dionatos}}{2006}]{grillmair_dionatos2006}
{Grillmair} C.~J.,  {Dionatos} O.,  2006, \mn@doi [\apjl] {10.1086/505111},
  \href {https://ui.adsabs.harvard.edu/abs/2006ApJ...643L..17G} {643, L17}

\bibitem[\protect\citeauthoryear{{Gunn} et~al.,}{{Gunn}
  et~al.}{2006}]{gunn2006}
{Gunn} J.~E.,  et~al., 2006, \mn@doi [\aj] {10.1086/500975}, \href
  {https://ui.adsabs.harvard.edu/abs/2006AJ....131.2332G} {131, 2332}

\bibitem[\protect\citeauthoryear{{Hansen} et~al.,}{{Hansen}
  et~al.}{2017}]{hansen2017}
{Hansen} T.~T.,  et~al., 2017, \mn@doi [\apj] {10.3847/1538-4357/aa634a}, \href
  {https://ui.adsabs.harvard.edu/abs/2017ApJ...838...44H} {838, 44}

\bibitem[\protect\citeauthoryear{{Hargreaves}, {Gilmore}, {Irwin}  \&
  {Carter}}{{Hargreaves} et~al.}{1994}]{hargreaves1994}
{Hargreaves} J.~C.,  {Gilmore} G.,  {Irwin} M.~J.,   {Carter} D.,  1994,
  \mn@doi [\mnras] {10.1093/mnras/271.3.693}, \href
  {https://ui.adsabs.harvard.edu/abs/1994MNRAS.271..693H} {271, 693}

\bibitem[\protect\citeauthoryear{{Harris}}{{Harris}}{1996}]{harris1996}
{Harris} W.~E.,  1996, \mn@doi [\aj] {10.1086/118116}, \href
  {https://ui.adsabs.harvard.edu/abs/1996AJ....112.1487H} {112, 1487}

\bibitem[\protect\citeauthoryear{{Hayes} et~al.,}{{Hayes}
  et~al.}{2023}]{hayes2023}
{Hayes} C.~R.,  et~al., 2023, \mn@doi [arXiv e-prints]
  {10.48550/arXiv.2306.04804}, \href
  {https://ui.adsabs.harvard.edu/abs/2023arXiv230604804H} {p. arXiv:2306.04804}

\bibitem[\protect\citeauthoryear{{Helmi}}{{Helmi}}{2020}]{helmi2020}
{Helmi} A.,  2020, \mn@doi [\araa] {10.1146/annurev-astro-032620-021917}, \href
  {https://ui.adsabs.harvard.edu/abs/2020ARA&A..58..205H} {58, 205}

\bibitem[\protect\citeauthoryear{{Higgs}, {McConnachie}, {Annau}, {Irwin},
  {Battaglia}, {C{\^o}t{\'e}}, {Lewis}  \& {Venn}}{{Higgs}
  et~al.}{2021}]{higgs2021}
{Higgs} C.~R.,  {McConnachie} A.~W.,  {Annau} N.,  {Irwin} M.,  {Battaglia} G.,
   {C{\^o}t{\'e}} P.,  {Lewis} G.~F.,   {Venn} K.,  2021, \mn@doi [\mnras]
  {10.1093/mnras/stab002}, \href
  {https://ui.adsabs.harvard.edu/abs/2021MNRAS.503..176H} {503, 176}

\bibitem[\protect\citeauthoryear{{Hill} et~al.,}{{Hill}
  et~al.}{2019}]{hill2019}
{Hill} V.,  et~al., 2019, \mn@doi [\aap] {10.1051/0004-6361/201833950}, \href
  {https://ui.adsabs.harvard.edu/abs/2019A&A...626A..15H} {626, A15}

\bibitem[\protect\citeauthoryear{{Ibata}, {Bellazzini}, {Thomas}, {Malhan},
  {Martin}, {Famaey}  \& {Siebert}}{{Ibata} et~al.}{2020}]{ibata2020}
{Ibata} R.,  {Bellazzini} M.,  {Thomas} G.,  {Malhan} K.,  {Martin} N.,
  {Famaey} B.,   {Siebert} A.,  2020, \mn@doi [\apjl]
  {10.3847/2041-8213/ab77c7}, \href
  {https://ui.adsabs.harvard.edu/abs/2020ApJ...891L..19I} {891, L19}

\bibitem[\protect\citeauthoryear{{Iorio}, {Nipoti}, {Battaglia}  \&
  {Sollima}}{{Iorio} et~al.}{2019}]{iorio2019}
{Iorio} G.,  {Nipoti} C.,  {Battaglia} G.,   {Sollima} A.,  2019, \mn@doi
  [\mnras] {10.1093/mnras/stz1342}, \href
  {https://ui.adsabs.harvard.edu/abs/2019MNRAS.487.5692I} {487, 5692}

\bibitem[\protect\citeauthoryear{{Irwin} \& {Hatzidimitriou}}{{Irwin} \&
  {Hatzidimitriou}}{1995}]{irwin_hatzidimitriou1995}
{Irwin} M.,  {Hatzidimitriou} D.,  1995, \mn@doi [\mnras]
  {10.1093/mnras/277.4.1354}, \href
  {https://ui.adsabs.harvard.edu/abs/1995MNRAS.277.1354I} {277, 1354}

\bibitem[\protect\citeauthoryear{{Jensen} et~al.,}{{Jensen}
  et~al.}{2021}]{jensen2021}
{Jensen} J.,  et~al., 2021, \mn@doi [\mnras] {10.1093/mnras/stab2325}, \href
  {https://ui.adsabs.harvard.edu/abs/2021MNRAS.507.1923J} {507, 1923}

\bibitem[\protect\citeauthoryear{{Ji}, {Frebel}, {Simon}  \& {Geha}}{{Ji}
  et~al.}{2016a}]{ji2016_boo2}
{Ji} A.~P.,  {Frebel} A.,  {Simon} J.~D.,   {Geha} M.,  2016a, \mn@doi [\apj]
  {10.3847/0004-637X/817/1/41}, \href
  {https://ui.adsabs.harvard.edu/abs/2016ApJ...817...41J} {817, 41}

\bibitem[\protect\citeauthoryear{{Ji}, {Frebel}, {Ezzeddine}  \& {Casey}}{{Ji}
  et~al.}{2016b}]{ji2016}
{Ji} A.~P.,  {Frebel} A.,  {Ezzeddine} R.,   {Casey} A.~R.,  2016b, \mn@doi
  [\apjl] {10.3847/2041-8205/832/1/L3}, \href
  {https://ui.adsabs.harvard.edu/abs/2016ApJ...832L...3J} {832, L3}

\bibitem[\protect\citeauthoryear{{Kacharov} et~al.,}{{Kacharov}
  et~al.}{2017}]{kacharov2017}
{Kacharov} N.,  et~al., 2017, \mn@doi [\mnras] {10.1093/mnras/stw3188}, \href
  {https://ui.adsabs.harvard.edu/abs/2017MNRAS.466.2006K} {466, 2006}

\bibitem[\protect\citeauthoryear{{Kallivayalil}, {van der Marel}, {Besla},
  {Anderson}  \& {Alcock}}{{Kallivayalil} et~al.}{2013}]{kallivayalil2013}
{Kallivayalil} N.,  {van der Marel} R.~P.,  {Besla} G.,  {Anderson} J.,
  {Alcock} C.,  2013, \mn@doi [\apj] {10.1088/0004-637X/764/2/161}, \href
  {https://ui.adsabs.harvard.edu/abs/2013ApJ...764..161K} {764, 161}

\bibitem[\protect\citeauthoryear{{Kirby}, {Boylan-Kolchin}, {Cohen}, {Geha},
  {Bullock}  \& {Kaplinghat}}{{Kirby} et~al.}{2013}]{kirby2013}
{Kirby} E.~N.,  {Boylan-Kolchin} M.,  {Cohen} J.~G.,  {Geha} M.,  {Bullock}
  J.~S.,   {Kaplinghat} M.,  2013, \mn@doi [\apj] {10.1088/0004-637X/770/1/16},
  \href {https://ui.adsabs.harvard.edu/abs/2013ApJ...770...16K} {770, 16}

\bibitem[\protect\citeauthoryear{{Kirby}, {Simon}  \& {Cohen}}{{Kirby}
  et~al.}{2015a}]{kirby2015_hydra_pisc}
{Kirby} E.~N.,  {Simon} J.~D.,   {Cohen} J.~G.,  2015a, \mn@doi [\apj]
  {10.1088/0004-637X/810/1/56}, \href
  {https://ui.adsabs.harvard.edu/abs/2015ApJ...810...56K} {810, 56}

\bibitem[\protect\citeauthoryear{{Kirby}, {Cohen}, {Simon}  \&
  {Guhathakurta}}{{Kirby} et~al.}{2015b}]{kirby2015}
{Kirby} E.~N.,  {Cohen} J.~G.,  {Simon} J.~D.,   {Guhathakurta} P.,  2015b,
  \mn@doi [\apjl] {10.1088/2041-8205/814/1/L7}, \href
  {https://ui.adsabs.harvard.edu/abs/2015ApJ...814L...7K} {814, L7}

\bibitem[\protect\citeauthoryear{{Kirby}, {Cohen}, {Simon}, {Guhathakurta},
  {Thygesen}  \& {Duggan}}{{Kirby} et~al.}{2017}]{kirby2017}
{Kirby} E.~N.,  {Cohen} J.~G.,  {Simon} J.~D.,  {Guhathakurta} P.,  {Thygesen}
  A.~O.,   {Duggan} G.~E.,  2017, \mn@doi [\apj] {10.3847/1538-4357/aa6570},
  \href {https://ui.adsabs.harvard.edu/abs/2017ApJ...838...83K} {838, 83}

\bibitem[\protect\citeauthoryear{{Kleyna}, {Wilkinson}, {Evans}  \&
  {Gilmore}}{{Kleyna} et~al.}{2005}]{kleyna2005}
{Kleyna} J.~T.,  {Wilkinson} M.~I.,  {Evans} N.~W.,   {Gilmore} G.,  2005,
  \mn@doi [\apjl] {10.1086/491654}, \href
  {https://ui.adsabs.harvard.edu/abs/2005ApJ...630L.141K} {630, L141}

\bibitem[\protect\citeauthoryear{{Koch} et~al.,}{{Koch}
  et~al.}{2009}]{koch2009}
{Koch} A.,  et~al., 2009, \mn@doi [\apj] {10.1088/0004-637X/690/1/453}, \href
  {https://ui.adsabs.harvard.edu/abs/2009ApJ...690..453K} {690, 453}

\bibitem[\protect\citeauthoryear{{Koposov} et~al.,}{{Koposov}
  et~al.}{2011}]{koposov2011}
{Koposov} S.~E.,  et~al., 2011, \mn@doi [\apj] {10.1088/0004-637X/736/2/146},
  \href {https://ui.adsabs.harvard.edu/abs/2011ApJ...736..146K} {736, 146}

\bibitem[\protect\citeauthoryear{{Koposov}, {Belokurov}, {Torrealba}  \&
  {Evans}}{{Koposov} et~al.}{2015}]{koposov2015}
{Koposov} S.~E.,  {Belokurov} V.,  {Torrealba} G.,   {Evans} N.~W.,  2015,
  \mn@doi [\apj] {10.1088/0004-637X/805/2/130}, \href
  {https://ui.adsabs.harvard.edu/abs/2015ApJ...805..130K} {805, 130}

\bibitem[\protect\citeauthoryear{{Koposov} et~al.,}{{Koposov}
  et~al.}{2018}]{koposov2018}
{Koposov} S.~E.,  et~al., 2018, \mn@doi [\mnras] {10.1093/mnras/sty1772}, \href
  {https://ui.adsabs.harvard.edu/abs/2018MNRAS.479.5343K} {479, 5343}

\bibitem[\protect\citeauthoryear{{Koposov} et~al.,}{{Koposov}
  et~al.}{2019}]{koposov2019}
{Koposov} S.~E.,  et~al., 2019, \mn@doi [\mnras] {10.1093/mnras/stz457}, \href
  {https://ui.adsabs.harvard.edu/abs/2019MNRAS.485.4726K} {485, 4726}

\bibitem[\protect\citeauthoryear{{Koposov} et~al.,}{{Koposov}
  et~al.}{2022}]{koposov2022}
{Koposov} S.~E.,  et~al., 2022, \mn@doi [arXiv e-prints]
  {10.48550/arXiv.2211.04495}, \href
  {https://ui.adsabs.harvard.edu/abs/2022arXiv221104495K} {p. arXiv:2211.04495}

\bibitem[\protect\citeauthoryear{{Laevens} et~al.,}{{Laevens}
  et~al.}{2015}]{laevens2015}
{Laevens} B. P.~M.,  et~al., 2015, \mn@doi [\apj] {10.1088/0004-637X/813/1/44},
  \href {https://ui.adsabs.harvard.edu/abs/2015ApJ...813...44L} {813, 44}

\bibitem[\protect\citeauthoryear{{Li} et~al.,}{{Li} et~al.}{2017}]{li2017}
{Li} T.~S.,  et~al., 2017, \mn@doi [\apj] {10.3847/1538-4357/aa6113}, \href
  {https://ui.adsabs.harvard.edu/abs/2017ApJ...838....8L} {838, 8}

\bibitem[\protect\citeauthoryear{{Li} et~al.,}{{Li}
  et~al.}{2018a}]{li2018_car_UFDs}
{Li} T.~S.,  et~al., 2018a, \mn@doi [\apj] {10.3847/1538-4357/aab666}, \href
  {https://ui.adsabs.harvard.edu/abs/2018ApJ...857..145L} {857, 145}

\bibitem[\protect\citeauthoryear{{Li} et~al.,}{{Li} et~al.}{2018b}]{li2018}
{Li} T.~S.,  et~al., 2018b, \mn@doi [\apj] {10.3847/1538-4357/aadf91}, \href
  {https://ui.adsabs.harvard.edu/abs/2018ApJ...866...22L} {866, 22}

\bibitem[\protect\citeauthoryear{{Lindegren} et~al.,}{{Lindegren}
  et~al.}{2018}]{lindegren2018}
{Lindegren} L.,  et~al., 2018, \mn@doi [\aap] {10.1051/0004-6361/201832727},
  \href {https://ui.adsabs.harvard.edu/abs/2018A&A...616A...2L} {616, A2}

\bibitem[\protect\citeauthoryear{{Lindegren} et~al.,}{{Lindegren}
  et~al.}{2021a}]{lindegren2021_astrometricsolution}
{Lindegren} L.,  et~al., 2021a, \mn@doi [\aap] {10.1051/0004-6361/202039709},
  \href {https://ui.adsabs.harvard.edu/abs/2021A&A...649A...2L} {649, A2}

\bibitem[\protect\citeauthoryear{{Lindegren} et~al.,}{{Lindegren}
  et~al.}{2021b}]{lindegren2021}
{Lindegren} L.,  et~al., 2021b, \mn@doi [\aap] {10.1051/0004-6361/202039653},
  \href {https://ui.adsabs.harvard.edu/abs/2021A&A...649A...4L} {649, A4}

\bibitem[\protect\citeauthoryear{{Longeard} et~al.,}{{Longeard}
  et~al.}{2018}]{longeard2018}
{Longeard} N.,  et~al., 2018, \mn@doi [\mnras] {10.1093/mnras/sty1986}, \href
  {https://ui.adsabs.harvard.edu/abs/2018MNRAS.480.2609L} {480, 2609}

\bibitem[\protect\citeauthoryear{{Longeard} et~al.,}{{Longeard}
  et~al.}{2020}]{longeard2020}
{Longeard} N.,  et~al., 2020, \mn@doi [\mnras] {10.1093/mnras/stz2854}, \href
  {https://ui.adsabs.harvard.edu/abs/2020MNRAS.491..356L} {491, 356}

\bibitem[\protect\citeauthoryear{{Longeard} et~al.,}{{Longeard}
  et~al.}{2021}]{longeard2021}
{Longeard} N.,  et~al., 2021, \mn@doi [\mnras] {10.1093/mnras/stab604}, \href
  {https://ui.adsabs.harvard.edu/abs/2021MNRAS.503.2754L} {503, 2754}

\bibitem[\protect\citeauthoryear{{Longeard} et~al.,}{{Longeard}
  et~al.}{2022}]{longeard2022}
{Longeard} N.,  et~al., 2022, \mn@doi [\mnras] {10.1093/mnras/stac1827}, \href
  {https://ui.adsabs.harvard.edu/abs/2022MNRAS.516.2348L} {516, 2348}

\bibitem[\protect\citeauthoryear{{Majewski} et~al.,}{{Majewski}
  et~al.}{2017}]{APOGEE_majewski}
{Majewski} S.~R.,  et~al., 2017, \mn@doi [\aj] {10.3847/1538-3881/aa784d},
  \href {https://ui.adsabs.harvard.edu/abs/2017AJ....154...94M} {154, 94}

\bibitem[\protect\citeauthoryear{{Marshall} et~al.,}{{Marshall}
  et~al.}{2019}]{marshall2019}
{Marshall} J.~L.,  et~al., 2019, \mn@doi [\apj] {10.3847/1538-4357/ab3653},
  \href {https://ui.adsabs.harvard.edu/abs/2019ApJ...882..177M} {882, 177}

\bibitem[\protect\citeauthoryear{{Martin}, {Ibata}, {Chapman}, {Irwin}  \&
  {Lewis}}{{Martin} et~al.}{2007}]{martin2007}
{Martin} N.~F.,  {Ibata} R.~A.,  {Chapman} S.~C.,  {Irwin} M.,   {Lewis} G.~F.,
   2007, \mn@doi [\mnras] {10.1111/j.1365-2966.2007.12055.x}, \href
  {https://ui.adsabs.harvard.edu/abs/2007MNRAS.380..281M} {380, 281}

\bibitem[\protect\citeauthoryear{{Martin}, {de Jong}  \& {Rix}}{{Martin}
  et~al.}{2008}]{martin2008}
{Martin} N.~F.,  {de Jong} J. T.~A.,   {Rix} H.-W.,  2008, \mn@doi [\apj]
  {10.1086/590336}, \href
  {https://ui.adsabs.harvard.edu/abs/2008ApJ...684.1075M} {684, 1075}

\bibitem[\protect\citeauthoryear{{Martin} et~al.,}{{Martin}
  et~al.}{2016a}]{martin2016}
{Martin} N.~F.,  et~al., 2016a, \mn@doi [\mnras] {10.1093/mnrasl/slw013}, \href
  {https://ui.adsabs.harvard.edu/abs/2016MNRAS.458L..59M} {458, L59}

\bibitem[\protect\citeauthoryear{{Martin} et~al.,}{{Martin}
  et~al.}{2016b}]{martin2016_tri2}
{Martin} N.~F.,  et~al., 2016b, \mn@doi [\apj] {10.3847/0004-637X/818/1/40},
  \href {https://ui.adsabs.harvard.edu/abs/2016ApJ...818...40M} {818, 40}

\bibitem[\protect\citeauthoryear{{Mart{\'\i}nez-Garc{\'\i}a}, {del Pino}  \&
  {Aparicio}}{{Mart{\'\i}nez-Garc{\'\i}a} et~al.}{2022}]{martinez-garcia2022}
{Mart{\'\i}nez-Garc{\'\i}a} A.~M.,  {del Pino} A.,   {Aparicio} A.,  2022,
  arXiv e-prints, \href {https://ui.adsabs.harvard.edu/abs/2022arXiv220606339M}
  {p. arXiv:2206.06339}

\bibitem[\protect\citeauthoryear{{Mart{\'\i}nez-V{\'a}zquez}
  et~al.,}{{Mart{\'\i}nez-V{\'a}zquez} et~al.}{2019}]{martinez-vazquez2019}
{Mart{\'\i}nez-V{\'a}zquez} C.~E.,  et~al., 2019, \mn@doi [\mnras]
  {10.1093/mnras/stz2609}, \href
  {https://ui.adsabs.harvard.edu/abs/2019MNRAS.490.2183M} {490, 2183}

\bibitem[\protect\citeauthoryear{{Martinez}, {Minor}, {Bullock}, {Kaplinghat},
  {Simon}  \& {Geha}}{{Martinez} et~al.}{2011}]{martinez2011}
{Martinez} G.~D.,  {Minor} Q.~E.,  {Bullock} J.,  {Kaplinghat} M.,  {Simon}
  J.~D.,   {Geha} M.,  2011, \mn@doi [\apj] {10.1088/0004-637X/738/1/55}, \href
  {https://ui.adsabs.harvard.edu/abs/2011ApJ...738...55M} {738, 55}

\bibitem[\protect\citeauthoryear{{Mateo}, {Olszewski}  \& {Walker}}{{Mateo}
  et~al.}{2008}]{mateo2008}
{Mateo} M.,  {Olszewski} E.~W.,   {Walker} M.~G.,  2008, \mn@doi [\apj]
  {10.1086/522326}, \href
  {https://ui.adsabs.harvard.edu/abs/2008ApJ...675..201M} {675, 201}

\bibitem[\protect\citeauthoryear{{Mateu}}{{Mateu}}{2023}]{mateu2023}
{Mateu} C.,  2023, \mn@doi [\mnras] {10.1093/mnras/stad321}, \href
  {https://ui.adsabs.harvard.edu/abs/2023MNRAS.tmp..322M} {}

\bibitem[\protect\citeauthoryear{{McConnachie}}{{McConnachie}}{2012}]{mcconnachie2012}
{McConnachie} A.~W.,  2012, \mn@doi [\aj] {10.1088/0004-6256/144/1/4}, \href
  {https://ui.adsabs.harvard.edu/abs/2012AJ....144....4M} {144, 4}

\bibitem[\protect\citeauthoryear{{McConnachie} \& {Venn}}{{McConnachie} \&
  {Venn}}{2020a}]{McVenn2020_updated}
{McConnachie} A.~W.,  {Venn} K.~A.,  2020a, \mn@doi [Research Notes of the
  American Astronomical Society] {10.3847/2515-5172/abd18b}, \href
  {https://ui.adsabs.harvard.edu/abs/2020RNAAS...4..229M} {4, 229}

\bibitem[\protect\citeauthoryear{{McConnachie} \& {Venn}}{{McConnachie} \&
  {Venn}}{2020b}]{McVenn2020}
{McConnachie} A.~W.,  {Venn} K.~A.,  2020b, \mn@doi [\aj]
  {10.3847/1538-3881/aba4ab}, \href
  {https://ui.adsabs.harvard.edu/abs/2020AJ....160..124M} {160, 124}

\bibitem[\protect\citeauthoryear{{Mo}, {van den Bosch}  \& {White}}{{Mo}
  et~al.}{2010}]{mo2010}
{Mo} H.,  {van den Bosch} F.~C.,   {White} S.,  2010, {Galaxy Formation and
  Evolution}

\bibitem[\protect\citeauthoryear{{Moehler}, {Landsman}, {Sweigart}  \&
  {Grundahl}}{{Moehler} et~al.}{2003}]{moehler2003}
{Moehler} S.,  {Landsman} W.~B.,  {Sweigart} A.~V.,   {Grundahl} F.,  2003,
  \mn@doi [\aap] {10.1051/0004-6361:20030622}, \href
  {https://ui.adsabs.harvard.edu/abs/2003A&A...405..135M} {405, 135}

\bibitem[\protect\citeauthoryear{{Moskowitz} \& {Walker}}{{Moskowitz} \&
  {Walker}}{2020}]{moskowitz2020}
{Moskowitz} A.~G.,  {Walker} M.~G.,  2020, \mn@doi [\apj]
  {10.3847/1538-4357/ab7459}, \href
  {https://ui.adsabs.harvard.edu/abs/2020ApJ...892...27M} {892, 27}

\bibitem[\protect\citeauthoryear{{Mu{\~n}oz}, {Carlin}, {Frinchaboy},
  {Nidever}, {Majewski}  \& {Patterson}}{{Mu{\~n}oz} et~al.}{2006}]{munoz2006}
{Mu{\~n}oz} R.~R.,  {Carlin} J.~L.,  {Frinchaboy} P.~M.,  {Nidever} D.~L.,
  {Majewski} S.~R.,   {Patterson} R.~J.,  2006, \mn@doi [\apjl]
  {10.1086/508685}, \href
  {https://ui.adsabs.harvard.edu/abs/2006ApJ...650L..51M} {650, L51}

\bibitem[\protect\citeauthoryear{{Mu{\~n}oz}, {C{\^o}t{\'e}}, {Santana},
  {Geha}, {Simon}, {Oyarz{\'u}n}, {Stetson}  \& {Djorgovski}}{{Mu{\~n}oz}
  et~al.}{2018}]{munoz2018}
{Mu{\~n}oz} R.~R.,  {C{\^o}t{\'e}} P.,  {Santana} F.~A.,  {Geha} M.,  {Simon}
  J.~D.,  {Oyarz{\'u}n} G.~A.,  {Stetson} P.~B.,   {Djorgovski} S.~G.,  2018,
  \mn@doi [\apj] {10.3847/1538-4357/aac16b}, \href
  {https://ui.adsabs.harvard.edu/abs/2018ApJ...860...66M} {860, 66}

\bibitem[\protect\citeauthoryear{{Nagasawa} et~al.,}{{Nagasawa}
  et~al.}{2018}]{nagasawa2018}
{Nagasawa} D.~Q.,  et~al., 2018, \mn@doi [\apj] {10.3847/1538-4357/aaa01d},
  \href {https://ui.adsabs.harvard.edu/abs/2018ApJ...852...99N} {852, 99}

\bibitem[\protect\citeauthoryear{{Nidever} et~al.,}{{Nidever}
  et~al.}{2015}]{nidever2015}
{Nidever} D.~L.,  et~al., 2015, \mn@doi [\aj] {10.1088/0004-6256/150/6/173},
  \href {https://ui.adsabs.harvard.edu/abs/2015AJ....150..173N} {150, 173}

\bibitem[\protect\citeauthoryear{{Niederste-Ostholt}, {Belokurov}, {Evans},
  {Gilmore}, {Wyse}  \& {Norris}}{{Niederste-Ostholt}
  et~al.}{2009}]{niederste-osthold2009}
{Niederste-Ostholt} M.,  {Belokurov} V.,  {Evans} N.~W.,  {Gilmore} G.,  {Wyse}
  R.~F.~G.,   {Norris} J.~E.,  2009, \mn@doi [\mnras]
  {10.1111/j.1365-2966.2009.15287.x}, \href
  {https://ui.adsabs.harvard.edu/abs/2009MNRAS.398.1771N} {398, 1771}

\bibitem[\protect\citeauthoryear{{Norris}, {Gilmore}, {Wyse}, {Wilkinson},
  {Belokurov}, {Evans}  \& {Zucker}}{{Norris} et~al.}{2008}]{norris2008}
{Norris} J.~E.,  {Gilmore} G.,  {Wyse} R. F.~G.,  {Wilkinson} M.~I.,
  {Belokurov} V.,  {Evans} N.~W.,   {Zucker} D.~B.,  2008, \mn@doi [\apjl]
  {10.1086/595962}, \href
  {https://ui.adsabs.harvard.edu/abs/2008ApJ...689L.113N} {689, L113}

\bibitem[\protect\citeauthoryear{{Norris}, {Wyse}, {Gilmore}, {Yong}, {Frebel},
  {Wilkinson}, {Belokurov}  \& {Zucker}}{{Norris}
  et~al.}{2010}]{norris2010_boo1_seg1}
{Norris} J.~E.,  {Wyse} R. F.~G.,  {Gilmore} G.,  {Yong} D.,  {Frebel} A.,
  {Wilkinson} M.~I.,  {Belokurov} V.,   {Zucker} D.~B.,  2010, \mn@doi [\apj]
  {10.1088/0004-637X/723/2/1632}, \href
  {https://ui.adsabs.harvard.edu/abs/2010ApJ...723.1632N} {723, 1632}

\bibitem[\protect\citeauthoryear{{Olszewski} \& {Aaronson}}{{Olszewski} \&
  {Aaronson}}{1985}]{olszewski_aaronson1985}
{Olszewski} E.~W.,  {Aaronson} M.,  1985, \mn@doi [\aj] {10.1086/113925}, \href
  {https://ui.adsabs.harvard.edu/abs/1985AJ.....90.2221O} {90, 2221}

\bibitem[\protect\citeauthoryear{{Pace} \& {Li}}{{Pace} \&
  {Li}}{2019}]{pace_li2019}
{Pace} A.~B.,  {Li} T.~S.,  2019, \mn@doi [\apj] {10.3847/1538-4357/ab0aee},
  \href {https://ui.adsabs.harvard.edu/abs/2019ApJ...875...77P} {875, 77}

\bibitem[\protect\citeauthoryear{{Pace} et~al.,}{{Pace}
  et~al.}{2020}]{pace2020}
{Pace} A.~B.,  et~al., 2020, \mn@doi [\mnras] {10.1093/mnras/staa1419}, \href
  {https://ui.adsabs.harvard.edu/abs/2020MNRAS.495.3022P} {495, 3022}

\bibitem[\protect\citeauthoryear{{Pace}, {Erkal}  \& {Li}}{{Pace}
  et~al.}{2022}]{pace2022}
{Pace} A.~B.,  {Erkal} D.,   {Li} T.~S.,  2022, arXiv e-prints, \href
  {https://ui.adsabs.harvard.edu/abs/2022arXiv220505699P} {p. arXiv:2205.05699}

\bibitem[\protect\citeauthoryear{{Palma}, {Majewski}, {Siegel}, {Patterson},
  {Ostheimer}  \& {Link}}{{Palma} et~al.}{2003}]{palma2003}
{Palma} C.,  {Majewski} S.~R.,  {Siegel} M.~H.,  {Patterson} R.~J.,
  {Ostheimer} J.~C.,   {Link} R.,  2003, \mn@doi [\aj] {10.1086/367594}, \href
  {https://ui.adsabs.harvard.edu/abs/2003AJ....125.1352P} {125, 1352}

\bibitem[\protect\citeauthoryear{{Pazder}, {Burley}, {Ireland}, {Robertson},
  {Sheinis}  \& {Zhelem}}{{Pazder} et~al.}{2016}]{pazder2016}
{Pazder} J.,  {Burley} G.,  {Ireland} M.~J.,  {Robertson} G.,  {Sheinis} A.,
  {Zhelem} R.,  2016, in {Evans} C.~J.,  {Simard} L.,   {Takami} H.,  eds,
  Society of Photo-Optical Instrumentation Engineers (SPIE) Conference Series
  Vol. 9908, Ground-based and Airborne Instrumentation for Astronomy VI. p.
  99087F, \mn@doi{10.1117/12.2234366}

\bibitem[\protect\citeauthoryear{{Pe{\~n}arrubia}, {Navarro}  \&
  {McConnachie}}{{Pe{\~n}arrubia} et~al.}{2008}]{penarrubia2008}
{Pe{\~n}arrubia} J.,  {Navarro} J.~F.,   {McConnachie} A.~W.,  2008, \mn@doi
  [\apj] {10.1086/523686}, \href
  {https://ui.adsabs.harvard.edu/abs/2008ApJ...673..226P} {673, 226}

\bibitem[\protect\citeauthoryear{{Piatek}, {Pryor}, {Bristow}, {Olszewski},
  {Harris}, {Mateo}, {Minniti}  \& {Tinney}}{{Piatek}
  et~al.}{2005}]{piatek2005}
{Piatek} S.,  {Pryor} C.,  {Bristow} P.,  {Olszewski} E.~W.,  {Harris} H.~C.,
  {Mateo} M.,  {Minniti} D.,   {Tinney} C.~G.,  2005, \mn@doi [\aj]
  {10.1086/430532}, \href
  {https://ui.adsabs.harvard.edu/abs/2005AJ....130...95P} {130, 95}

\bibitem[\protect\citeauthoryear{{Qi}, {Zivick}, {Pace}, {Riley}  \&
  {Strigari}}{{Qi} et~al.}{2022}]{qi2022}
{Qi} Y.,  {Zivick} P.,  {Pace} A.~B.,  {Riley} A.~H.,   {Strigari} L.~E.,
  2022, \mn@doi [\mnras] {10.1093/mnras/stac805}, \href
  {https://ui.adsabs.harvard.edu/abs/2022MNRAS.512.5601Q} {512, 5601}

\bibitem[\protect\citeauthoryear{{Ramos} et~al.,}{{Ramos}
  et~al.}{2022}]{ramos2022}
{Ramos} P.,  et~al., 2022, \mn@doi [\aap] {10.1051/0004-6361/202142830}, \href
  {https://ui.adsabs.harvard.edu/abs/2022A&A...666A..64R} {666, A64}

\bibitem[\protect\citeauthoryear{{Read}, {Wilkinson}, {Evans}, {Gilmore}  \&
  {Kleyna}}{{Read} et~al.}{2006}]{read2006}
{Read} J.~I.,  {Wilkinson} M.~I.,  {Evans} N.~W.,  {Gilmore} G.,   {Kleyna}
  J.~T.,  2006, \mn@doi [\mnras] {10.1111/j.1365-2966.2005.09959.x}, \href
  {https://ui.adsabs.harvard.edu/abs/2006MNRAS.367..387R} {367, 387}

\bibitem[\protect\citeauthoryear{{Roderick}, {Mackey}, {Jerjen}  \& {Da
  Costa}}{{Roderick} et~al.}{2016}]{roderick2016}
{Roderick} T.~A.,  {Mackey} A.~D.,  {Jerjen} H.,   {Da Costa} G.~S.,  2016,
  \mn@doi [\mnras] {10.1093/mnras/stw1541}, \href
  {https://ui.adsabs.harvard.edu/abs/2016MNRAS.461.3702R} {461, 3702}

\bibitem[\protect\citeauthoryear{{Roederer}, {Pace}, {Placco}, {Caldwell},
  {Koposov}, {Mateo}, {Olszewski}  \& {Walker}}{{Roederer}
  et~al.}{2023}]{roederer2023}
{Roederer} I.~U.,  {Pace} A.~B.,  {Placco} V.~M.,  {Caldwell} N.,  {Koposov}
  S.~E.,  {Mateo} M.,  {Olszewski} E.~W.,   {Walker} M.~G.,  2023, \mn@doi
  [arXiv e-prints] {10.48550/arXiv.2307.02585}, \href
  {https://ui.adsabs.harvard.edu/abs/2023arXiv230702585R} {p. arXiv:2307.02585}

\bibitem[\protect\citeauthoryear{{Santana} et~al.,}{{Santana}
  et~al.}{2021}]{santana2021}
{Santana} F.~A.,  et~al., 2021, \mn@doi [\aj] {10.3847/1538-3881/ac2cbc}, \href
  {https://ui.adsabs.harvard.edu/abs/2021AJ....162..303S} {162, 303}

\bibitem[\protect\citeauthoryear{{Santos-Santos}, {Fattahi}, {Sales}  \&
  {Navarro}}{{Santos-Santos} et~al.}{2021}]{santos-santos2021}
{Santos-Santos} I. M.~E.,  {Fattahi} A.,  {Sales} L.~V.,   {Navarro} J.~F.,
  2021, \mn@doi [\mnras] {10.1093/mnras/stab1020}, \href
  {https://ui.adsabs.harvard.edu/abs/2021MNRAS.504.4551S} {504, 4551}

\bibitem[\protect\citeauthoryear{{Schlegel}, {Finkbeiner}  \&
  {Davis}}{{Schlegel} et~al.}{1998}]{schlegel1998}
{Schlegel} D.~J.,  {Finkbeiner} D.~P.,   {Davis} M.,  1998, \mn@doi [\apj]
  {10.1086/305772}, \href
  {https://ui.adsabs.harvard.edu/abs/1998ApJ...500..525S} {500, 525}

\bibitem[\protect\citeauthoryear{{Sestito}, {Roediger}, {Navarro}, {Jensen},
  {Venn}, {Smith}, {Hayes}  \& {McConnachie}}{{Sestito}
  et~al.}{2023a}]{sestito2023_scul}
{Sestito} F.,  {Roediger} J.,  {Navarro} J.~F.,  {Jensen} J.,  {Venn} K.~A.,
  {Smith} S. E.~T.,  {Hayes} C.,   {McConnachie} A.~W.,  2023a, \mn@doi
  [\mnras] {10.1093/mnras/stad1417}, \href
  {https://ui.adsabs.harvard.edu/abs/2023MNRAS.tmp.1367S} {}

\bibitem[\protect\citeauthoryear{{Sestito} et~al.,}{{Sestito}
  et~al.}{2023b}]{sestito2023_umi}
{Sestito} F.,  et~al., 2023b, \mn@doi [arXiv e-prints]
  {10.48550/arXiv.2301.13214}, \href
  {https://ui.adsabs.harvard.edu/abs/2023arXiv230113214S} {p. arXiv:2301.13214}

\bibitem[\protect\citeauthoryear{{Shapley}}{{Shapley}}{1938}]{shapley1938}
{Shapley} H.,  1938, Harvard College Observatory Bulletin, \href
  {https://ui.adsabs.harvard.edu/abs/1938BHarO.908....1S} {908, 1}

\bibitem[\protect\citeauthoryear{{Shipp} et~al.,}{{Shipp}
  et~al.}{2018}]{shipp2018}
{Shipp} N.,  et~al., 2018, \mn@doi [\apj] {10.3847/1538-4357/aacdab}, \href
  {https://ui.adsabs.harvard.edu/abs/2018ApJ...862..114S} {862, 114}

\bibitem[\protect\citeauthoryear{{Simon}}{{Simon}}{2019}]{simon2019}
{Simon} J.~D.,  2019, \mn@doi [\araa] {10.1146/annurev-astro-091918-104453},
  \href {https://ui.adsabs.harvard.edu/abs/2019ARA&A..57..375S} {57, 375}

\bibitem[\protect\citeauthoryear{{Simon} \& {Geha}}{{Simon} \&
  {Geha}}{2007}]{simon_geha2007}
{Simon} J.~D.,  {Geha} M.,  2007, \mn@doi [\apj] {10.1086/521816}, \href
  {https://ui.adsabs.harvard.edu/abs/2007ApJ...670..313S} {670, 313}

\bibitem[\protect\citeauthoryear{{Simon} et~al.,}{{Simon}
  et~al.}{2011}]{simon2011}
{Simon} J.~D.,  et~al., 2011, \mn@doi [\apj] {10.1088/0004-637X/733/1/46},
  \href {https://ui.adsabs.harvard.edu/abs/2011ApJ...733...46S} {733, 46}

\bibitem[\protect\citeauthoryear{{Simon} et~al.,}{{Simon}
  et~al.}{2015}]{simon2015}
{Simon} J.~D.,  et~al., 2015, \mn@doi [\apj] {10.1088/0004-637X/808/1/95},
  \href {https://ui.adsabs.harvard.edu/abs/2015ApJ...808...95S} {808, 95}

\bibitem[\protect\citeauthoryear{{Simon} et~al.,}{{Simon}
  et~al.}{2017}]{simon2017}
{Simon} J.~D.,  et~al., 2017, \mn@doi [\apj] {10.3847/1538-4357/aa5be7}, \href
  {https://ui.adsabs.harvard.edu/abs/2017ApJ...838...11S} {838, 11}

\bibitem[\protect\citeauthoryear{{Simon} et~al.,}{{Simon}
  et~al.}{2020}]{simon2020}
{Simon} J.~D.,  et~al., 2020, \mn@doi [\apj] {10.3847/1538-4357/ab7ccb}, \href
  {https://ui.adsabs.harvard.edu/abs/2020ApJ...892..137S} {892, 137}

\bibitem[\protect\citeauthoryear{{Smith} et~al.,}{{Smith}
  et~al.}{2022}]{smith2022}
{Smith} S. E.~T.,  et~al., 2022, arXiv e-prints, \href
  {https://ui.adsabs.harvard.edu/abs/2022arXiv220908242S} {p. arXiv:2209.08242}

\bibitem[\protect\citeauthoryear{{Spencer}, {Mateo}, {Walker}  \&
  {Olszewski}}{{Spencer} et~al.}{2017}]{spencer2017}
{Spencer} M.~E.,  {Mateo} M.,  {Walker} M.~G.,   {Olszewski} E.~W.,  2017,
  \mn@doi [\apj] {10.3847/1538-4357/836/2/202}, \href
  {https://ui.adsabs.harvard.edu/abs/2017ApJ...836..202S} {836, 202}

\bibitem[\protect\citeauthoryear{{Spencer}, {Mateo}, {Olszewski}, {Walker},
  {McConnachie}  \& {Kirby}}{{Spencer} et~al.}{2018}]{spencer2018}
{Spencer} M.~E.,  {Mateo} M.,  {Olszewski} E.~W.,  {Walker} M.~G.,
  {McConnachie} A.~W.,   {Kirby} E.~N.,  2018, \mn@doi [\aj]
  {10.3847/1538-3881/aae3e4}, \href
  {https://ui.adsabs.harvard.edu/abs/2018AJ....156..257S} {156, 257}

\bibitem[\protect\citeauthoryear{{Tarumi}, {Yoshida}  \& {Frebel}}{{Tarumi}
  et~al.}{2021}]{tarumi2021}
{Tarumi} Y.,  {Yoshida} N.,   {Frebel} A.,  2021, \mn@doi [\apjl]
  {10.3847/2041-8213/ac024e}, \href
  {https://ui.adsabs.harvard.edu/abs/2021ApJ...914L..10T} {914, L10}

\bibitem[\protect\citeauthoryear{{Tolstoy} et~al.,}{{Tolstoy}
  et~al.}{2004}]{tolstoy2004}
{Tolstoy} E.,  et~al., 2004, \mn@doi [\apjl] {10.1086/427388}, \href
  {https://ui.adsabs.harvard.edu/abs/2004ApJ...617L.119T} {617, L119}

\bibitem[\protect\citeauthoryear{{Tolstoy} et~al.,}{{Tolstoy}
  et~al.}{2023}]{tolstoy2023}
{Tolstoy} E.,  et~al., 2023, \mn@doi [arXiv e-prints]
  {10.48550/arXiv.2304.11980}, \href
  {https://ui.adsabs.harvard.edu/abs/2023arXiv230411980T} {p. arXiv:2304.11980}

\bibitem[\protect\citeauthoryear{{Torrealba} et~al.,}{{Torrealba}
  et~al.}{2016}]{torrealba2016}
{Torrealba} G.,  et~al., 2016, \mn@doi [\mnras] {10.1093/mnras/stw2051}, \href
  {https://ui.adsabs.harvard.edu/abs/2016MNRAS.463..712T} {463, 712}

\bibitem[\protect\citeauthoryear{{Torrealba} et~al.,}{{Torrealba}
  et~al.}{2019}]{torrealba2019}
{Torrealba} G.,  et~al., 2019, \mn@doi [\mnras] {10.1093/mnras/stz1624}, \href
  {https://ui.adsabs.harvard.edu/abs/2019MNRAS.488.2743T} {488, 2743}

\bibitem[\protect\citeauthoryear{{Vargas}, {Geha}, {Kirby}  \&
  {Simon}}{{Vargas} et~al.}{2013}]{vargas2013}
{Vargas} L.~C.,  {Geha} M.,  {Kirby} E.~N.,   {Simon} J.~D.,  2013, \mn@doi
  [\apj] {10.1088/0004-637X/767/2/134}, \href
  {https://ui.adsabs.harvard.edu/abs/2013ApJ...767..134V} {767, 134}

\bibitem[\protect\citeauthoryear{{Walker}, {Mateo}  \& {Olszewski}}{{Walker}
  et~al.}{2009a}]{walker2009}
{Walker} M.~G.,  {Mateo} M.,   {Olszewski} E.~W.,  2009a, \mn@doi [\aj]
  {10.1088/0004-6256/137/2/3100}, \href
  {https://ui.adsabs.harvard.edu/abs/2009AJ....137.3100W} {137, 3100}

\bibitem[\protect\citeauthoryear{{Walker}, {Belokurov}, {Evans}, {Irwin},
  {Mateo}, {Olszewski}  \& {Gilmore}}{{Walker} et~al.}{2009b}]{walker2009_leo5}
{Walker} M.~G.,  {Belokurov} V.,  {Evans} N.~W.,  {Irwin} M.~J.,  {Mateo} M.,
  {Olszewski} E.~W.,   {Gilmore} G.,  2009b, \mn@doi [\apjl]
  {10.1088/0004-637X/694/2/L144}, \href
  {https://ui.adsabs.harvard.edu/abs/2009ApJ...694L.144W} {694, L144}

\bibitem[\protect\citeauthoryear{{Walker}, {Mateo}, {Olszewski}, {Bailey},
  {Koposov}, {Belokurov}  \& {Evans}}{{Walker} et~al.}{2015}]{walker2015}
{Walker} M.~G.,  {Mateo} M.,  {Olszewski} E.~W.,  {Bailey} John~I. I.,
  {Koposov} S.~E.,  {Belokurov} V.,   {Evans} N.~W.,  2015, \mn@doi [\apj]
  {10.1088/0004-637X/808/2/108}, \href
  {https://ui.adsabs.harvard.edu/abs/2015ApJ...808..108W} {808, 108}

\bibitem[\protect\citeauthoryear{{Walker} et~al.,}{{Walker}
  et~al.}{2016}]{walker2016}
{Walker} M.~G.,  et~al., 2016, \mn@doi [\apj] {10.3847/0004-637X/819/1/53},
  \href {https://ui.adsabs.harvard.edu/abs/2016ApJ...819...53W} {819, 53}

\bibitem[\protect\citeauthoryear{{Waller} et~al.,}{{Waller}
  et~al.}{2023}]{waller2023}
{Waller} F.,  et~al., 2023, \mn@doi [\mnras] {10.1093/mnras/stac3563}, \href
  {https://ui.adsabs.harvard.edu/abs/2023MNRAS.519.1349W} {519, 1349}

\bibitem[\protect\citeauthoryear{{Wan} et~al.,}{{Wan} et~al.}{2020}]{wan2020}
{Wan} Z.,  et~al., 2020, \mn@doi [\nat] {10.1038/s41586-020-2483-6}, \href
  {https://ui.adsabs.harvard.edu/abs/2020Natur.583..768W} {583, 768}

\bibitem[\protect\citeauthoryear{{Webster}, {Frebel}  \&
  {Bland-Hawthorn}}{{Webster} et~al.}{2016}]{webster2016}
{Webster} D.,  {Frebel} A.,   {Bland-Hawthorn} J.,  2016, \mn@doi [\apj]
  {10.3847/0004-637X/818/1/80}, \href
  {https://ui.adsabs.harvard.edu/abs/2016ApJ...818...80W} {818, 80}

\bibitem[\protect\citeauthoryear{{Weiler}}{{Weiler}}{2018}]{weiler2018}
{Weiler} M.,  2018, \mn@doi [\aap] {10.1051/0004-6361/201833462}, \href
  {https://ui.adsabs.harvard.edu/abs/2018A&A...617A.138W} {617, A138}

\bibitem[\protect\citeauthoryear{{Wheeler} et~al.,}{{Wheeler}
  et~al.}{2019}]{wheeler2019}
{Wheeler} C.,  et~al., 2019, \mn@doi [\mnras] {10.1093/mnras/stz2887}, \href
  {https://ui.adsabs.harvard.edu/abs/2019MNRAS.490.4447W} {490, 4447}

\bibitem[\protect\citeauthoryear{{White} \& {Rees}}{{White} \&
  {Rees}}{1978}]{white_rees1978}
{White} S.~D.~M.,  {Rees} M.~J.,  1978, \mn@doi [\mnras]
  {10.1093/mnras/183.3.341}, \href
  {https://ui.adsabs.harvard.edu/abs/1978MNRAS.183..341W} {183, 341}

\bibitem[\protect\citeauthoryear{{Willman}, {Geha}, {Strader}, {Strigari},
  {Simon}, {Kirby}, {Ho}  \& {Warres}}{{Willman} et~al.}{2011}]{willman2011}
{Willman} B.,  {Geha} M.,  {Strader} J.,  {Strigari} L.~E.,  {Simon} J.~D.,
  {Kirby} E.,  {Ho} N.,   {Warres} A.,  2011, \mn@doi [\aj]
  {10.1088/0004-6256/142/4/128}, \href
  {https://ui.adsabs.harvard.edu/abs/2011AJ....142..128W} {142, 128}

\bibitem[\protect\citeauthoryear{{Wilson}}{{Wilson}}{1955}]{wilson1955}
{Wilson} A.~G.,  1955, \mn@doi [\pasp] {10.1086/126754}, \href
  {https://ui.adsabs.harvard.edu/abs/1955PASP...67...27W} {67, 27}

\bibitem[\protect\citeauthoryear{{Wilson} et~al.,}{{Wilson}
  et~al.}{2019}]{wilson2019}
{Wilson} J.~C.,  et~al., 2019, \mn@doi [\pasp] {10.1088/1538-3873/ab0075},
  \href {https://ui.adsabs.harvard.edu/abs/2019PASP..131e5001W} {131, 055001}

\bibitem[\protect\citeauthoryear{{Yang}, {Hammer}, {Jiao}  \&
  {Pawlowski}}{{Yang} et~al.}{2022}]{yang2022}
{Yang} Y.,  {Hammer} F.,  {Jiao} Y.,   {Pawlowski} M.~S.,  2022, \mn@doi
  [\mnras] {10.1093/mnras/stac644}, \href
  {https://ui.adsabs.harvard.edu/abs/2022MNRAS.512.4171Y} {512, 4171}

\bibitem[\protect\citeauthoryear{{Zasowski} et~al.,}{{Zasowski}
  et~al.}{2013}]{zasowski2013}
{Zasowski} G.,  et~al., 2013, \mn@doi [\aj] {10.1088/0004-6256/146/4/81}, \href
  {https://ui.adsabs.harvard.edu/abs/2013AJ....146...81Z} {146, 81}

\bibitem[\protect\citeauthoryear{{Zasowski} et~al.,}{{Zasowski}
  et~al.}{2017}]{zasowski2017}
{Zasowski} G.,  et~al., 2017, \mn@doi [\aj] {10.3847/1538-3881/aa8df9}, \href
  {https://ui.adsabs.harvard.edu/abs/2017AJ....154..198Z} {154, 198}

\bibitem[\protect\citeauthoryear{{Zoutendijk} et~al.,}{{Zoutendijk}
  et~al.}{2020}]{zoutendijk2020}
{Zoutendijk} S.~L.,  et~al., 2020, \mn@doi [\aap]
  {10.1051/0004-6361/201936155}, \href
  {https://ui.adsabs.harvard.edu/abs/2020A&A...635A.107Z} {635, A107}

\bibitem[\protect\citeauthoryear{{de los Reyes}, {Kirby}, {Ji}  \&
  {Nu{\~n}ez}}{{de los Reyes} et~al.}{2022}]{delosreyes2022}
{de los Reyes} M. A.~C.,  {Kirby} E.~N.,  {Ji} A.~P.,   {Nu{\~n}ez} E.~H.,
  2022, \mn@doi [\apj] {10.3847/1538-4357/ac332b}, \href
  {https://ui.adsabs.harvard.edu/abs/2022ApJ...925...66D} {925, 66}

\makeatother
\end{thebibliography}





\appendix

\section{Appendix}

\begin{table*}
    \centering
    \renewcommand{\arraystretch}{1.1}
    \begin{tabular}{c|cccccccccc}
         Galaxy & RA ($^{\circ}$) & DEC ($^{\circ}$) & ($m - M$)$_{0}$ & $r_{h}$ (arcmin) & $e = 1 - \frac{b}{a}$ & $\theta^{\circ}$ & $v_{h} (km s^{-1})$ & $\sigma_{v} (km s^{-1})$ & $<[Fe/H]>$ & References \\
         \hline
         Boo3 & 209.300 & 26.800 & 18.34$^{+0.09}_{-0.10}$ & 33.03 $\pm$ 2.50 & 0.33 $\pm$ 0.09 & -81.0 $\pm$ 8.0 & 197.5 $\pm$ 3.8 & 14.0 $\pm$ 3.2 & $-$2.1 $\pm$ 0.2  & (1) \\ 
         
         *Boo5 & 213.911 & 32.912 & 20.0$^{+0.40}_{-0.48}$ &  0.92$^{+0.26}_{-0.19}$ & 0.35$^{+0.17}_{-0.21}$ & 29.8$^{+22.8}_{-21.8}$  & $-$ & $-$ & $-$ & (2); (3) \\

         *DELVE2 & 28.772 & -68.253 & 19.26 $\pm$  0.1 & 1.02$^{+0.18}_{-0.15}$ & 0.03$^{+0.15}_{-0.03}$ & 74$^{+84}_{-40}$ & $-$ & $-$ & $-$ & (4) \\
         
         *Eri4 & 76.438 & $-$9.515 & 19.43$^{+0.1}_{-0.2}$ & 4.9$^{+1.1}_{-0.8}$ &  0.54$^{+0.10}_{-0.14}$  & 65.0$^{+9.0}_{-8.0}$ & $-$ & $-$ & $-$ & (5) \\
         
         Gru1 & 344.177 & $-$50.163 & 20.4$^{+0.2}_{-0.2}$ & 4.16$^{+0.54}_{-0.87}$ & 0.44$^{+0.08}_{-0.1}$ & 153.0$^{+8.0}_{-7.0}$ & $-$140.5$^{+2.4}_{-1.6}$  & $<$9.8 &  $-$1.42$^{+0.55}_{-0.42}$ & (6) \\
         
         *LeoMi & 164.261 & 28.875 & 19.56$^{+0.11}_{-0.19}$ & 1.09$^{+0.37}_{-0.35}$ & $-$ & $-$  & $-$ & $-$ & $-$ & (3) \\
         
         Peg4 & 328.539 & 26.620 & 19.77 $\pm$ 0.1 & 1.6$^{+0.29}_{-0.25}$ & $-$ & $-$ & $-$273.6$^{+1.6}_{-1.5}$ & 3.3$^{+1.7}_{-1.1}$ & $-$2.63$^{+0.26}_{-0.25}$ & (7) \\
         
         *Vir2 & 225.059 & 5.909 & 19.3 $\pm$ 0.22 &  0.74$^{+0.13}_{-0.11}$ & $-$ & $-$ & $-$ & $-$ & $-$ & (3) \\
    \hline         
    \end{tabular}
    \caption{Relevant structural parameters, radial velocities, and metallicities for the newly discovered MW satellites. Updated parameters for Boo3 and Gru1 are additionally provided. Galaxies are marked with * if no systemic RV is reported for the dwarf. Citations: (1) \citet{moskowitz2020}, (2) \citet{smith2022}, (3) \citet{cerny2022}, (4) \citet{cerny2021_delve2}, (5) \citet{cerny2021}, (6) \citet{cantu2021}, (7) \citet{cerny2023}.}
    \label{tab:updates}
\end{table*}

\begin{figure*}
    \centering
    \includegraphics[width=0.9\textwidth]{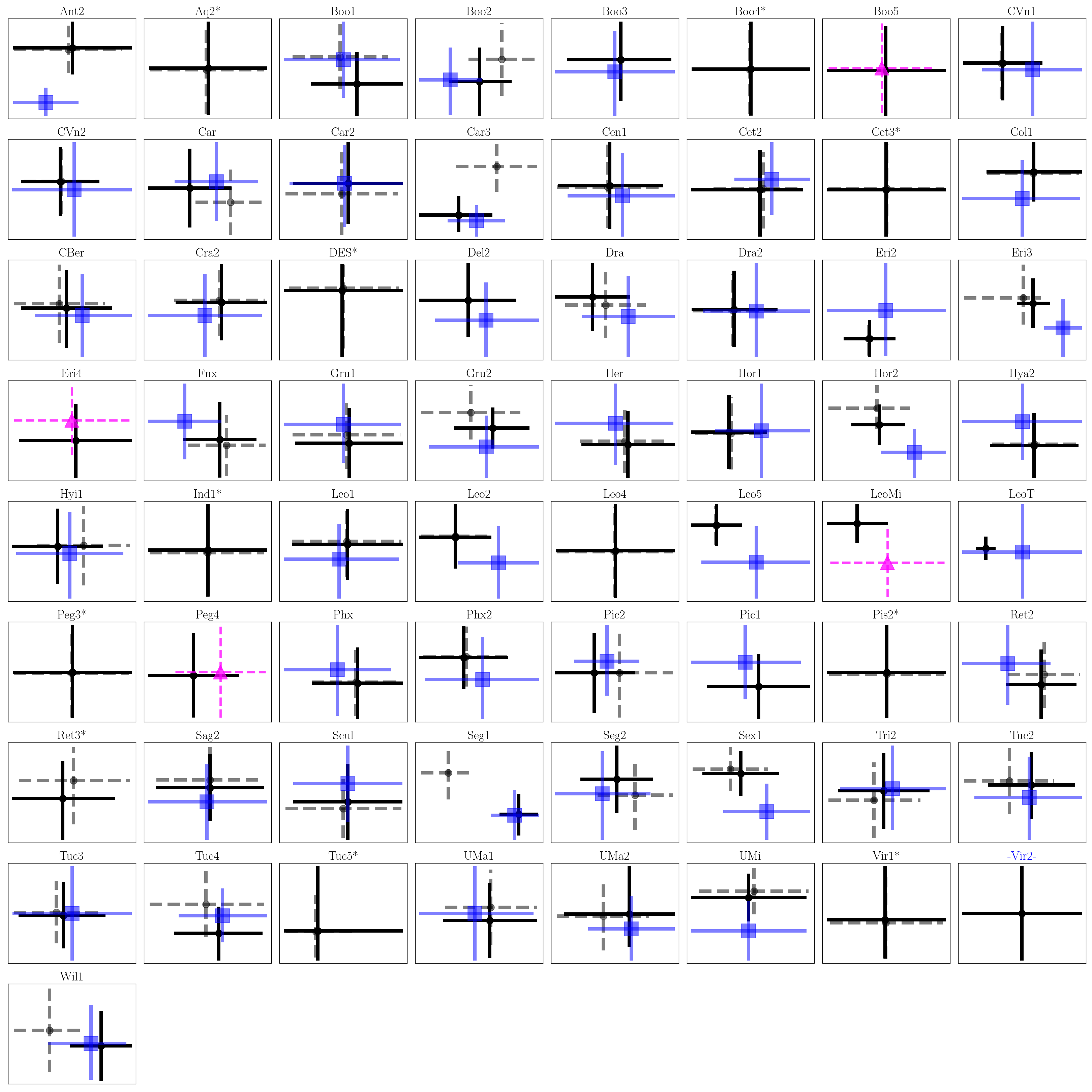}
    \caption{Comparison of the estimated proper motions of each dwarf used in this paper. To confirm internal consistency, we compare our 1-component PM estimates from this work (black) to those reported in \citetalias{McVenn2020_updated} (grey dashed) and additionally to estimates from \citet{battaglia2022} (blue square). All measurements show good consistency, and are within 3-$\sigma$. All errors in figure are reported as 1-$\sigma$ errors. Some systems that are excluded from the \citet{battaglia2022} dataset are those reported to have poor PM estimates (provided with a '*' in the title). For the recently discovered systems which have independently established proper motion estimates, we add the PM measurements from their relevant discovery paper (pink triangle).}
    \label{fig:pm_comparisons}
\end{figure*}

\begin{figure*}
    \centering
    \includegraphics[width=0.8\textwidth]{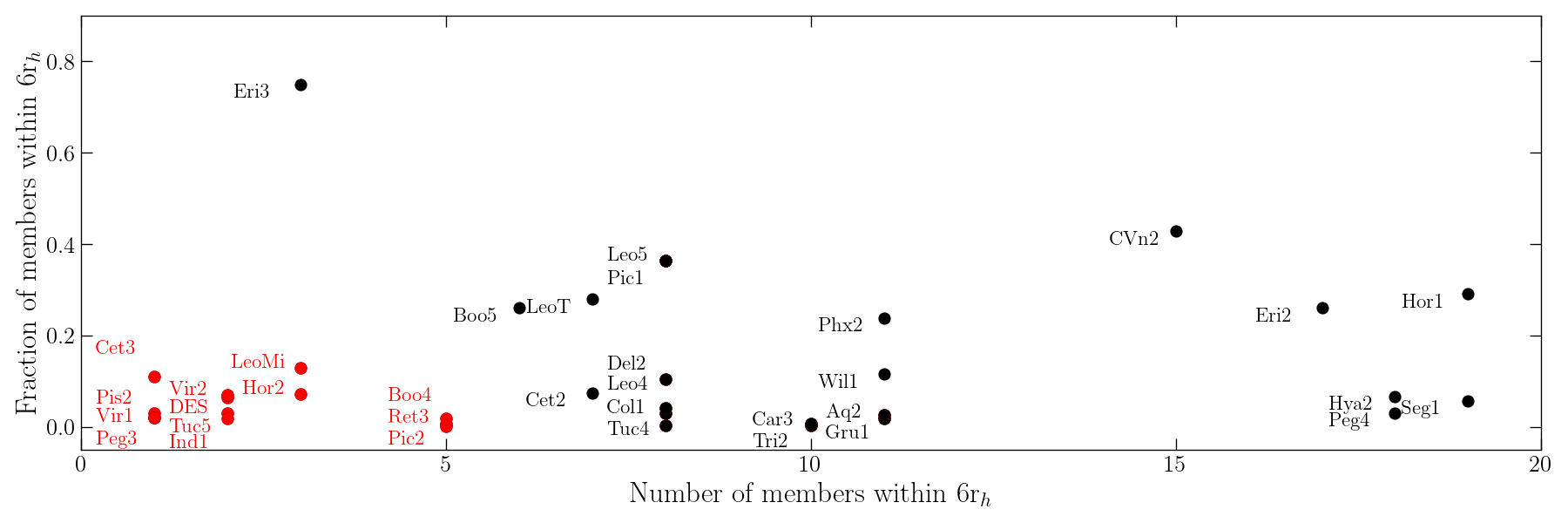}
    \caption{A comparison of the dwarf signal versus the MW foreground contamination, for systems with less than 20 high probability members ($P_{sat}$~$>$~50\%) in the 1-component runs. We find that the addition of free-parameters in the spatial likelihood results in a failure to converge in the 2-component runs due to a) the small number of member stars to begin with ($\lesssim$5), and b) the signal in the field dominates by a factor of $>$5. The systems where the 2-component algorithm fails are highlighted in red text. These dwarfs are excluded from the rest of the analysis.}
    \label{fig:bad_systems}
\end{figure*}

\bsp	
\label{lastpage}
\end{document}